\documentclass[12pt,a4paper]{article}
\usepackage{amsmath}
\usepackage{setspace}
\usepackage[font={footnotesize,it}]{caption}
\usepackage[top=1.4in, bottom=1.4in, left=1.2in, right=1.2in]{geometry}
\usepackage{pdflscape}
\usepackage[ruled,vlined]{algorithm2e}
\DeclareCaptionStyle{italic}[justification=centering]{labelfont={bf},textfont={it},labelsep=colon}
\usepackage{graphicx,psfrag,epsf}
\usepackage{enumerate,bbm,multirow,afterpage}
\usepackage{natbib}
\usepackage{url} 
\usepackage{booktabs, subfig, bm, paralist,mathpazo,tikz,todonotes,longtable,microtype,rotating}
\usepackage[pdftex,colorlinks=true]{hyperref}
\definecolor{darkblue}{rgb}{0,0,.6}
  \definecolor{ao(english)}{rgb}{0.0, 0.5, 0.0}
\hypersetup{citecolor=darkblue,linkcolor=darkblue,urlcolor=darkblue}
\DeclareMathAlphabet\mathbfcal{OMS}{cmsy}{b}{n}
\newcommand{\blind}{1}

\usepackage{authblk}
\usepackage{blindtext}
\graphicspath{{plots/}}

\newsavebox\CBox

\newcommand{\bb}{\boldsymbol}
\newcommand\overmat[2]{
  \makebox[0pt][l]{$\smash{\overbrace{\phantom{%
    \begin{matrix}#2\end{matrix}}}^{\text{$#1$}}}$}#2}

\date{\today}

\begin{document}

\def\spacingset#1{\renewcommand{\baselinestretch}
{#1}\small\normalsize} \spacingset{1}

\if1\blind
{
   \title{\bf Clustering and Forecasting Multiple Functional Time Series} 
   \author[1]{ Chen Tang\thanks{Postal address: Research School of Finance, Actuarial Studies and Statistics, Level 4, Building 26C, Australian National University, Kingsley Street, Acton, Canberra, ACT 2601, Australia; Email: chen.tang@anu.edu.au.}}
   \author[2]{Han Lin Shang}
   \author[1]{Yanrong Yang}
\affil[1]{\small Australian National University}
\affil[2]{\small Macquarie University}
    \date{}
   \maketitle
} \fi

\begin{abstract}
\vspace{.12in}
\spacingset{1.2}
Modeling and forecasting homogeneous age-specific mortality rates of multiple countries could lead to improvements in long-term forecasting. Data fed into joint models are often grouped according to nominal attributes, such as geographic regions, ethnic groups, and socioeconomic status, which may still contain heterogeneity and deteriorate the forecast results. Our paper proposes a novel clustering technique to pursue homogeneity among multiple functional time series based on functional panel data modeling to address this issue. Using a functional panel data model with fixed effects, we can extract common functional time series features. These common features could be decomposed into two components: the functional time trend and the mode of variations of functions (functional pattern). The functional time trend reflects the dynamics across time, while the functional pattern captures the fluctuations within curves. The proposed clustering method searches for homogeneous age-specific mortality rates of multiple countries by accounting for both the modes of variations and the temporal dynamics among curves. We demonstrate that the proposed clustering technique outperforms other existing methods through a Monte Carlo simulation and could handle complicated cases with slow decaying eigenvalues. In empirical data analysis, we find that the clustering results of age-specific mortality rates can be explained by the combination of geographic region, ethnic groups, and socioeconomic status. We further show that our model produces more accurate forecasts than several benchmark methods in forecasting age-specific mortality rates.    
\\

\noindent {\bf\textit{Keywords:}} Functional panel data; multilevel functional data; functional time series; functional principal component analysis; age-specific mortality forecasting.
\end{abstract}

\newpage
\spacingset{1.68}

\section{Introduction}

In actuarial science, accurate forecasting of mortality rates is paramount to insurance companies and governments for pricing, reserving, policy-making, and longevity risk management. Therefore, modeling and forecasting age-specific mortality rates have been an endeavor of many scholars for centuries [A thorough review can be found in \cite{currie2004smoothing}, \cite{girosi2008demographic} and \cite{booth2008mortality}]. Among these, \cite{lee1992modeling} stood out as a milestone, and many extensions were derived \citep{renshaw2003lee,hyndman2007robust,girosi2008demographic,li2013extending,wisniowski2015bayesian}. However, these works focused on forecasting the mortality of a single population. Many scholars, such as \cite{pampel2005forecasting}; \cite{li2005coherent} and \cite{li2013poisson}, have criticized the individual forecasts for increasing divergence in mortality rates in the long-run. In this vein, joint modeling mortality for multiple populations simultaneously has begun to gain popularity in the literature due to the merits of improving forecast accuracy by exploring additional common information from other populations \cite[see, e.g.,][]{shang2016mortality}. This work is motivated by the pursuit of forecast accuracy of mortality through a joint modeling framework. 

However, in the literature for joint modeling mortality data, multiple populations are grouped based on some naive attributes, such as sex, state, ethnic group, and socioeconomic status, which is not convincing from the statistical point of view. The subgroups may still exhibit heterogeneity. According to \cite{boivin2006more}, heterogeneity would deteriorate prediction accuracy as forecasting accuracy relies heavily on efficient modeling and estimation. As a consequence, heterogeneity in the data poses challenges in model efficiency. Therefore, common feature extraction is key in improving forecast accuracy, reducing model variation. This results in the need to search for homogeneous subgroups among multiple populations, where clustering analysis occurs. Cluster analysis, which aims to group homogeneous objects without labeled responses, can be broadly classified into two categories, namely, partitioning \citep[e.g.,][$k$-means]{macqueen1967} and hierarchical clustering \citep[e.g.,][]{ward1963hierarchical}. 

When clustering data with very large or even infinite dimensions (functions), cluster analysis is often coupled with dimension reduction techniques to ease the problem of the ``curse of dimensionality''. Consult \cite{jacques2014functional} for a thorough review of the categorization of existing clustering methods. There has been much literature on clustering discrete functional data, where values of functions are observed over a certain time frame \citep[see][]{muller2005functional}. Many scholars approximated the original data using fewer bases and then applied conventional clustering methods on the fitted coefficients \citep[e.g.][]{abraham2003unsupervised,garcia2005proposal,tarpey2003clustering,serban2005cats}. However, such approaches assume the same basis functions for all clusters, which are problematic as proper basis functions are required to ensure these coefficients reflect the cluster differences adequately. To avoid this problem, \cite{chiou2007functional} proposed $k$-centers functional clustering. This non-parametric clustering method iteratively predicts and updates cluster membership based on estimated cluster structure (both the mean and the mode of variation). \cite{bouveyron2015discriminative} proposed a functional mixture model-based clustering method (funFEM) to identify the common patterns between and within different bike-sharing systems.

Despite the practical usefulness and implementation ease of $k$-centers functional clustering and funFEM, they are only suitable for univariate functional data. While for the multivariate functional data, \cite{bouveyron2011model} extended the high-dimensional data clustering algorithm of \cite{bouveyron2007high} to functional case (funHDDC) and \cite{jacques2014model} proposed an extension of \cite{jacques2012model} to multivariate functional data, both of them assumed a certain Gaussian distribution for the principal component scores. \cite{slimen2018model} proposed a co-clustering algorithm (funLBM) based on the latent block model using a Gaussian model for the functional principal components, which assumes that data into a block are independent and identically distributed. All these works are extensive on the FPC scores and, hence, suffer from assuming the same basis functions for all clusters. 

We aim to improve mortality forecasting accuracy using a clustering approach designed for multiple sets of functional time series. The clustering approach requires us to extract common features representing homogeneous countries and maintain the forecasting ability of the original data as much as we can. This has created the challenging problem of modeling cross-sectional functional time series where the common time trends for all cross-sections are of functional form. From the multilevel functional data model \citep{di2009multilevel,crainiceanu2009generalized,crainiceanu2010bayesian,greven2011longitudinal}, \cite{shang2016mortality} forecast age-specific mortality and life expectancy at birth for a group of populations, where multilevel functional data model captures the common trend and the population-specific trend. However, this model may not be adequate in our clustering analysis from two aspects. Firstly, the common features include the common time trend and the common functional pattern. Secondly, to achieve the forecasting goal, we need to maintain the temporal dependence among the curves within each population.

To this end, we propose a novel functional panel data model with fixed effects to model multiple sets of functional time series, which allows us to use this model to carry out clustering analysis, i.e., search for homogeneous subgroups by extracting the common features. The common features can be further decomposed into the deviation of the country-specific mean from the overall mean, common functional patterns (mode of variations), common time trends, and country-specific time trends. Specifically, the common time trend and population-specific time trend preserve the temporal dynamics among curves in the original data. In precluding the "curse of dimensionality," as well as capturing the temporal dynamics among curves, we incorporate the dynamic version of functional principal component analysis (FPCA) into our model \citep[see, e.g.,][]{hormann2015dynamic,rice2017plug}. Our paper can be seen as an extension of the work of \cite{chiou2007functional} to multiple functional time series that may be intercorrelated. To the best of our knowledge, this has not been pursued so far. As demonstrated in the simulation studies, the proposed clustering method can group multiple functional time series with pre-known labels more accurately than other competing clustering methods.

The rest of the paper is organized as follows. Section~\ref{sec:2} presents the mortality data and shows how they motivate us to develop the proposed model. Section~\ref{sec:3} introduces the functional panel data model with fixed effects. Section~\ref{sec:4} proposes the model-based clustering method. Section~\ref{sec:5} presents the simulation studies to show the robustness and superiority of our proposed method in improving clustering quality. Section~\ref{sec:6} presents an application of the proposed methodology to multi-country mortality data. Our proposed method can achieve more reliable clustering results and, thus, better long-term forecasts. Section~\ref{sec:7} concludes this paper. Our clustering algorithm is included in the \texttt{ftsa} package of \cite{HS21} in R.

\section{Mortality data analysis}\label{sec:2}

In this work, we view mortality rates as functional time series. Age-specific mortality rates have been studied as functional time series by many scholars \citep[e.g.,][]{hyndman2007robust,hyndman2009forecasting,shang2016mortality}. Functional time series consist of a set of random functions observed at regular time intervals. There are two broad categories of functional time series. One is a segmentation of an almost continuous time record into consecutive natural intervals, such as days, months, or quarters, where the continuum of each function is a time variable \citep[e.g.,][]{hormann2012functional}. At the same time, the other type arises when each of the observations in a period represents a continuous function, where the continuum is a variable other than time \citep[e.g.,][]{chiou2009modeling}. The age-specific mortality rate is an example of the latter. Each year, it can be regarded as a function with age being the continuum, and a series of such functions are obtained over a certain time frame. 
\begin{figure}[!htbp]
    \centering
    \subfloat[Female unsmoothed mortality rates in Austria]{{\includegraphics[width=7.34cm]{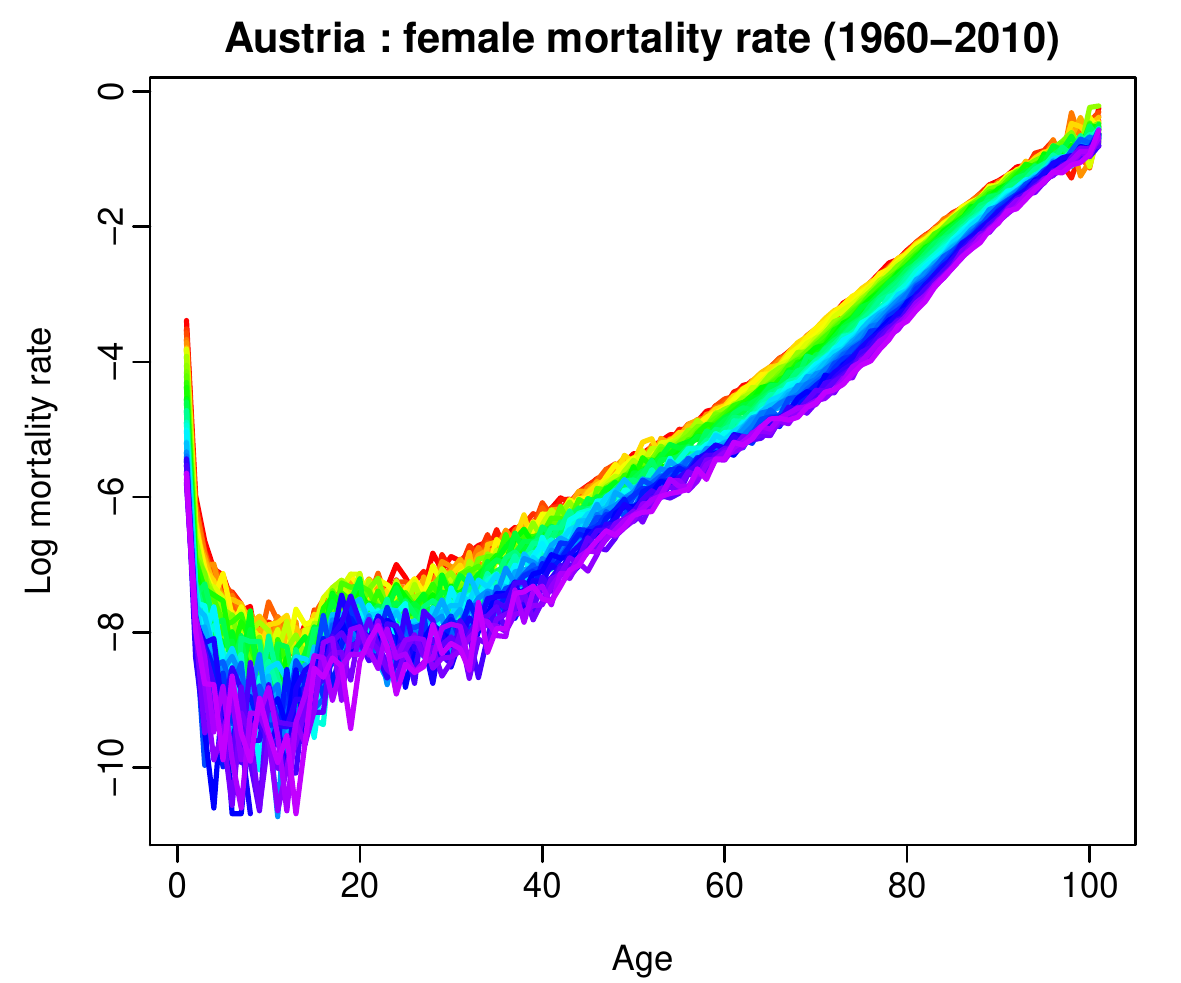}}}
    \quad
    \subfloat[Female smoothed mortality rates in Austria]{{\includegraphics[width=7.34cm]{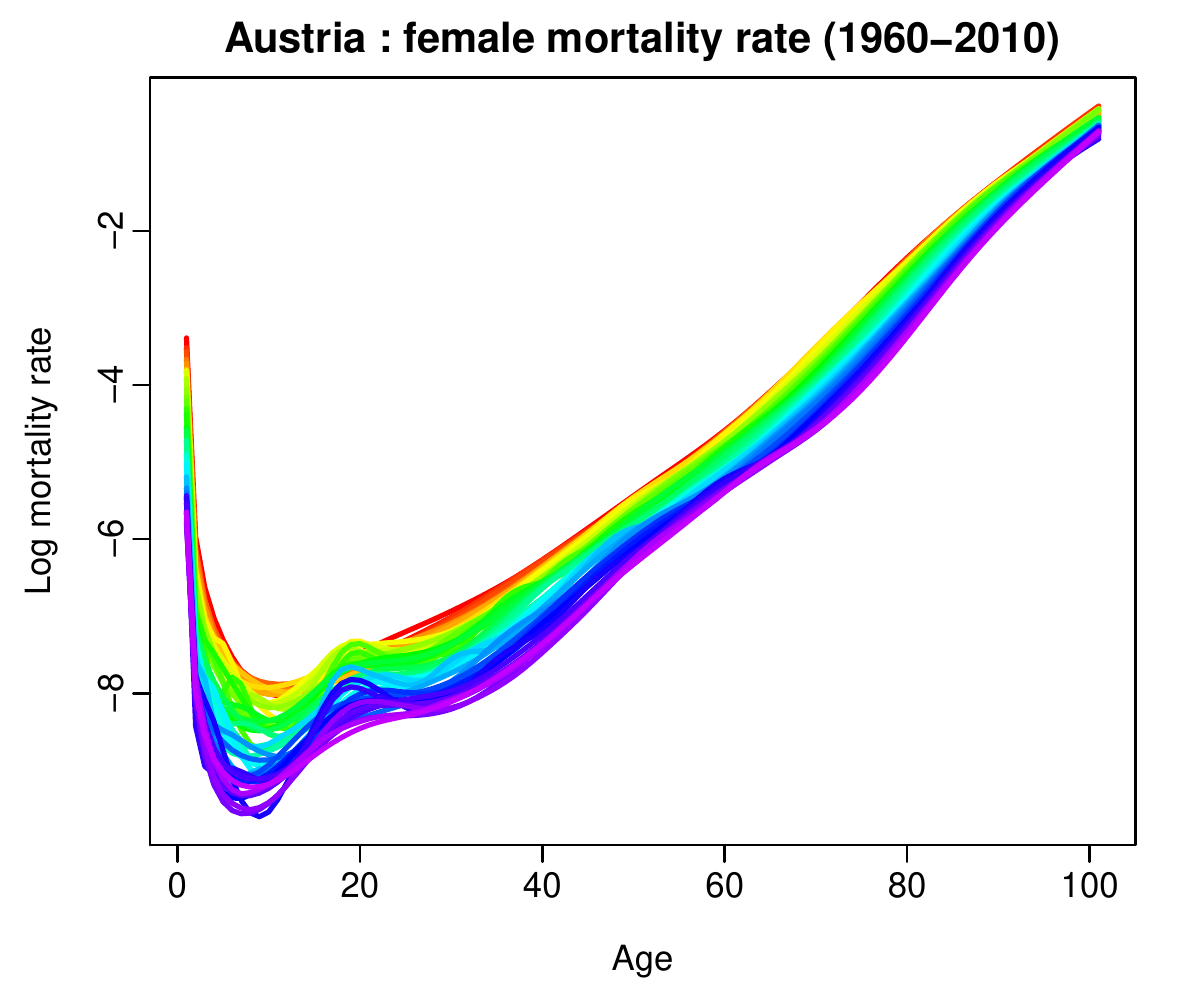} }}
    \caption{Functional time series displays for unsmoothed and smoothed age-specific log-mortality rates in Austria}%
    \label{fig:1}
\end{figure}

\cite{hyndman2010rainbow} developed rainbow plots to show the time ordering of functions in the color order of a rainbow, where the functions from earlier times are in red, and the more recent functions are in purple. Figure~\ref{fig:1} depicts the rainbow plots of unsmoothed and smoothed logarithm of the age-specific central mortality rates for females in Austria from $1960$ to $2010$ side-by-side. As we can see, the patterns are relatively difficult to observe from the unsmoothed rates as the noise masks them. However, if the mortality rates are adequately smoothed, the patterns are obvious.

\begin{figure}[!htbp]
    \centering
    \subfloat[Female smoothed mortality rates in Australia]{{\includegraphics[width=7.15cm]{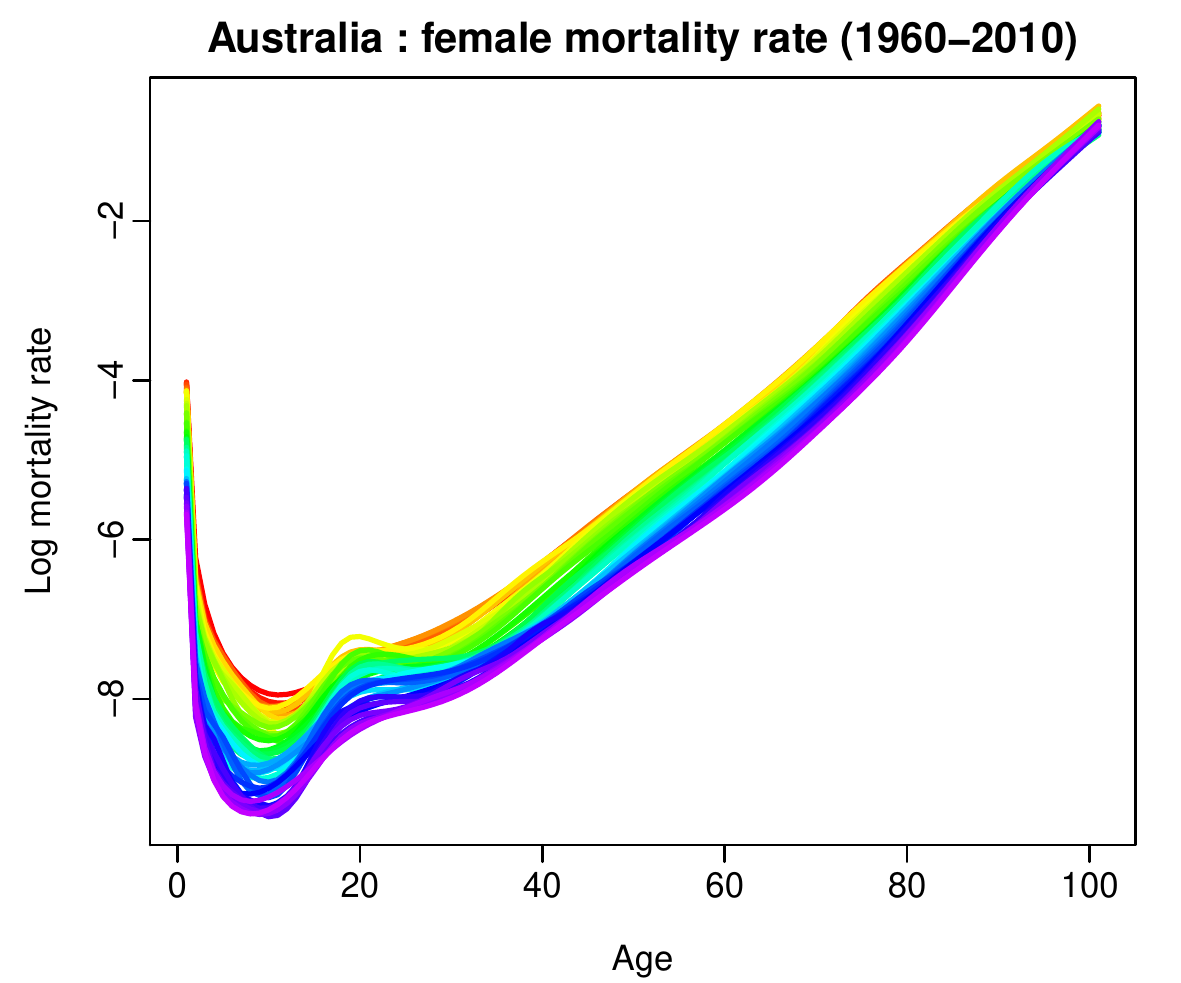} }}
    \qquad
    \subfloat[Female smoothed mortality rates in Austria]{{\includegraphics[width=7.15cm]{autfemale} }}
    \qquad
    \subfloat[Female smoothed mortality rates in Russia]{{\includegraphics[width=7.15cm]{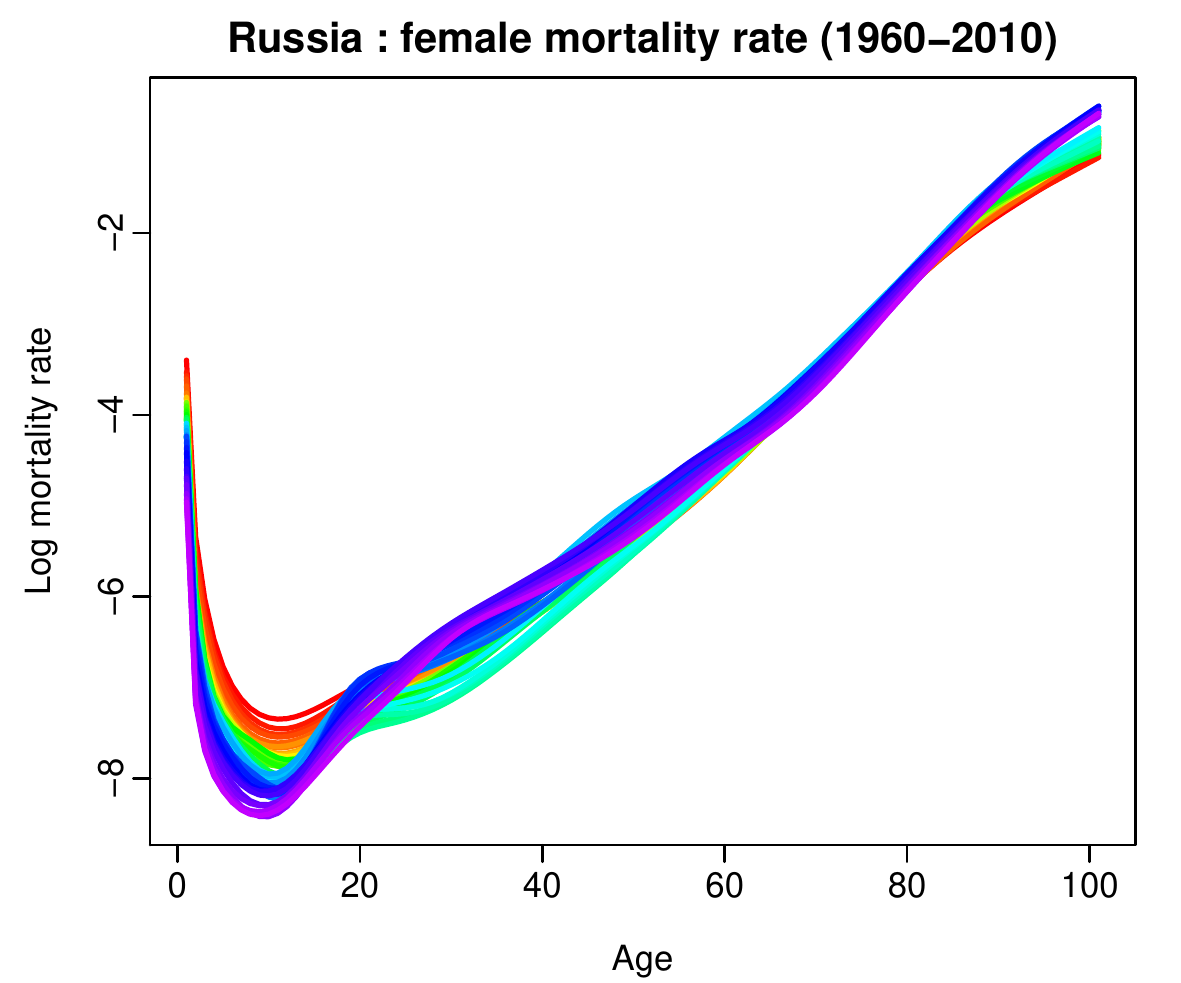} }}
    \qquad
    \subfloat[Female smoothed mortality rates in Ukraine]{{\includegraphics[width=7.15cm]{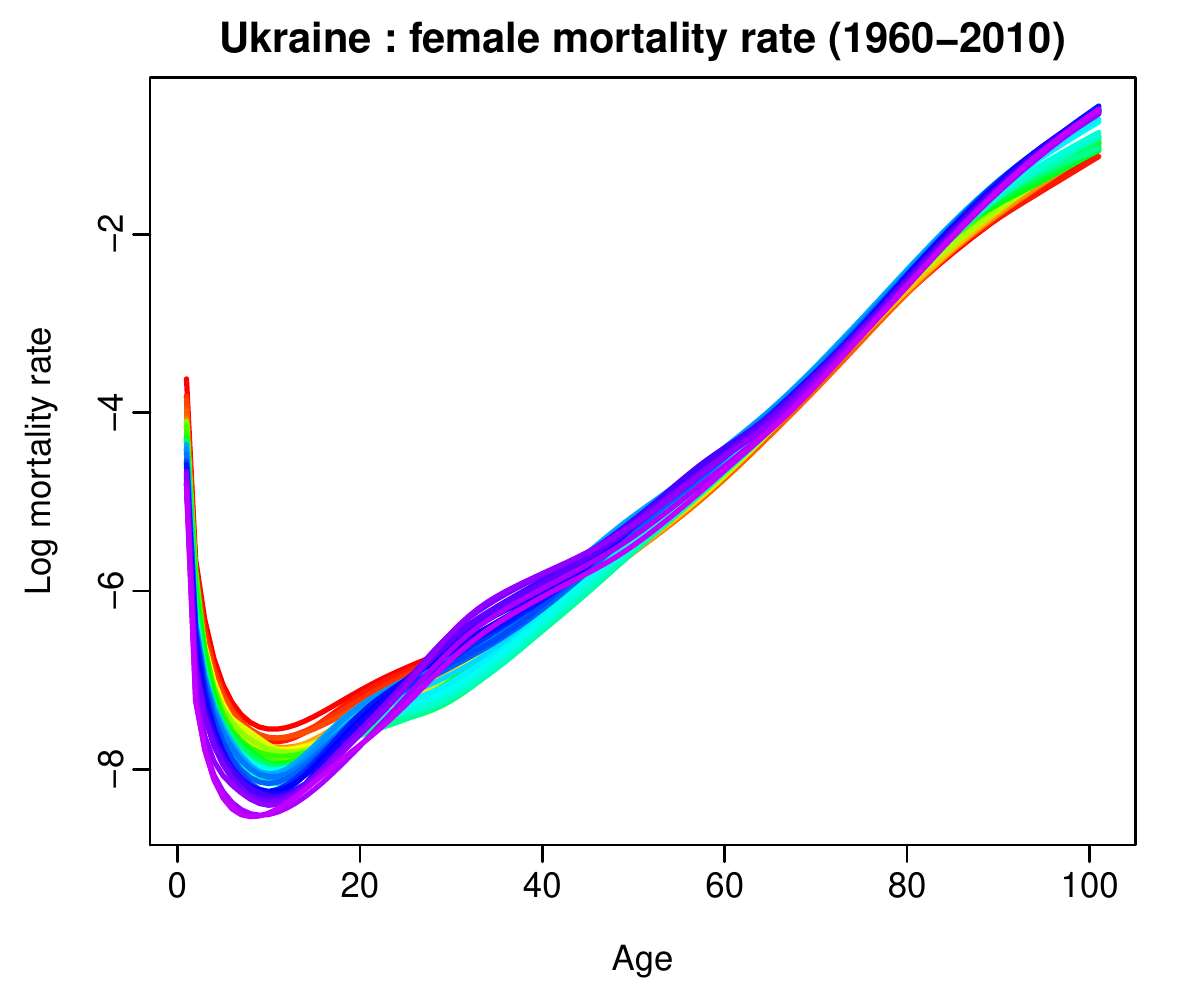} }}
    \caption{Functional time series displays for smoothed age-specific mortality rates of selected countries}%
    \label{fig:2}
\end{figure}

Figure~\ref{fig:2} displays the rainbow plots of the smoothed female age-specific log mortality rates of four selected countries from $1960$ to $2010$. To illustrate the necessity and advantage of clustering analysis on multi-country mortality data, we analyze the mortality data from the same cluster and different clusters (from our clustering results later).  
For demonstration purposes, these four countries are selected from two groups: Australia and Austria (upper panel) from one cluster and Russia and Ukraine (lower panel) from the other. The rainbow plots' patterns in the upper panel differ significantly from those in the lower panel. For Australia and Austria, the color spread is more dispersed, showing that the mortality is improving over time for all ages. The curves are more concentrated for Russia and Ukraine, and the uniform decrease in mortality over time disappears. Despite the general peaks and troughs being similar for all mortality rates, there are a slight upward trend for earlier years and a downward trend for later years at age $30$ in the Russian and Ukrainian mortality rates. These were not observed in the mortality rates of Australia and Austria.

\begin{figure}[!htbp]
\centering
\includegraphics[width=10.2cm]{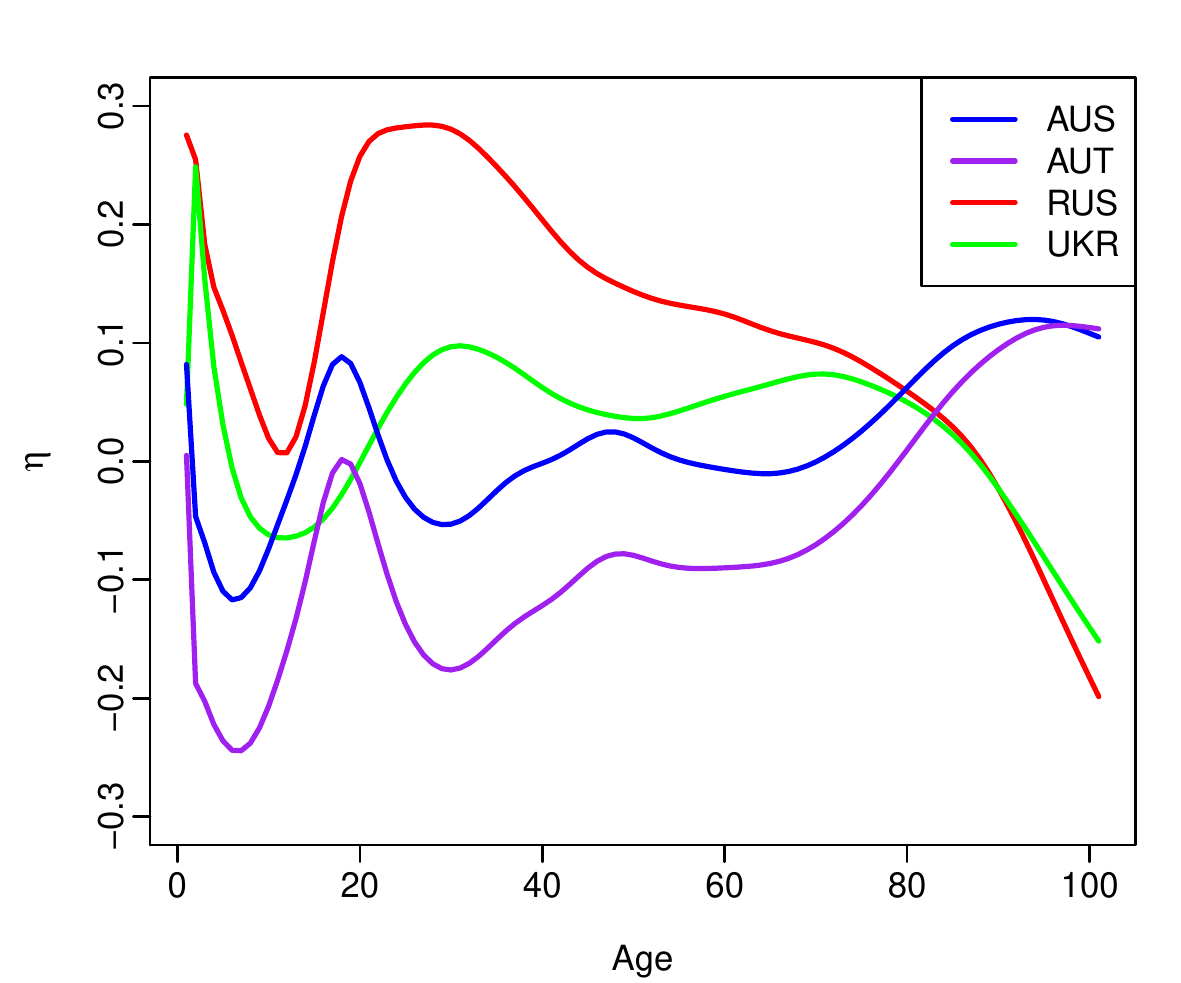}
\caption{Deviation of the country-specific mean from the grand mean}
\label{fig:3}
\end{figure}

\begin{figure}[!htbp]
    \centering
    \subfloat[Common time trend of Australia and Austria]{{\includegraphics[width=7.2cm]{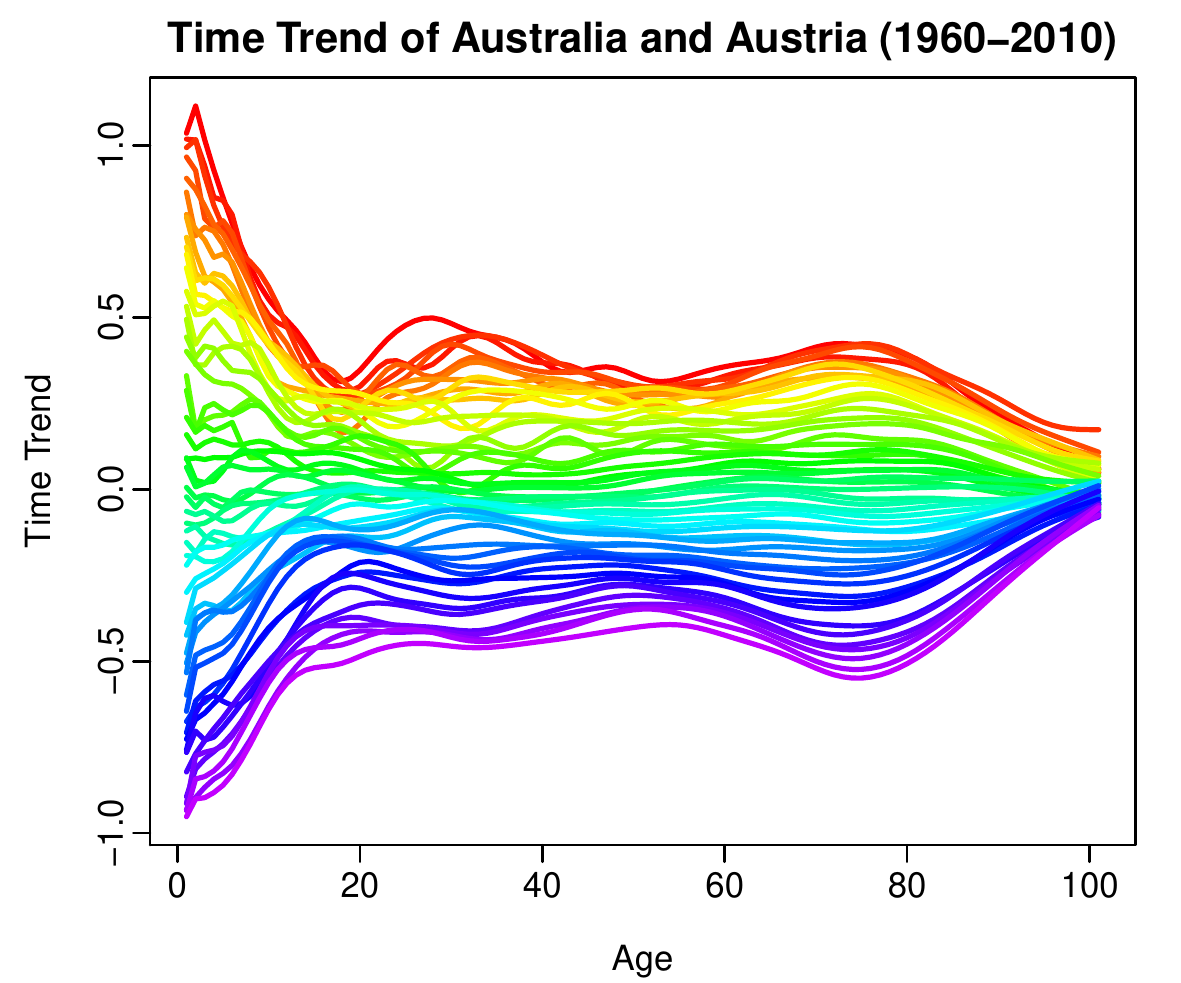}}}
    \qquad
    \subfloat[Common time trend of Russia and Ukraine]{{\includegraphics[width=7.2cm]{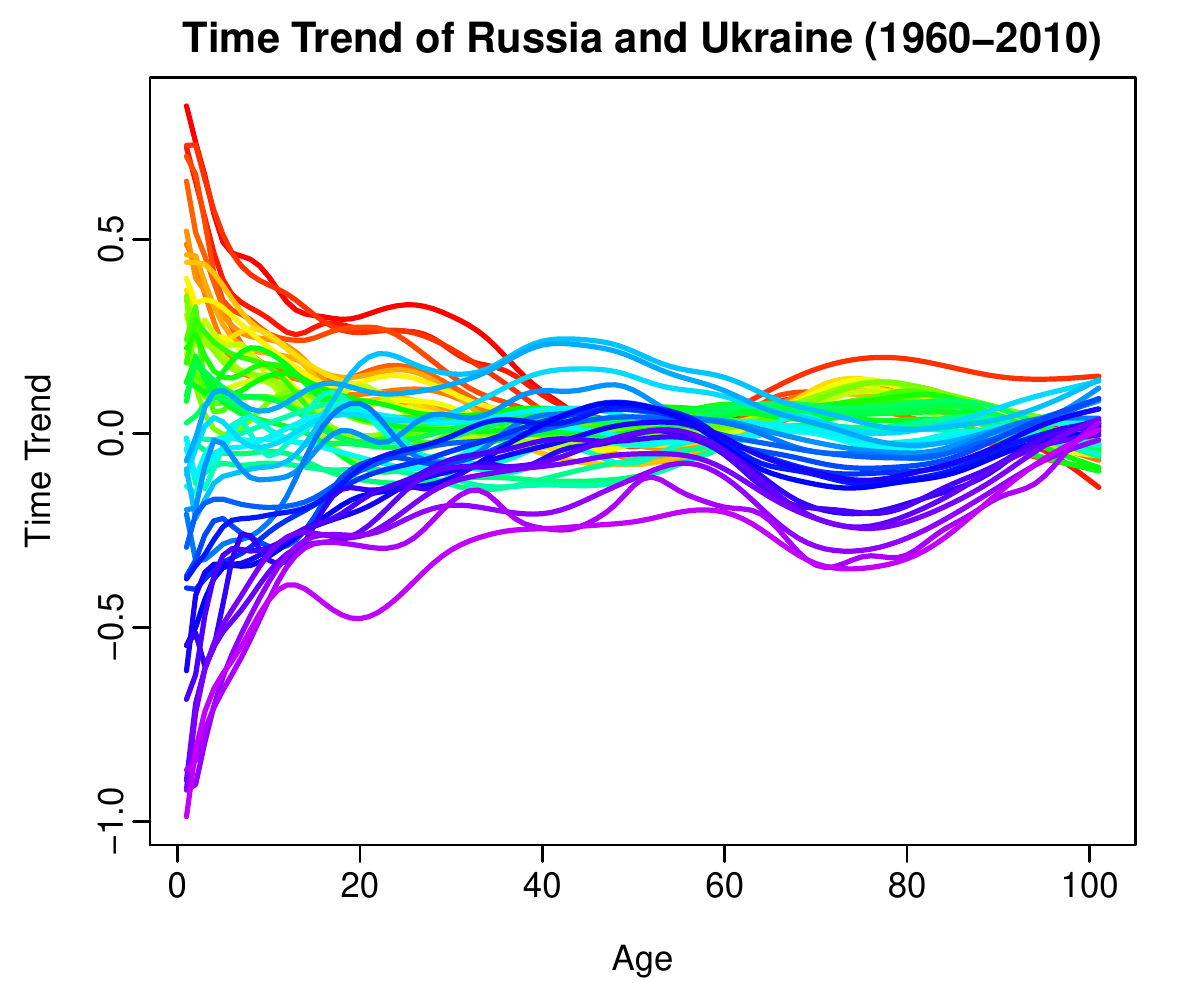} }}
    \caption{Rainbow plot of common time trend of two different types of countries}
    \label{fig:4}
\end{figure}

\begin{figure}[!htbp]
    \centering
    \subfloat[Residual time trend of Australia]{{\includegraphics[width=7.15cm]{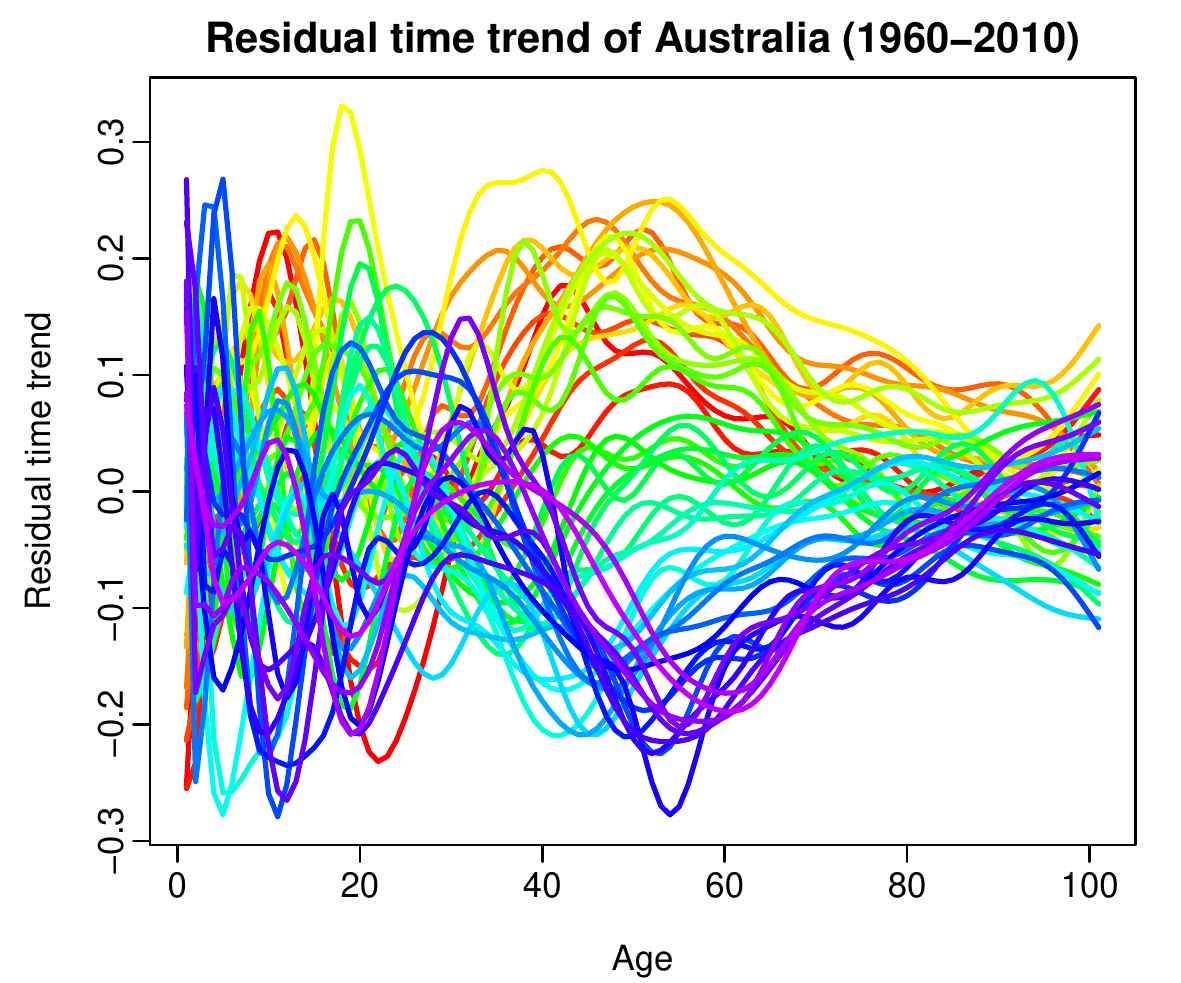} }}
    \qquad
    \subfloat[Residual time trend of Austria]{{\includegraphics[width=7.15cm]{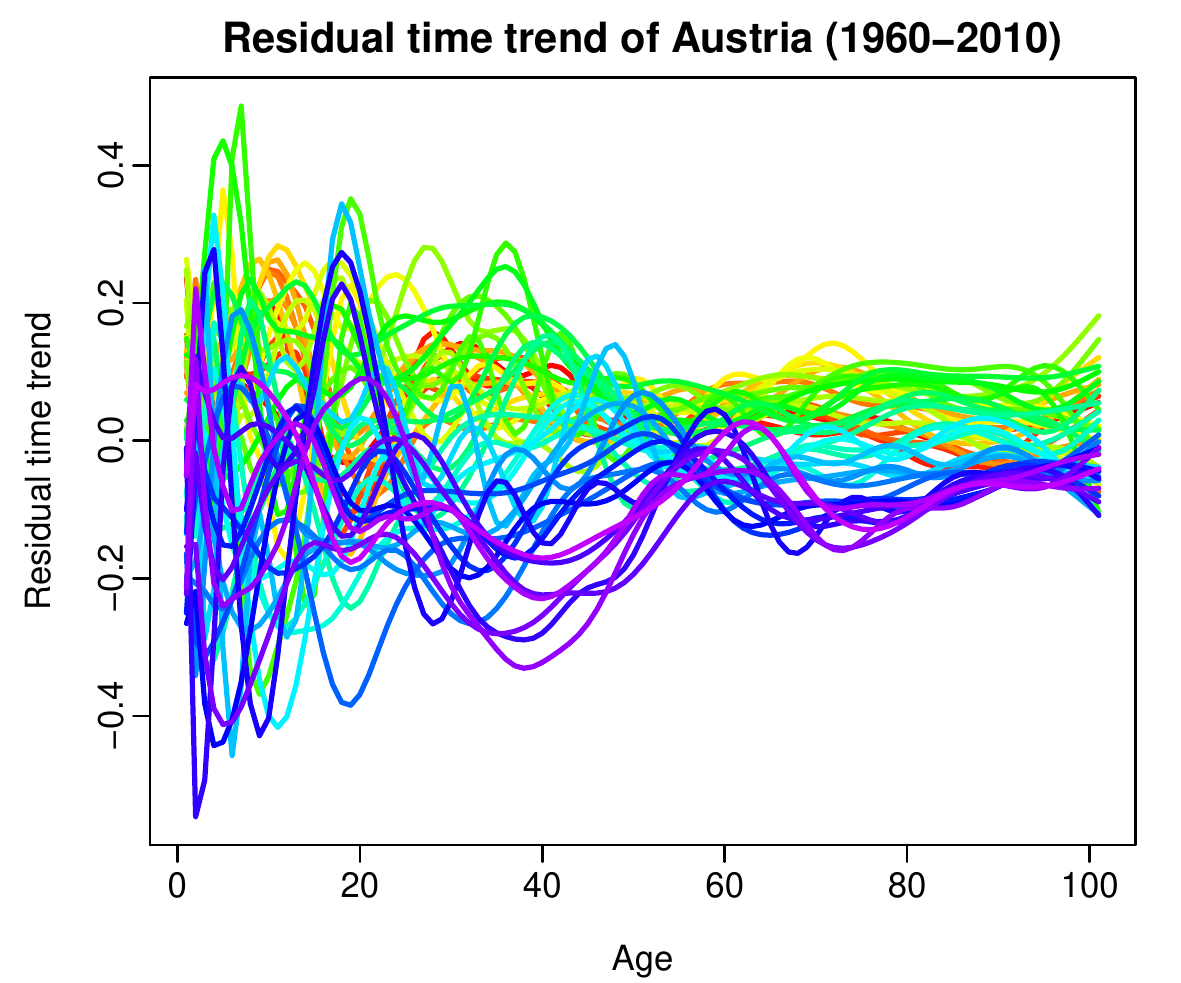} }}
    \qquad
    \subfloat[Residual time trend of Russia]{{\includegraphics[width=7.15cm]{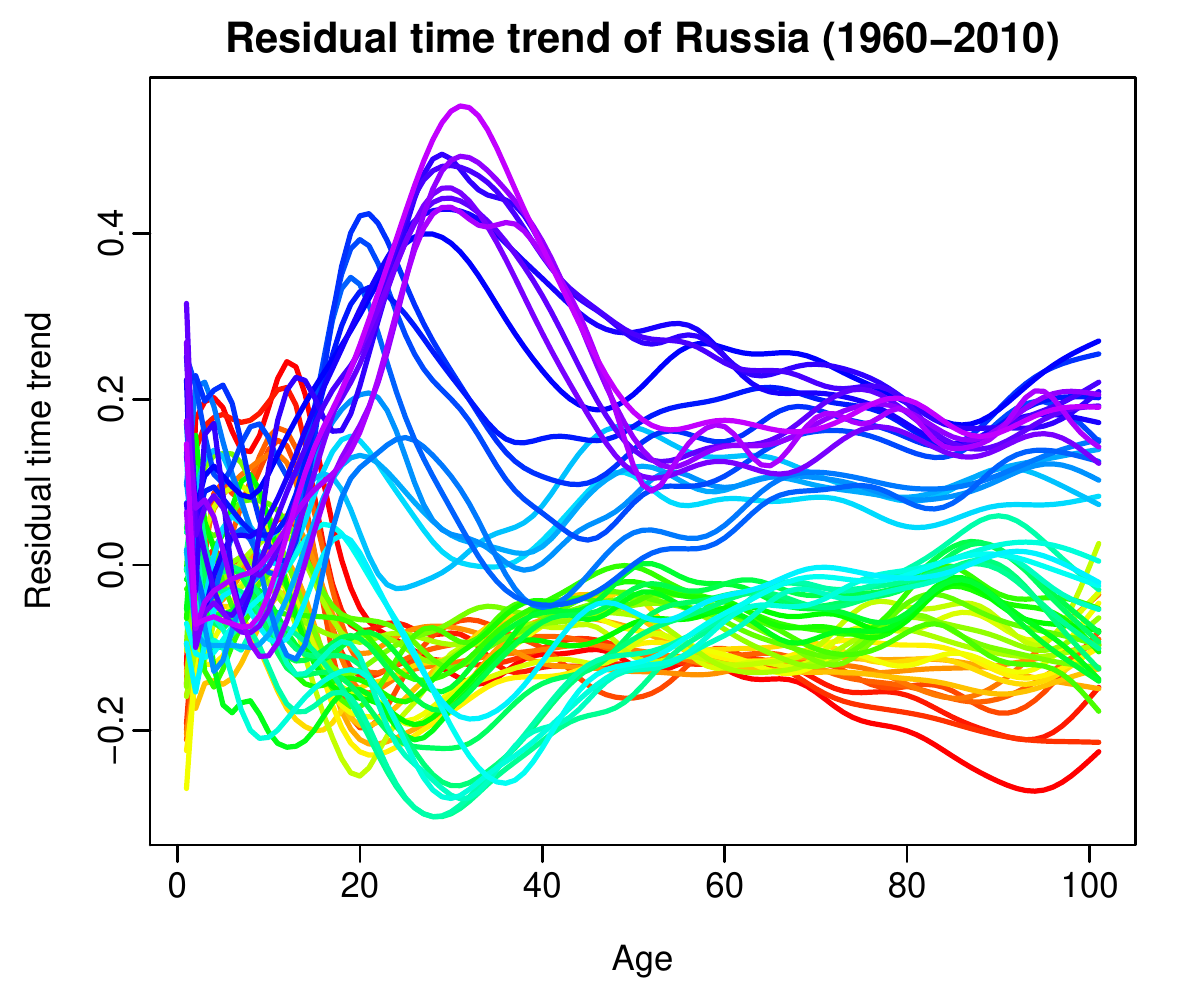} }}
    \qquad
    \subfloat[Residual time trend of Ukraine]{{\includegraphics[width=7.15cm]{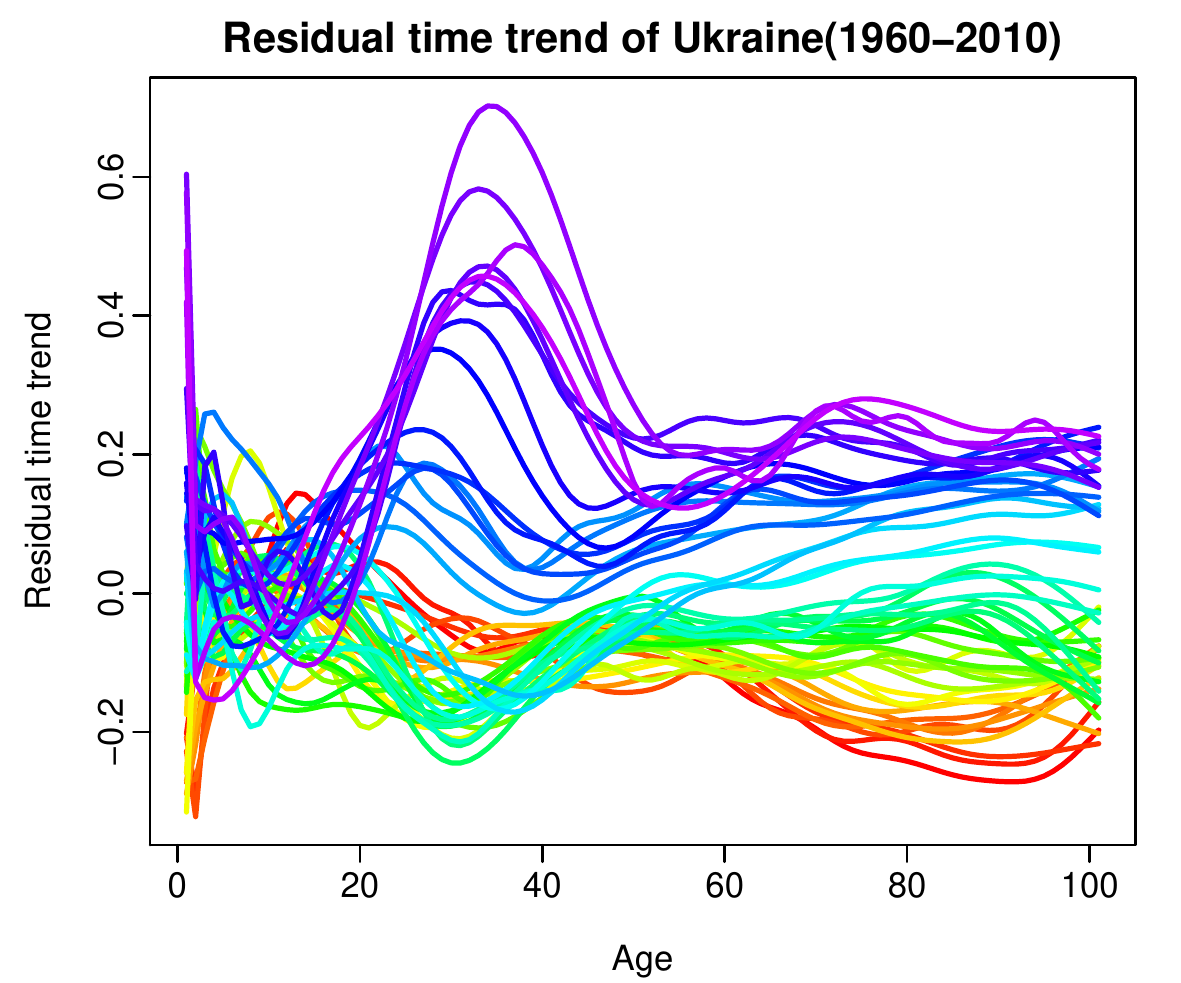} }}
    \caption{Residual time trend of four different countries}%
    \label{fig:5}
\end{figure}

The rainbow plots' patterns in Figure~\ref{fig:2} differ from each other in two aspects: the color pattern of the curves and the shape of the curves. After taking out the overall mean, the smoothed log mortality of each country can be decomposed into three components, namely, the deviation of the population-specific mean from the overall mean (Figure~\ref{fig:3}), the common time trend (Figure~\ref{fig:4}), and the population-specific time trend (Figure~\ref{fig:5}). As we can see from Figure~\ref{fig:3}, the deviations of the population-specific mean from the overall mean of Australia and Austria are similar, and those of Russia and Ukraine are alike; these different patterns in the curves reflect the major differences in the magnitude of peaks and troughs discussed earlier.

As for Figure~\ref{fig:4}, the common time trends of two clusters are displayed. There are fewer overlaps for the color spread in Australia and Austria than in Russia and Ukraine, confirming our temporal dynamics observation. Hence, it is reasonable to believe that the common time trend strongly characterizes the time ordering.

Figure~\ref{fig:5} plots the population-specific time trend, which captures the residual trend of the functional time series after taking out the overall mean, the deviation of the population-specific mean from the overall mean, and the common time trend. Similar patterns could still be observed for each cluster but with a relatively high degree of variation. Therefore, the population-specific time trend could be a relatively weaker characterization of the time ordering than the common trend, which could supplement it. 

The color pattern reflects the time ordering of the curves, which corresponds to the common time trend and the population-specific time trend. In contrast, the shape of the curves corresponds to the population-specific mean deviation from the overall mean. It is easy to observe that all these three components determine the functional time series characteristics. Therefore, our clustering method aims to group homogeneous functional time series based on their common features, i.e., the common functional time trends and the common functional patterns reflected in the rainbow plots. It is important to model these two types of patterns while maintaining the temporal dynamics in the homogeneous mortality rates. Our ultimate goal of accurate forecasts can be achieved. This has motivated the development of our method.

The mortality data are obtained from \cite{HMD}. Our dataset covers the period from $1960$ to $2010$, with $32$ out of $38$ countries or areas with sufficient data to use. Table~\ref{tab:1} shows a list of these countries and the corresponded ISO Alpha-3 codes. 
\begin{table}[!htbp] 
\centering 
  \caption{List of selected countries and corresponded ISO Alpha-3 codes} 
  \label{tab:1} 
  \setlength{\tabcolsep}{7.1pt}
\renewcommand{\arraystretch}{1.2}
\begin{tabular}{@{} lclclclc@{}} 
\toprule
Country &  Code & Country & Code & Country  &  Code & Country  &  Code \\
\midrule
\texttt{Australia} & \texttt{AUS} & \texttt{Estonia} & \texttt{EST} & \texttt{Lithuania}& \texttt{LTU} & \texttt{Russia} & \texttt{RUS}\\
\texttt{Austria} & \texttt{AUT} & \texttt{Finland} & \texttt{FIN} & \texttt{Latvia} & \texttt{LVA} & \texttt{Slovakia} & \texttt{SVK}\\
\texttt{Belgium} & \texttt{BEL} & \texttt{France} & \texttt{FRA} & \texttt{Luxembourg} & \texttt{LUX} & \texttt{Spain} & \texttt{ESP}\\
\texttt{Belarus} & \texttt{BLR} & \texttt{Hungary} & \texttt{HUN} & \texttt{Norway} & \texttt{NOR} & \texttt{Sweden} & \texttt{SWE}\\
\texttt{Bulgaria} & \texttt{BGR} & \texttt{Iceland} & \texttt{ISL} & \texttt{Portugal} & \texttt{PRT} & \texttt{Switzerland} & \texttt{CHE}\\
\texttt{Canada} & \texttt{CAN}  & \texttt{Ireland} & \texttt{IRE} & \texttt{Poland} & \texttt{POL} & \texttt{Great Britain} & \texttt{GBR}\\
\texttt{Denmark} & \texttt{DNK}  & \texttt{Italy} & \texttt{ITA} & \texttt{Netherlands} & \texttt{NLD} & \texttt{United States} & \texttt{USA}\\
\texttt{Czech Republic} & \texttt{CZE} & \texttt{Japan} & \texttt{JPN} & \texttt{New Zealand} & \texttt{NZL} & \texttt{Ukraine} & \texttt{UKR} \\
\bottomrule
\end{tabular}
\end{table} 

\section{Model and estimation}\label{sec:3}

We propose a functional panel data model with fixed effects to model all the components mentioned above in multiple mortality rates. Classical panel data models with fixed effects are discussed in \cite{Wooldridge2010} and \cite{Hsiao2014}, we herein extend it to functional data. Combining a panel data model with functional data techniques, such as FPCA, simultaneously finds common time-trend features and common functional patterns. The model interpretation and estimation are provided in detail in this section. To better utilize the forecasting ability of the model, dynamic FPCA is incorporated into the model estimation. 

\subsection{Functional panel data model with fixed effects}

Let $f_{it}(x)$ be a function measured with errors over a continuous variable $x$ for observation $t$ within subject $i$. In our application on the age-specific mortality rates, $f_{it}(x_j)$ is the $\log_{e}$ central mortality rate \footnote{The log central mortality rate for each country is calculated by $\ln(\frac{d_x}{E_x})$, where $d_x$ is the number of deaths in this country during each year and $E_x$ is the average number alive in this country during each year.} observed at the beginning of each year $t = 1, 2, \ldots, T$ for ages $x_1, x_2, \ldots, x_J$, where $J$ is the number of ages and $i = 1, 2, \ldots, I $ denotes the country index. 

Under the functional data framework, it can be assumed that there is an underlying continuous and smooth function, $Y_{it}(x)$ observed at discrete data points with an error such that
\begin{equation*}
  f_{it}(x_j) = Y_{it}(x_j) + \delta_{it}(x_j)\epsilon_{it,j},
\end{equation*}
where $x_j$ represents the center of each age or age group for $j = 1, 2, \ldots, J$, $\{\epsilon_{it,j}: i=1, 2, \ldots, I\}$ are independent and identically distributed (i.i.d) random variables for each age $j$ in year t, and $\delta_{it}(x_j)$ measures the variability in mortality at each age in year $t$ for the $i^{\textsuperscript{th}}$ population. The multiplication of $\delta_{it}(x_j)$ and $\epsilon_{it,j}$ represents the smoothing error. The technical details no smoothing are provided in Appendix~\ref{appna}. One should note that the pre-smoothing step is to smooth out the measurement error to obtain smooth trajectories. Alternatively, one can incorporate a smoothness penalty while extracting latent functional principal components \citep[see, e.g.,][]{reiss2007functional}.

We assume that $\bb{Y_i(x)} = (Y_{i1}(x), Y_{i2}(x), \ldots, Y_{iT}(x))^{\top}$ is a set of random smooth functions that represents each of the functional time series objects in the study. Each random function $Y_{it}(x)$ is defined in $L^{2}(\mathcal{I})$, a Hilbert space of square integrable functions on a real interval $\mathcal{I} \in [a,b]$. The inner product of two functions $f$ and $g$ is defined by $\langle f, g\rangle = \int f(x)g(x)dx$ with the norm $\|\cdot\| = \langle \cdot, \cdot\rangle ^{\frac{1}{2}}$.

Consider the functional panel data model with fixed effects
\begin{equation}\label{eq:1}
  Y_{it} (x)= \mu(x) + \eta_i(x) + R_t(x) + U_{it}(x), 
\end{equation}
where $x$ is age, the continuum of the random functions; $\mu(x)$ is the grand mean of the mortality of all countries and years; $\eta_i(x)$ is the country-specific individual effect; $R_t(x)$ is the time trend common to all countries; and $U_{it}(x)$ is the country-specific time trend. Here, $\mu(x)$ is a deterministic function, while $\eta_i(x)$, $R_t(x)$ and $U_{it}(x)$ are mean zero random functions.  

The functional patterns are reflected in the country-specific mean deviation from the grand mean, $\eta_i(x)$ for the functional panel data model with fixed effects. The time trend is captured by the combination of $R_t(x)$, the common time trend (a strong characterization) and $U_{it}(x)$, the country-specific time trend (a weak characterization). Our primary goal is to identify homogeneous subgroups of countries based on similar mortality structures and model these subgroups to reduce model variation, thus improving forecasting. As discussed earlier in Section~\ref{sec:2}, we need to rely on the four terms of the right-hand side of~\eqref{eq:1} to perform the cluster analysis. Once homogeneous subgroups are identified, we utilize both $R_t(x)$ and $U_{it}(x)$ in forecasting $Y_{it}(x)$.

\subsection{Dimension reduction using functional principal component analysis}

The difficulty in clustering and forecasting using the functional panel data model with fixed effects is that all four components in this model are of functional forms. To avoid the ``curse of dimensionality" incurred by the functional aspect, we adopt a dimension-reduction technique for functional data -- FPCA. The basic idea of classical FPCA is to decompose the functions into principal directions of variation, based on Karhunen-Lo\`eve (KL) expansion (\citealp{karhunen1946spektraltheorie,loeve1955probability}). This paper utilizes classical FPCA and dynamic FPCA on various terms to attain more efficient dimension reduction for better forecasting results. Next, we first introduce the preliminary for dynamic FPCA and then apply it on the terms $R_t(x)$ and $U_{it}(x)$, respectively. For the term $\eta_i(x)$, we still adopt the classical FPCA as it is not a temporal sequence. 

\subsubsection{Dynamic functional principal component analysis}\label{subsubsec:3}

The classical FPCA reduces dimensions by maximizing the variance explained. It is not an adequate dimension-reduction technique for functional time series data as it fails to account for the essential information provided by the time serial dependence structure \citep{hormann2015dynamic}. Since the ultimate goal is to make forecasts on mortality rates, the original functional time series' forecasting ability cannot be preserved fully by conventional FPCA. Hence, it is of great interest to reduce the dimension of functional time series and simultaneously preserve the temporal dynamics among functions to still enjoy the benefits of producing forecasts from time series. Therefore, we adopt a dynamic version of FPCA. In analyzing functional time series, \cite{horvath2013estimation} and \cite{panaretos2013fourier} defined smoothed periodogram type estimates of the long-run covariance and spectral density operators for functional time series. \cite{hormann2015dynamic} used the spectral density operator to create functional filters to construct mutually uncorrelated dynamic FPCs. 

\cite{rice2017plug} proposed a bandwidth selection method for estimates of the long-run covariance function based on finite-order weight functions that aim to minimize the estimator's asymptotic mean-squared normed error. Following their work, for a given functional time series $\{X_t(u), t\in{1,\ldots,T}\}$, the long-run covariance function is defined as
\begin{equation*}
  c(u,v) = \sum_{q=-\infty}^{\infty}\gamma_q(u,v), \text{ where } \gamma_q(u, v) = \text{cov}\big(X_t(u), X_{t+q}(v)\big).
\end{equation*}
Define a Hilbert-Schmidt integral operator $C$ on $L^{2}(\mathcal{I})$, such that 
\begin{equation}\label{y2}
  C(f)(u) = \int_{\mathcal{I}}c(u,v)f(v)dv.
\end{equation}
It is noteworthy that the long-run covariance function $c(u, v)$ incorporates auto-covariance functions $\gamma_q(u, v)$ with all time-lags $q\in\mathbb{Z}$, by way of summing them together. The dynamic FPCA defined by \cite{hormann2015dynamic} also accumulates all auto-covariance functions with various time-lags but in a different way that adopts the spectral density operator.  

In the literature for functional data analysis, the long-run covariance function $c(u,v)$ is estimated by a kernel estimator  
\begin{equation*}
  \widehat c(u,v) = \sum_{q=-\infty}^{\infty}K\Big(\frac{q}{h}\Big) \widehat \gamma_q(u,v),
\end{equation*}
where
\begin{align*}
\widehat \gamma_q(u,v) = \left\{ \begin{array}{cc} 
                \frac{1}{T-q}\sum_{t=1}^{T-q}\Big(X_t(u) - \bar X(u)\Big)\Big(X_{t+q}(v) - \bar X(v)\Big), & q\geq 0;\\
                \frac{1}{T-q}\sum_{t=1-q}^{T}\Big(X_t(u) - \bar X(u)\Big)\Big(X_{t+q}(v) - \bar X(v)\Big),&  q< 0,\\
                \end{array} \right.
\end{align*}
with $\overline X(u) = \frac{1}{T}\sum_{t=1}^{T}X_t(u)$ and $K\Big(\frac{q}{h}\Big)$ is the kernel function which assigns different weights to the auto-covariance functions with different lags, and $h$ is the bandwidth.

There are various kernel functions in use: \cite{hansen1982large} and \cite{white2014asymptotic} used the truncated kernel; \cite{newey1986simple} used the Bartlett kernel; \cite{gallant2009nonlinear} used the Parzen kernel, and \cite{andrews1991heteroskedasticity} used the Quadratic Spectral (QS) kernel. However, all these kernel functions' common nature is to assign more weights to the auto-covariance functions with smaller lags and fewer weights to the auto-covariance functions with larger lags. Infinite-order ``flat-top'' kernels have gained popularity, as they give a reduced bias and faster rates of convergence \citep{politis1996flat,politis1999multivariate}. Flat-top kernels are of the following form
\begin{align*}
  K\Big(\frac{q}{h}\Big) = \left\{ \begin{array}{lll}
                1, & 0\leq|\frac{q}{h}|< k; \\ 
                 \frac{|\frac{q}{h}|-1}{k-1}, & k\leq|\frac{q}{h}|< 1;\\
                0, &  |\frac{q}{h}|\geq1,\\
                \end{array} \right.
\end{align*}
where $k<1$ is a thresholding parameter. 

Hence, the long-run covariance function is a weighted average of all lags' auto-covariance functions, which contains the full information of the family of covariance operators. However, the choice of bandwidth can greatly affect its performance on a finite sample. In this paper, we apply the adaptive bandwidth selection procedure of \cite{rice2017plug} to estimate the long-run covariance of functional time series. Dynamic FPCA is implemented through the eigenanalysis on the long-run covariance function.

\subsubsection{Mercer's theorem and the Karhunen-Lo\`eve expansion}

To facilitate the clustering procedure, we need to represent the three components that characterize the functional time series (i.e., $\eta_i(x), R_t(x)$ and $U_{it}(x)$) via the Karhunen-Lo\`eve expansion.

Since $\eta_i(x)$ reflects the general mode of variations of each country, define the covariance operator of $\eta_i(x)$ as $C_{\eta}$ on $L^{2}(\mathcal{I})$, such that 
\begin{equation*}
  C_{\eta}(f)(u) = \int_{\mathcal{I}}c_{\eta}(u,v)f(v)dv,
\end{equation*}
where $c_{\eta}(u, v) = \text{cov}\big(\eta_i(u), \eta_i(v)\big)$. Note that $\eta_i(x)$ across $i=1, 2, \ldots, I$ are assumed to share the same covariance function $c_{\eta}(u, v)$. By Mercer's theorem, the operator $C_{\eta}$ admits an eigen-decomposition
\begin{equation*}
  C_{\eta}(f) = \sum_{m=1}^{\infty}\lambda_m^{\eta}\langle f, \phi_m\rangle \phi_m,
\end{equation*}
where $\lambda_m^{\eta}$ is the $m^\text{th}$ largest eigenvalue of $C_{\eta}$ with $m=1, 2, \ldots$  and $\phi_m$ is the corresponding eigenfunction. Based the Karhunen-Lo\`eve expansion, $\eta_i(x)$ can be represented with 
\begin{equation*}
  \eta_i(x) = \sum_{m=1}^{\infty}\gamma_{im}\phi_m(x),
\end{equation*}
where $\gamma_{im} = \langle \eta_i, \phi_m\rangle$, is the $m^\text{th}$ principal component score for $\eta_i(x)$. 

As $R_t(x)$ captures the common time trend, we need to maintain the time serial dependence structure. Define the long-run covariance operator $C_R$ as (\ref{y2}) in Section \ref{subsubsec:3}, with the long-run covariance kernel being 
\begin{equation*}
  c_R(u,v) = \sum_{q=-\infty}^{\infty}\gamma_q^{R}(u,v),
\end{equation*}
where $\gamma_q^{R}(u, v) = \text{cov}\big(R_t(u), R_{t+q}(v)\big)$. Similarly, let $\rho_k(x)$ be the eigenfunction associated with the $k^\text{th}$ eigenvalue of $C_R$ in descending order, $\lambda_1^{R}\geq\lambda_2^{R}\geq\ldots\geq 0$, then $R_t(x)$ can be represented as
\begin{equation*}
  R_t(x) = \sum_{k=1}^{\infty}\xi_{tk}\rho_k(x),
\end{equation*}
where $\xi_{tk} = \langle R_t, \rho_k\rangle$, is $k^\text{th}$ the principal component score for $R_t(x)$.

As for country-specific time trend $U_{it}$, since we assume $\{U_{it}: i=1, 2, \ldots, I\}$ are indepedent across $i$, define a long-run covariance operator $C_U$ as (\ref{y2}) in Section \ref{subsubsec:3}, with the long-run covariance kernel being 
\begin{equation*}
  c_{U}(u,v) = \sum_{q=-\infty}^{\infty}\gamma_q^{U}(u,v),
\end{equation*}
where $\gamma_q^{U}(u, v) = \text{cov}\big(U_{it}(u), U_{i(t+q)}(v)\big)$. Let $\psi_l(x)$ be the eigenfunction associated with the $l^\text{th}$ eigenvalue of $C_U$ in descending order, $\lambda_1^{U}\geq\lambda_2^{U}\geq\ldots\geq 0$, then $U_{it}(x)$ can be represented as
\begin{equation*}
  U_{it}(x) = \sum_{l=1}^{\infty}\zeta_{itl}\psi_l(x),
\end{equation*}
where $\zeta_{itl} = \langle U_{it}, \psi_l\rangle$, is the $l^\text{th}$ principal component score for $U_{it}(x)$.

Consequently, our model in~\eqref{eq:1} can be written as
\begin{equation}\label{eq:2}
  Y_{it} (x)= \mu(x) + \sum_{m=1}^{\infty}\gamma_{im}\phi_m(x) + \sum_{k=1}^{\infty}\xi_{tk}\rho_k(x) + \sum_{l=1}^{\infty}\zeta_{itl}\psi_l(x),
\end{equation}
where $\gamma_{im}$, $\xi_{tk}$ and $\zeta_{itl}$ are the principal component scores for $\eta_i(x)$, $R_t(x)$ and $U_{it}(x)$, respectively; $\phi_m(x)$, $\rho_k(x)$ and $\psi_l(x)$ are the corresponding eigenfunctions. 

In this model, the temporal dynamics of original functional time series are reflected in $\xi_{tk}$ and $\zeta_{itl}$; while the eigenfunctions, $\phi_m(x)$, $\rho_k(x)$ and $\psi_l(x)$, capture the functional patterns. From Figures~\ref{fig:4} and ~\ref{fig:5}, it is clear to see that $R_{t}(x)$ is a strong characterization, and the residual term $U_{it}$ reflects the additional time trend that cannot be captured by $R_{t}(x)$, which could serve as a supplement. 

Note that our model setup in~\eqref{eq:1} is similar to that of \cite{di2009multilevel}, where a two-way ANOVA model was employed to extract core intra- and inter-subject components of multiple sets of functional data, which motivated \cite{shang2016mortality} in modeling male and female mortality rates jointly. However, our model differs from the multilevel FPCA in the following aspects. Firstly, the proposed model utilizes $\eta_i(x)$, $R_t(x)$ and $U_{it}(x)$ to extract common features, in which we perform FPCA for these three components. Secondly, to maintain the forecasting ability in the functional time series, we use dynamic FPCA instead of conventional FPCA. Thirdly, we adopt the functional panel data model framework, which allows us to estimate all these three components, $\eta_i(x)$, $R_t(x)$ and $U_{it}(x)$ directly. The functional panel data model brings two advantages: 
\begin{inparaenum}
\item[1)] it enables us to estimate the long-run covariance to accommodate the temporal dynamics, and 
\item[2)] in the estimation, we do not rely on the assumption that $R_t(x)$ and $U_{it}(x)$ are uncorrelated as in their work. 
\end{inparaenum}

\cite{li2013selecting} pointed out that, due to the difficulties in estimating and interpreting the infinite terms, the conventional treatment of this is to truncate them at finite sums. The last three terms on the right-hand side of~\eqref{eq:2} can be truncated, and the optimal numbers of components retained, $M$, $N_1$ and $N_2$, respectively
\begin{equation}\label{eq:3}
  Y_{it} (x)= \mu(x) + \sum_{m=1}^{M}\gamma_{im}\phi_m(x) + \sum_{k=1}^{N_1}\xi_{tk}\rho_k(x) + \sum_{l=1}^{N_2}\zeta_{itl}\psi_l(x) + \upsilon_{it}(x),
\end{equation}
where $\upsilon_{it}(x)$ is the residual term due to truncation.

\subsection{Estimation method}

Since the true values of some terms in the model are unknown in practice, we need to estimate them from the realizations of the smoothed function $Y_{it}(x)$. Suppose the realization of $Y_{it}(x)$ is $y_{it}(x)$, and we have $I$ countries and $T$ years, then the estimations of $\mu(x)$ and $\eta_i(x)$ are as follows
\begin{align*}
  \widehat\mu (x) &= \overline y_{\cdot \cdot} = \frac{1}{IT}\sum_{i=1}^{I}\sum_{t=1}^{T}y_{it}(x), \\
  \widehat\eta_i (x) &= \overline y_{i\cdot} - \widehat\mu (x) = \frac{1}{T}\sum_{t=1}^{T}y_{it}(x) - \widehat\mu (x).
\end{align*}
By averaging~\eqref{eq:1} across $i$, we obtain
\begin{equation*}
  \overline y_{\cdot t} (x) = \mu(x) + \overline \eta_{\cdot}(x) + R_t(x) + \overline U_{\cdot t} (x).
\end{equation*}
Since $\mathbb{E}\left[\eta_i(x)\right] = 0$ and $\mathbb{E}[U_{it}(x)] = 0$, it is obvious that
\begin{equation*}
  \widehat R_t(x) = \overline y_{\cdot t}(x) - \widehat \mu(x) = \frac{1}{I}\sum_{i=1}^{I}y_{it}(x) - \widehat \mu(x).
\end{equation*}
Then $U_{it}$ can be estimated as
\begin{equation*}
  \begin{split}
    \widehat U_{it}(x) & = y_{it}(x) - \widehat\mu(x) - \widehat\eta_i (x) - \widehat R_t(x)\\
    & = y_{it}(x) - \overline y_{i\cdot} - \overline y_{\cdot t}(x) + \widehat \mu(x)\\
    & = y_{it}(x) - \frac{1}{T}\sum_{t=1}^{T}y_{it}(x) - \frac{1}{I}\sum_{i=1}^{I}y_{it}(x) + \frac{1}{IT}\sum_{i=1}^{I}\sum_{t=1}^{T}y_{it}(x).
  \end{split}
\end{equation*}
Once $\widehat \eta_i(x)$, $\widehat R_t(x)$ and $\widehat U_{it}(x)$ are obtained, they can be used to calculate the estimators of $C_\eta$, $C_R$ and $C_U$, which are denoted by $\widehat C_\eta$, $\widehat C_R$ and $\widehat C_U$, respectively. The basis functions $\phi_m(x)$, $\rho_k(x)$ and $\psi_l(x)$ can be estimated by performing eigen-decomposition on $\widehat C_\eta$, $\widehat C_R$ and $\widehat C_U$, respectively, with associated eigenvalues being $\widehat\lambda_m^{\eta}$, $\widehat\lambda_k^{R}$ and $\widehat\lambda_l^{U}$ correspondingly. The FPC scores can be estimated as $\widehat\gamma_{im} = \langle \widehat\eta_i, \widehat\phi_m\rangle$, $\widehat\xi_{tk} = \langle \widehat R_{t}, \widehat\rho_k\rangle$ and $\widehat\zeta_{itl} = \langle \widehat U_{it}, \widehat\psi_l\rangle$.

Selecting the optimal numbers of functional principal components, $M$, $N_{1}$ and $N_{2}$, has been well studied in the literature. \cite{rice1991estimating} used cross-validation, \cite{yao2005functional} proposed an Akaike's information criterion (AIC) approach, \cite{hall2006assessing} employed a bootstrap method, while \cite{chiou2012dynamical} used the proportion of variance explained method. All these methods were designed for independent data. Later, \cite{hormann2015note} proposed a choice of the number of functional principal components for dependent data based on the ``bias variance trade-off", which is a data-driven approach that accounts for the sample size information. The number of functional principal components, $N$, was selected based on $\widehat{N} = \underset{N\geq 1}{\operatorname{argmax}}\{\widehat\lambda_1/\widehat\lambda_N \leq m_n\}$, where $n$ is the sample size, $m_{n} \rightarrow \infty$ and $m_n = o(\sqrt{n})$. A choice of $m_{n}$ suggested by \cite{hormann2015note} is ${n^{1/2}}/{\log_{10}(n)}$. They found out that this choice of number of functional principal components performed reasonably well compared with the cross-validation method. In practice, all these method could be applied to dependent functional time series. 

In this paper, the numbers of principal components used, $M$, $N_{1}$ and $N_{2}$, are determined by the combination of the cumulative percentage of variance method and the data-driven approach, such that
\begin{align*}
  \widehat{M} &= \text{max}\Big\{ \underset{M : M \geq 1}{\operatorname{argmin}}\Big(\sum_{m=1}^{M}\widehat\lambda_m^{\eta}\Big/\sum_{m=1}^{\infty}\widehat\lambda_m^{\eta} \mathbbm{1}\{\widehat\lambda_m^{\eta}>0\}\geq P_1\Big),\quad \underset{M : M \geq 1}{\operatorname{argmax}}\Big(\frac{\widehat{\lambda}_1^{\eta}}{\widehat{\lambda}_{M}^{\eta}}\leq \frac{\sqrt{I}}{\log_{10}(I)}\Big)\Big\}, \\
  \widehat{N}_1 &= \text{max}\Big\{ \underset{N_1 : N_1 \geq 1}{\operatorname{argmin}}\Big(\sum_{k=1}^{N_1}\widehat\lambda_k^{R}\Big/\sum_{k=1}^{\infty}\widehat\lambda_k^{R} \mathbbm{1}\{\widehat\lambda_k^{R}>0\}\geq P_2\Big),\quad \underset{N_1 : N_1 \geq 1}{\operatorname{argmax}}\Big(\frac{\widehat{\lambda}_1^{R}}{\widehat{\lambda}_{N_1}^{R}}\leq \frac{\sqrt{T}}{\log_{10}(T)}\Big)\Big\}, \\
 \widehat{N}_2 &= \text{max}\Big\{\underset{N_2 : N_2 \geq 1}{\operatorname{argmin}}\Big(\sum_{l=1}^{N_2}\widehat\lambda_l^{U}\Big/\sum_{l=1}^{\infty}\widehat\lambda_l^{U} \mathbbm{1}\{\widehat\lambda_l^{U}>0\}\geq P_3\Big), \quad \underset{N_2 : N_2 \geq 1}{\operatorname{argmax}}\Big(\frac{\widehat{\lambda}_1^{U}}{\widehat{\lambda}_{N_2}^{U}}\leq \frac{\sqrt{IT}}{\log_{10}(IT)}\Big)\Big\},
\end{align*}
where $\mathbbm{1}\{\cdot\}$ is an indicator function and $P_1$, $P_2$ and $P_3$ are all chosen to be 0.9 \citep{horvath2012inference}, $I$ is the number of countries, and $T$ is the number of years.

\section{Model-based functional clustering}\label{sec:4}

We propose a clustering procedure based on the functional panel data model to divide multiple functional time series into different clusters, using the similarity in functional time trends and functional patterns derived from the model. Similar to \cite{chiou2007functional}, the clustering procedure consists of two steps: an initial clustering step involving a classical clustering method as a starting point and an iterative membership updating step. In this section, the clustering procedure is outlined in detail. 

\subsection{Initial clustering step}

The classical clustering approach, $k$-means, is used in the initial step. Given our data's nature, to avoid the problem of the ``curse of dimensionality'', we first apply FPCA to the original data before applying the $k$-means clustering method. To ensure that the FPC scores for all curves are meaningful and directly comparable in that the features extracted are common to all curves, we apply the FPCA to a combined set of all standardized curves from the multiple functional time series. By combing all standardized curves of these $I$ sets of functional time series, we obtain a combined set of $I\times T$ curves. Given a large number of curves, we apply the conventional FPCA to the combined set of all curves. 

Let $Q$ be the number of FPC selected, the original $I$ sets of functional time series are reduced to $I$ score matrices of dimension $T\times Q$ . For mortality data, the value of $Q$ is around $2$-$4$.  The $k$-means method is then applied to these $I$ score matrices. In this study, $I$ is the number of the countries under study. 

Since cluster analysis is unsupervised by nature, the number of clusters is unknown in advance and needs to be determined before clustering. There are different approaches to selecting the optimal number of clusters (see \citealp{kodinariya2013review} for an intensive review). Here we adopt the information-theoretic approach of \cite{sugar2003finding}. The information-theoretic approach is based on ``distortion'', which is a measure of within-cluster dispersion. It is not hard to see that distortion is decreasing with the number of clusters.
Then a scree plot of distortion can be used to search for the optimal number of clusters. The optimal number is when the scree plot of distortions of all possible numbers of clusters levels off. Where the distortion levels off is where the biggest jump in negative-power transformed distortion occurs.

Therefore the optimal number of clusters, $K_{\text{opt}}$ can be found as
\begin{equation}\label{eq:9}
  K_{\text{opt}} = \underset{k \in (2, 3, \ldots, K)}{\operatorname{argmax}}\left(\widehat d_k^{-\frac{I}{2}}-\widehat d_{k-1}^{-\frac{I}{2}}\right),
\end{equation}
where $\widehat d_k$ is the empirical distortion when the cluster number is $k$, which can be calculated as the total within-cluster sum of squares per cluster, and $I$ is the number of sets of the functional time series. The term $(\widehat d_k^{-\frac{I}{2}}-\widehat d_{k-1}^{-\frac{I}{2}})$ represents the jump in negative-power transformed distortion of cluster $k$. Then $K_{\text{opt}}$ is used as the number of clusters for the initial step. 

In practice, we use the function \texttt{kmeans} in \texttt{R} to implement the $k$-mean clustering, where the input is the pairwise distance of the score matrices. 

\subsection{Iterative reclassification}

We apply the proposed functional panel data model to reclassify each functional time series object using the initial clustering result as a starting point. Here we follow the similar idea of the leave-one-out prediction approach of \cite{chiou2007functional} recursively to reclassify the functional time series objects. 

\subsubsection{Leave-one-out estimation based on cluster characteristic}

Let $c_i^{(l)} \in \{1, 2, \ldots, K\}$ be the label of cluster membership for the $i^{\textsuperscript{th}}$ object at the $l^{\textsuperscript{th}}$ iteration. Given the clustering results, $\mathcal C^{(l)} = \{c_i^{(l)}, i = 1, 2, \ldots, I\}$, by excluding the $i^{\text{th}}$ object we can obtain the leave-one-out estimated structure components of a given cluster $c$, $\widehat\mu^{(c)}_{(-i)} (x)$, $\widehat\phi^{(c)}_{m(-i)}(x)$, $\widehat\rho^{(c)}_{k(-i)}(x)$, and $\widehat\psi^{(c)}_{l(-i)}(x)$ according the model in~\eqref{eq:3}. Note that if the cluster $c$ does not contain object $i$, then the leave-one-out estimated structure components is simply $\widehat\mu^{(c)}(x)$, $\widehat\phi^{(c)}_{m}(x)$, $\widehat\rho^{(c)}_{k}(x)$, and $\widehat\psi^{(c)}_{l}(x)$. Then we can get the leave-one-out predicted $t^{\textsuperscript{th}}$ curve of the $i^{\textsuperscript{th}}$ object at the $l^{\textsuperscript{th}}$ iteration for each cluster $c, c=1,\ldots, K$,
\begin{equation*}
  \widehat y_{it}^{(c)(l)}(x) = \widehat\mu^{(c)}_{(-i)} (x) + \sum_{m=1}^{M^{(c)}}\widehat\gamma_{im}^{(c)} \widehat\phi^{(c)}_{m(-i)} (x) + \sum_{k=1}^{N_1^{(c)}}\widehat\xi_{tk}^{(c)} \widehat\rho^{(c)}_{{}k(-i)} (x) + \sum_{l=1}^{N_2^{(c)}}\widehat\zeta_{itl}^{(c)} \widehat\psi^{(c)}_{l(-i)} (x),
\end{equation*}
where the leave-one-out estimates $\widehat\mu^{(c)}_{(-i)}$, $\widehat\phi^{(c)}_{m(-i)}$, $\widehat\rho^{(c)}_{k(-i)}$, and $\widehat\psi^{(c)}_{l(-i)}$ for each cluster $c$, can be obtained using the estimation procedure introduced in Section~\ref{sec:3}. $\widehat\gamma_{im}^{(c)}$ is calculated as $ \langle \overline y_{i\cdot} - \widehat\mu^{(c)}_{(-i)} , \widehat\phi^{(c)}_{m(-i)} \rangle$. 

The estimated functional principal component scores $\widehat\xi_{tk}^{(c)}$ and $\widehat\zeta_{itl}^{(c)}$ cannot be estimated directly from $\widehat R_{t(-i)}^{(c)}$ and $\widehat U_{it}^{(c)} = y_{it} - \widehat\mu_i - \widehat{R}_{t(-i)}^{(c)}$. This is because for any cluster $c$, if we calculated the score directly from $\widehat R_{t(-i)}^{(c)}$ and $\widehat U_{it}^{(c)} = y_{it} - \widehat\mu_i - \widehat{R}_{t(-i)}^{(c)}$, without any truncation of the infinite terms in the FPCA step, the leave-one-out prediction becomes $\widehat\mu_i + \widehat R_{t(-i)}^{(c)} + y_{it} - \widehat\mu_i - \widehat{R}_{t(-i)}^{(c)}$, which is just $y_{it}$. This means that all the leave-one-out predictions for any clusters $c\in\{1, 2, \ldots, K\}$ are $y_{it}$, which results in the incapability to cluster $y_{it}$. 

The estimated functional principal component scores $\widehat\xi_{tk}^{(c)}$ and $\widehat\zeta_{itl}^{(c)}$ could be calculated based on the demeaned curve, $y_{it} (x) - \widehat\mu_i (x)  = \widehat R_{t(-i)}^{(c)} + \widehat U_{it}^{(c)}$. However, $\widehat\rho^{(c)}_{k(-i)}$, and $\widehat\psi^{(c)}_{l(-i)}$ may not be orthogonal, direct numerical integration would not work. \cite{di2009multilevel} propose a projection method to address this issue. We here use the projection method in conjunction with the least-squares multivariate linear regression approach to calculate $\widehat\xi_{tk}^{(c)}$ and $\widehat\zeta_{itl}^{(c)}$. Technical details are provided in Appendix~\ref{appnb}. 

\subsubsection{Iterative cluster membership updating}

Once the leave-one-out estimates for all curves of the $i^{\textsuperscript{th}}$ object at the $l^{\textsuperscript{th}}$ iteration for each cluster are obtained, then the clustering membership for the $l+1^{\textsuperscript{th}}$ iteration could be updated.

If the functional time series object $\bb{y_i} = (y_{i1}, y_{i2}, \ldots, y_{iT})^{\top}, t = 1, 2, \ldots, T$ belongs to a specific cluster $c$, then  $\widehat y_{it}^{(c)(l)}$ is very close to the observed smoothed curve, $y_{it}$. \cite{chiou2007functional} argued that for clustering problems, the goal is to identify the cluster that each object is most likely to belong to. The likelihood of cluster membership for any object can be connected to some distance measures. Based on the model setup assumption, the $L^{2}$-distance between curves is appropriate.

The $i^{\textsuperscript{th}}$ object is classified into cluster $c_i^{(l+1)}$ according to 
\begin{equation*}
  c_i^{(l+1)} = \underset{c\in \{1, 2, \ldots, K\}}{\operatorname{argmin}}\frac{1}{T}\sum_{t=1}^{T}\left\|y_{it} - \widehat y_{it}^{(c)(l)}\right\|.
\end{equation*}
The label of the membership is not important. As for clustering, we aim to find a subgroup for homogeneous objects, and the label is used to identify different clusters. Each reclassification step is performed for all $i, i = 1, 2, \ldots, I$ and at the end of the iteration, the cluster membership is updated to $\mathcal C^{(l+1)} = \{c_i^{(l+1)}, i = 1, 2, \ldots, I\}$.

The pseudo-code for the clustering procedure is summarized in Algorithm 1. 
\begin{algorithm}[!h]
\SetAlgoLined
\textbf{Input}: High-dimensional functional time series $\Big\{\bb{y}_i(u) = [{y}_{i1}, {y}_{i2}, \ldots, y_{iT}]^{\top}, i = 1, \ldots, I\Big\}$.\\
 \textbf{1. Initial Step}: 
 \begin{itemize}
   \item[1.1] Standardize the $I$ sets of functional time series $\bb{y}_i(u)$ separately for each $i$, and combine the standardized curves into a single combined set of $I\times T$ curves.
   \item [1.2] Apply standard FPCA to these $I\times T$ curves and select the optimal number of FPC, $Q$, so that each set of functional time series, $\bb{y}_i(u)$, can be represented by a $T\times Q$ FPC score matrix. 
   \item [1.3] Perform $k$-means clustering to the $I$ score matrices of dimension $T\times Q$ by varying the cluster number from $2$ to $K$.  
   \item [1.4] Calculate the total within-cluster sum of squares per cluster, $\widehat d_k$ for $k\in\{2, \ldots, K\}$.
   \item[1.4] Determine the optimal number of clusters, $K_{\text{opt}}$, using Equation~\eqref{eq:9};
   \item [1.5] Given $K_{\text{opt}}$, apply the conventional $k$-means clustering to these $I$ score matrices to obtain initial membership, $\mathcal C^{(1)} = \{c_i^{(1)}, i = 1, \ldots, I\}$, where $c_i^{(1)} \in \{1, \ldots, K_{\text{opt}}\}$. 
 \end{itemize}
 \textbf{2. Iterative Step}: 
 \begin{itemize}
   \item[2.1] Set $l = 1$, $K=K_{\text{opt}}$;
   \item[2.2] Based on the membership $\mathcal C^{(l)}$, excluding the $i^{\text{th}}$ object, obtain the estimated structure components of any cluster $c\in\{1, \ldots, K\}$: $\widehat\mu^{(c)}_{(-i)} (x)$, $\widehat\phi^{(c)}_{m(-i)}$, $\widehat\rho^{(c)}_{k(-i)}$, and $\widehat\psi^{(c)}_{l(-i)}$; 
   \item[2.3] Obtain the leave-one-out predicted $t^{\textsuperscript{th}}$ curve of the $i^{\textsuperscript{th}}$ object at the $l^{\textsuperscript{th}}$ step, $\widehat y_{it}^{(c)(l)}(x)$, based on the functional structure of cluster $c, c=1,\ldots, K$, as calculated in Step $2.2$;
   \item[2.4] Update membership of the $i^{\textsuperscript{th}}$ object to $c_i^{(l+1)}$, such that 
   \begin{equation*}
    c_i^{(l+1)} = \underset{c\in \{1, 2, \ldots, K\}}{\operatorname{argmin}}\frac{1}{T}\sum_{t=1}^{T}\left\|y_{it} - \widehat y_{it}^{(c)(l)}\right\|;
   \end{equation*}
   \item[2.5] Update $l = l+1$, while $\mathcal C^{(l)}\neq \mathcal C^{(l-1)}$, repeat from Step $2.2$ to Step $2.4$.
 \end{itemize}     
 \KwResult{$\mathcal C^{(l_0)} = \{c_i^{(l_0)}, i = 1, 2, \ldots, I\}$, where $l_0$ is the smallest step that satisfies $\mathcal{C}^{(l_0)}=\mathcal{C}^{(l_0-1)}$. }
  \caption{Multiple functional time series clustering procedures}
\end{algorithm}

\section{Simulation studies}\label{sec:5}

We evaluate the finite sample performance of the proposed clustering technique via Monte Carlo simulations. The actual cluster results will never be known for cluster analysis; however, we can generate objects from pre-known groups with simulations. The clustering quality of specific methods can then be measured. Two typical clustering quality measures are the correct classification rate (cRate) and the Rand index \citep{hubert1985comparing}. The cRate is the ratio of correctly classified objects to the total number of objects to be clustered. At the same time, the Rand index measures the similarity between the clustering result and the actual membership. The Rand index is altered to the adjusted Rand index (aRand) such that it has an expected value of 0 and is bounded by 1. A higher aRand value indicates greater similarity between two groups, i.e., higher cluster quality, while a value of $0$ suggests that the clustering method is purely a random guess. The simulation studies consist of two parts. In the first part, we demonstrate the importance of each component of the proposed model in improving the clustering performance. The second part illustrates the ability of the proposed clustering technique in handling a more complicated case.  

\subsection{A comparison of clustering quality}\label{subsec:5.1}

In this subsection, we evaluate the clustering performance by examining various combinations of the structure components (i.e., mean functions, eigenfunctions, and FPC scores) of the proposed model.

\subsubsection{Data generating process}

The functional time series data for cluster $c$, $c \in\{1,2\}$ are generated from the following model contaminated with measurement error $\upsilon_{it}^{(c)}(x)$:
\begin{equation*}
  Y_{it}^{(c)} (x)= \mu_i^{(c)}(x) + \sum_{k=1}^{2}\xi_{tk}^{(c)} \rho_k^{(c)} (x) + \sum_{l=1}^{2}\zeta_{itl}^{(c)} \psi_l^{(c)} (x) + \upsilon_{it}^{(c)} (x), 
\end{equation*}
with $\mu_i^{(c)}(x)$ being the mean function, which can be treated as the sum of $\mu^{(c)}(x)$ and $\eta_i^{(c)}(x)$ by definition, $\sum_{k=1}^{2}\xi_{tk}^{(c)} \rho_k^{(c)} (x)$ being the common time trend component and $\sum_{l=1}^{2}\zeta_{itl}^{(c)} \psi_l^{(c)} (x)$ being the country-specific time trend component. More specifically, the temporal dynamics are reflected in $\xi_{tk}^{(c)}$ and $\zeta_{itl}^{(c)}$, while the eigenfunctions for these two components, $\rho_k^{(c)}(x)$ and $\psi_l^{(c)}(x)$ represent the model of variation. 

The variates $\bb{\xi_{k}^{(c)}}=(\xi_{1k}^{(c)}, \xi_{2k}^{(c)}, \ldots, \xi_{Tk}^{(c)})^{\top}$ and $\bb{\zeta_{il}^{(c)}} = (\zeta_{i11}^{(c)}, \zeta_{i2l}^{(c)}, \ldots, \zeta_{iTl}^{(c)})^{\top}$ are generated from autoregressive of order 1 with parameters $\phi_{k}^{(c)}$ and $\tau_{l}^{(c)}$, respectively, and the measurement error $\upsilon_{it}$ for each continuum x is generated from independent and identically distributed $N(0, \sigma^2)$ and \{$x = \frac{m}{200} : m = 0,1,\ldots,200$\}.  The two candidates of the mean functions are $\mu^{(1)}(x) = -2(x-0.25)^{2} + 1.5$ and $\mu^{(2)}(x) = 4(x-0.6)^{2}+1$. We set the noise level to be moderate, with $\sigma = 0.2$.

We consider various scenarios by selecting a combination of eigenvalues and eigenfunctions for different clusters. We consider the following candidates of eigenvalues and eigenfunctions:
\begin{itemize}
  \item $\bb{\Phi_1} = (\phi_{11}, \phi_{12}) = (0.7, 0.6)$ and  $\bb{\Phi}_2 = (\phi_{21}, \phi_{22}) = (0.6, 0.5)$;
  \item $\bb{\tau_1} =(\tau_{11}, \tau_{12}) = (0.5, 0.4)$ and $\bb{\tau_2} =(\tau_{21}, \tau_{22}) = (0.3, 0.2)$;
  \item $\bb{E_1}$ = span($\nu_{11}$, $\nu_{12}$), where $\nu_{11} = \sqrt2\sin(\pi x)$ and $\nu_{12} = \sqrt2\cos(\pi x)$;
  \item $\bb{E_2}$ = span($\nu_{21}$, $\nu_{22}$), where $\nu_{21} = \sqrt2\sin(2\pi x)$ and $\nu_{22} = \sqrt2\cos(2\pi x)$;
  \item $\bb{E_3}$ = span($\nu_{31}$, $\nu_{32}$), where $\nu_{31} = \sqrt2\sin(3\pi x)$ and $\nu_{32} = \sqrt2\cos(3\pi x)$;
  \item $\bb{E_4}$ = span($\nu_{41}$, $\nu_{42}$), where $\nu_{41} = \sqrt2\sin(4\pi x)$ and $\nu_{42} = \sqrt2\cos(4\pi x)$,
\end{itemize} 
where $\bb{\Phi_.} = (\phi_{.1}, \phi_{.2})$ are the parameter candidates for $\phi_{k}^{(c)}$, for $k = 1, 2$ and $\bb{\tau_.} =(\tau_{.1}, \tau_{.2})$ are the parameter candidates for $\tau_l^{(c)}$, for $l = 1, 2$; $\bb{E_1}$ and $\bb{E_2}$ are the candidates for eigenfunctions of the functional common time trend component and $\bb{E_3}$ and $\bb{E_4}$ are the candidates for eigenfunctions of functional pattern component. 

To evaluate the importance of common time and country-specific time trends in clustering, firstly, we generate two clusters, where we use $\mu^{(1)}(x)$ for the mean function for both clusters. Then we allow for different combinations of scores and eigenfunctions from candidates listed above. Therefore, we have $16$ different combinations to consider. The various combinations are listed in Table~\ref{tab:2}. 
\begin{table}[!htbp] 
\centering 
\footnotesize{
  \caption{Simulation designs for different combinations of eigenvalues and eigenfunctions. In the top-left panel, the eigenfunctions are the same for the two components. In the top-right panel, the eigenfunctions for the common time trend component are different. In the bottom-left panel, the eigenfunctions for the country-specific time trend component are different. In the bottom-right panel, the eigenfunctions are different for two components. Scenarios from C1 to C4 are more and more distinct, and among each category, scenarios from $a$ to $d$ are more and more distinct.} 
  \label{tab:2} 
  \setlength{\tabcolsep}{20pt}
\renewcommand{\arraystretch}{1.4}
\begin{tabular}{@{}ccc@{}} 
\toprule
 & \multicolumn{2}{c}{Level 1 Eigenspaces}\\\cmidrule{2-3}\\ 
Level 2 Eigenspaces & $\bm{S_1}^{(1)} = \bm{S_1}^{(2)} $ & $\bm{S_1}^{(1)} \neq \bm{S_1}^{(2)} $\\
\midrule
\multirow{4}{*}{$\bm{S_2}^{(1)} = \bm{S_2}^{(2)}$} & C1a: $\bm{\Phi}^{(1)} = \bm{\Phi}^{(2)}$; $\bm{\tau}^{(1)} = \bm{\tau}^{(2)}$ & C3a: $\bm{\Phi}^{(1)} = \bm{\Phi}^{(2)}$; $\bm{\tau}^{(1)} = \bm{\tau}^{(2)} $ \\
&C1b: $\bm{\Phi}^{(1)} = \bm{\Phi}^{(2)}$; $\bm{\tau}^{(1)} \neq \bm{\tau}^{(2)}$ & C3b: $\bm{\Phi}^{(1)} = \bm{\Phi}^{(2)} $; $\bm{\tau}^{(1)} \neq \bm{\tau}^{(2)}$\\
&C1c:  $\bm{\Phi}^{(1)} \neq \bm{\Phi}^{(2)}$; $\bm{\tau}^{(1)} = \bm{\tau}^{(2)}$ & C3c: $\bm{\Phi}^{(1)} \neq \bm{\Phi}^{(2)}$; $\bm{\tau}^{(1)} =  \bm{\tau}^{(2)}$\\
&C1d: $\bm{\Phi}^{(1)} \neq \bm{\Phi}^{(2)} $; $\bm{\tau}^{(1)} \neq \bm{\tau}^{(2)}$ & C3d: $\bm{\Phi}^{(1)} \neq \bm{\Phi}^{(2)}$; $\bm{\tau}^{(1)} \neq \bm{\tau}^{(2)} $\\
\midrule

\multirow{4}{*}{$\bm{S_2}^{(1)} \neq \bm{S_2}^{(2)}$} &C2a: $\bm{\Phi}^{(1)} = \bm{\Phi}^{(2)} $; $\bm{\tau}^{(1)} =  \bm{\tau}^{(2)} $ & C4a: $\bm{\Phi}^{(1)} = \bm{\Phi}^{(2)} $; $\bm{\tau}^{(1)} = \bm{\tau}^{(2)}$ \\
&C2b: $\bm{\Phi}^{(1)} = \bm{\Phi}^{(2)} $; $\bm{\tau}^{(1)} \neq \bm{\tau}^{(2)}$ & C4b: $\bm{\Phi}^{(1)} = \bm{\Phi}^{(2)}$; $\bm{\tau}^{(1)} \neq \bm{\tau}^{(2)}$\\
&C2c: $\bm{\Phi}^{(1)} \neq \bm{\Phi}^{(2)} $; $\bm{\tau}^{(1)} = \bm{\tau}^{(2)} $ & C4c: $\bm{\Phi}^{(1)} \neq \bm{\Phi}^{(2)} $; $\bm{\tau}^{(1)} = \bm{\tau}^{(2)}$ \\
&C2d: $\bm{\Phi}^{(1)} \neq \bm{\Phi}^{(2)} $; $\bm{\tau}^{(1)} \neq \bm{\tau}^{(2)}$ & C4d: $\bm{\Phi}^{(1)} \neq \bm{\Phi}^{(2)}$; $\bm{\tau}^{(1)} \neq \bm{\tau}^{(2)} $ \\
\bottomrule \\[-1.8ex]  
\end{tabular} 
}
\end{table}

As proposed by our model, the common time trend is a stronger feature in clustering and the country-specific time trend serves as a compliment. It is clear that from designs C$1$ to C$4$, the two groups are more and more distinct. This is because, in design C$1$, the eigenfunctions for both components are the same. In design C$2$, the eigenfunctions for the country-specific time trend are different. In design C$3$, the eigenfunctions for the common-time trend (i.e., the strong characteristic) are different. While within each design, from scenarios $a$ to $d$, the scores are more and more dissimilar, which means that the scenarios are more and more distinct across $a$ to $d$. In general, scenarios in the top-left section of Table~\ref{tab:2} (design C$1$) represent the least distinct scenarios; while the top-right and bottom-left sections of Table~\ref{tab:2} (design C$3$ and design C$2$ correspondingly) represent moderately distinct scenarios; while the bottom-right section of Table~\ref{tab:2} corresponds to the most distinct scenarios.

To evaluate the importance of the deviation of the country-specific mean from the overall mean in clustering, we further create designs C$0$ with different mean functions, $\mu^{(1)}$ and $\mu^{(2)}$, for different groups, all other settings are the same as design C$1$. We generate $25$ subjects $\times$ $2$ groups $\times$ $61$ curves to evaluate the clustering quality. For each scenario, we conduct $100$ replications.

\subsubsection{Comparison of results}

To assess our proposed method's clustering quality, we compare the measures of the clustering results with other competitive methods. In general, the competitive methods can be classified into three categories: conventional clustering on dimension reduced data, such as $k$-means (kmeans) and hierarchical clustering (hclust) on the FPC scores; univariate functional clustering, such as the $k$-centers functional clustering ($k$CFC) of \cite{chiou2007functional} and discriminative functional mixture model clustering (funFEM) \citep{bouveyron2015discriminative}; and multivariate functional clustering, such as functional high-dimensional data clustering (funHDDC) \citep{bouveyron2011model}, functional latent block model clustering (funLBM) \citep{slimen2018model}, and the $k$-means clustering on multilevel FPCA ($\text{MFPCA}_k$) of \cite{serban2012multilevel}. When applying the univariate clustering method to the generated data, we calculate the functional median \citep{lopez2009concept}, which is the best representative of the set of functions. We calculate the averaged cRate, aRand, and the number of iterations before convergence (whenever possible) for the different methods. Table~\ref{tab:3} summarizes the comparisons of cRates and aRands for all the clustering methods. 

\afterpage{\begin{centering}
\renewcommand{\arraystretch}{0.7}
 \setlength{\tabcolsep}{3.6pt}
 \footnotesize{
\begin{longtable}{@{\extracolsep{0pt}}llllllllll@{}}
\caption{A comparison of cluster qualities of different methods for different scenarios.} \\\toprule
  \label{tab:3}
Scenarios & & $k$-means & hclust & funFEM & $k$CFC  &  funHDDC & funLBM & $\text{MFPCA}_k$  & MFTSC \\  
\midrule
\endfirsthead
\multicolumn{10}{c}%
{\textbf{\tablename}\ \textbf{\thetable:}\ \textit{ A comparison of cluster qualities of different methods for different scenarios (continued).}} \\
\\
\toprule

Scenarios & & $k$-means & hclust & funFEM & $k$CFC  &  funHDDC & funLBM & $\text{MFPCA}_k$  & MFTSC \\
\midrule
\endhead
\hline \multicolumn{10}{r}{\textit{Continued on next page}} \\
\endfoot
\hline
\endlastfoot
\multirow{3}{*}{C0a} & cRate & 0.566 & 0.546 & 0.581 & 0.580 & 0.577 & 0.557 & 0.566 & \textbf{0.907}  \\
  & aRand & 0.009 & -0.002 & 0.025 & 0.065 & 0.018 & 0.002 & 0.012 & \textbf{0.716}\\
  & Iter. No. & - & - & - & \textbf{6.87} & - & - & - & 9.79 \\
 \midrule
  \multirow{3}{*}{C0b} & cRate & 0.582 & 0.553 & 0.568 & 0.586 & 0.576 & 0.558 & 0.553 & \textbf{0.926}  \\
  & aRand & 0.020 & 0.002 & 0.013 & 0.069 & 0.022 & 0.000 & 0.000 & \textbf{0.774} \\
  & Iter. No. & - & - & - & \textbf{7.24} & - & - & - & 7.77 \\
\midrule
  \multirow{3}{*}{C0c} & cRate & 0.614 & 0.553 & 0.589 & 0.600 & 0.565 & 0.555 & 0.558 & \textbf{0.924}  \\
  & aRand & 0.051 & 0.003 & 0.033 & 0.097 & 0.009 & 0.000 & 0.003 & \textbf{0.769}  \\
  & Iter. No. & - & - & - & 7.46 & - & - & - & \textbf{3.53} \\
\midrule
  \multirow{3}{*}{C0d} & cRate & 0.680 & 0.566 & 0.577 & 0.620 & 0.562 & 0.565 & 0.562 & \textbf{0.952}   \\
  & aRand & 0.134 & 0.010 & 0.026& 0.129 & 0.008 & 0.006 & 0.008 & \textbf{0.847} \\
  & Iter. No. & - & - & - & 7.33 & - & - & - & \textbf{2.93} \\
\midrule
\multirow{3}{*}{C1a} & cRate & 0.560 & 0.553 & 0.556 & 0.515 & \textbf{0.569} & 0.544 & 0.564 & 0.503  \\
  & aRand & 0.004 & -0.001 & 0.001 & 0.001 & \textbf{0.009} & -0.005 & 0.006 & 0.000 \\
  & Iter. No. & - & - & - & \textbf{3.67} & - & - & - & 3.79 \\
 \midrule
  \multirow{3}{*}{C1b} & cRate & 0.572 & 0.548 & 0.559 & 0.517 & \textbf{0.580} & 0.550 & 0.571 & 0.505  \\
  & aRand & 0.011 & 0.001 & 0.003 & 0.001 & \textbf{0.021} & -0.004 & 0.010 & 0.000 \\
  & Iter. No. & - & - & - & \textbf{3.83} & - & - & - & 6.62 \\
\midrule
  \multirow{3}{*}{C1c} & cRate & \textbf{0.612} & 0.553 & 0.553 & 0.520 & 0.578 & 0.558 & 0.567 & 0.503  \\
  & aRand & \textbf{0.053} & 0.008 & 0.004 & 0.000 & 0.018 & 0.002 & 0.009 & 0.000  \\
  & Iter. No. & - & - & - & \textbf{3.10} & - & - & - & 4.21 \\
\midrule
  \multirow{3}{*}{C1d} & cRate & \textbf{0.711} & 0.556 & 0.558 & 0.522 & 0.567 & 0.550 & 0.548 & 0.504   \\
  & aRand & \textbf{0.179} & 0.006 & 0.006 & 0.000 & 0.011 & -0.003 & -0.002 & 0.000 \\
  & Iter. No. & - & - & - & \textbf{3.74} & - & - & - & 4.67 \\
\midrule
\multirow{3}{*}{C2a} & cRate & 0.553 & 0.549 & 0.554 & 0.720 & 0.560 & 0.557 & 0.550 & \textbf{0.801}  \\
  & aRand & 0.003 & -0.001 & -0.002 & 0.242 & 0.004 & 0.000 & -0.002 & \textbf{0.573}  \\
  & Iter. No. & - & - & - & 7.77 & - & - & - & \textbf{5.88} \\
 \midrule
  \multirow{3}{*}{C2b} & cRate & 0.571 & 0.544 & 0.560 & 0.705 & 0.577 & 0.555 & 0.551 & \textbf{0.827}  \\
  & aRand & 0.008 & -0.002 & 0.012 & 0.219 & 0.019 & 0.003 & 0.004 & \textbf{0.601} \\
  & Iter. No. & - & - & - & 7.46 & - & - & - & \textbf{5.78} \\
\midrule
  \multirow{3}{*}{C2c} & cRate & 0.621 & 0.546 & 0.554 & 0.708 & 0.580 & 0.549 & 0.548 & \textbf{0.874}  \\
  & aRand & 0.063 & -0.001 & 0.005 & 0.235 & 0.018 & -0.003 & 0.001 & \textbf{0.757}  \\
  & Iter. No. & - & - & - & 7.76 & - & - & - & \textbf{5.18}  \\
\midrule
  \multirow{3}{*}{C2d} & cRate & 0.687 & 0.561 & 0.564 & 0.717 & 0.563 & 0.545 & 0.568 & \textbf{0.898}  \\
  & aRand & 0.143 & 0.008 & 0.008 & 0.223 & 0.007 & -0.007 & 0.009 & \textbf{0.763} \\
  & Iter. No. & - & - & - & 6.86 & - & - & - & \textbf{5.92} \\
\midrule
\multirow{3}{*}{C3a} & cRate & 0.557 & 0.545 & 0.598 & 0.572 & 0.562 & 0.555 & 0.557 & \textbf{0.826}  \\
  & aRand & 0.001 & -0.001  & 0.042 & 0.046 & 0.007 & 0.000 & 0.005 & \textbf{0.637} \\
  & Iter. No. & - & - & - & 6.43 & - & - & - & \textbf{5.19} \\
 \midrule
  \multirow{3}{*}{C3b} & cRate & 0.576 & 0.551 & 0.599 & 0.575 & 0.563 & 0.560 & 0.558 & \textbf{0.863}  \\
  & aRand & 0.017 & 0.002 & 0.044 & 0.048 & 0.007 & 0.002 & 0.005 & \textbf{0.717} \\
  & Iter. No. & - & - & - & 7.38 & - & - & - & \textbf{4.85} \\
\midrule
  \multirow{3}{*}{C3c} & cRate & 0.599 & 0.551 & 0.602 & 0.565 & 0.575 & 0.553 & 0.550 & \textbf{0.891}   \\
  & aRand & 0.037 & -0.001 & 0.046 & 0.040 & 0.016 & -0.002 & 0.003 & \textbf{0.754} \\
  & Iter. No. & - & - & - & 6.08 & - & - & - & \textbf{3.93} \\
\midrule
  \multirow{3}{*}{C3d} & cRate & 0.690 & 0.566 & 0.602 & 0.588 & 0.564 & 0.557 & 0.557 & \textbf{0.904}   \\
  & aRand & 0.150 & 0.011 & 0.057 & 0.067 & 0.008 & 0.002 & 0.008 & \textbf{0.828} \\
  & Iter. No. & - & - & - & 6.74 & - & - & - & \textbf{3.66} \\
\midrule
\multirow{3}{*}{C4a} & cRate & 0.557 & 0.547 & 0.670 & 0.718 & 0.573 & 0.554  & 0.550 & \textbf{0.989}   \\
  & aRand & 0.002 & -0.001 & 0.161 & 0.265 & 0.014 & -0.002  & 0.004 & \textbf{0.966}  \\
  & Iter. No. & - & - & - & 6.53 & - & - & - & \textbf{5.01} \\
 \midrule
  \multirow{3}{*}{C4b} & cRate & 0.569 & 0.554 & 0.672 & 0.734 & 0.556 & 0.559 & 0.554 & \textbf{0.993}   \\
  & aRand & 0.012 & 0.004 & 0.163 & 0.258 & 0.003 & 0.005 & 0.007 & \textbf{0.967} \\
  & Iter. No. & - & - & - & 7.66 & - & - & - &\textbf{4.83}  \\
\midrule
  \multirow{3}{*}{C4c} & cRate & 0.617 & 0.531 & 0.666 & 0.652 & 0.559 & 0.559 & 0.546 & \textbf{0.996}  \\
  & aRand & 0.049 & -0.007 & 0.178 & 0.120 & 0.006 & 0.000 & 0.007 & \textbf{0.977} \\
  & Iter. No. & - & - & - & 6.75 & - & - & - & \textbf{4.23} \\
\midrule
  \multirow{3}{*}{C4d} & cRate & 0.701  & 0.564 & 0.697 & 0.742 & 0.568 & 0.554  &  0.553 & \textbf{0.996}  \\
  & aRand & 0.165 & 0.009 & 0.213 & 0.297 & 0.011  & 0.002 & 0.004 & \textbf{0.984} \\
  & Iter. No. & - & - & - & 6.70 & - & - & - & \textbf{3.77}\\
\bottomrule
\end{longtable}
}
\end{centering}
}

We observe that our proposed method, multiple functional time series clustering (MFTSC), performs the best in most cases, with very high cRates and aRands, except for the least distinct cases, where none of the other clustering methods performs well. More specifically, the clustering methods with an explicit membership updating step ($k$CFC and MFTSC) tend to perform better for the most distinct scenarios. In terms of convergence speed, our proposed method has a lower number of iterations until convergence than that of $k$CFC. However, for the least distinct case (designs C$1$), the clustering method based on FPC scores performs slightly better than the other methods (funHDCC for C$1$a, MFPCA$_k$ for C$1$b and $k$-means for C$1$c and d). This is because the eigenspaces for both clusters are the same under these cases. The inferior performance of the proposed method is that it relies on all the three components, namely, the deviation of the country-specific mean from the overall mean ($\eta_i$), the common time trend ($R_t$), and the country-specific time trend ($U_{it}$) to perform accurate clustering. For design C$1$, where the mean functions and the eigenfunctions for both levels are the same, data from two groups are too similar and are randomly assigned to either cluster. This is evident from the cRates, which are very close to $0.5$. This phenomenon indicates the clustering is almost a random guess. An investigation into the results for design C$0$ finds out that, despite the eigenfunctions being the same for both levels, the proposed method can outperform other methods if the mean functions are different. This confirms with the proposal that our model relies on all these three components to perform clustering.

Moreover, both the aRands and cRates for designs C$1$ to C$4$ and scenarios from $a$ to $d$ of all the designs C$0$ to C$4$ are increasing. This finding confirms with the proposed model that the common time trend, $R_t$, is the strong characterization, and the country-specific time trend, $U_{it}$, serves as a supplement.

\subsection{A more complicated case}
\begin{table}[!htbp] \centering 
  \caption{Candidates of simulation design for a more complicated case, where we assume six bases for the common time trend and three bases for the country-specific time trend} 
  \label{tab:9} 
  \setlength{\tabcolsep}{15pt}
\renewcommand{\arraystretch}{1.4}
\scriptsize{
\begin{tabular}{@{}lcc@{}} 
\toprule 
Components & Cluster 1 & Cluster 2\\
\midrule
$\mu^{(c)}(x)$ & $\mu^{(1)}(x) = -2(x-0.25)^{2} + 1.5$  & $\mu^{(2)}(x) = 4(x-0.6)^{2}+1$\\
\midrule
\multirow{3}{*}{$\rho_k^{(c)}(x)$} & $\rho_1^{(1)}(x) =\sqrt2\sin(\pi x)$; $\rho_2^{(1)}(x) =\sqrt2\sin(2\pi x)$ & $\rho_1^{(2)}(x) =\sqrt2\sin(5\pi x)$; $\rho_2^{(2)}(x) =\sqrt2\sin(6\pi x)$\\
& $\rho_3^{(1)}(x) =\sqrt2\sin(3\pi x)$; $\rho_4^{(1)}(x) =\sqrt2\sin(4\pi x)$ & $\rho_3^{(2)}(x) =\sqrt2\sin(7\pi x)$; $\rho_4^{(2)}(x) =\sqrt2\sin(8\pi x)$ \\
& $\rho_5^{(1)}(x) =\sqrt2\sin(5\pi x)$; $\rho_6^{(1)}(x) =\sqrt2\sin(6\pi x)$ & $\rho_5^{(2)}(x) =\sqrt2\sin(9\pi x)$; $\rho_6^{(2)}(x) =\sqrt2\sin(10\pi x)$\\
\midrule
\multirow{2}{*}{$\psi_l^{(c)}(x)$} & $\psi_1^{(1)}(x) = \sqrt2\sin(\pi x)$;$\psi_2^{(1)}(x) = \sqrt2\sin(2\pi x)$ & $\psi_1^{(2)}(x) = \sqrt2\sin(4\pi x)$; $\psi_2^{(2)}(x) = \sqrt2\sin(5\pi x)$\\

& $\psi_3^{(1)} = \sqrt2\sin(3\pi x)$ &  $\psi_3^{(2)} = \sqrt2\sin(6\pi x)$\\
\midrule
\multirow{3}{*}{$\xi_{tk}^{(c)}$} & $\xi_{t1}^{(1)} \sim AR(1), \phi_1 = 0.9$; $\xi_{t2}^{(1)} \sim AR(1), \phi_1 = 0.8$ & $\xi_{t1}^{(2)} \sim AR(1), \phi_1 = 0.8$; $\xi_{t2}^{(2)} \sim AR(1), \phi_1 = 0.7$ \\
& $\xi_{t3}^{(1)} \sim AR(1), \phi_1 = 0.7$; $\xi_{t4}^{(1)} \sim AR(1), \phi_1 = 0.6$ & $\xi_{t3}^{(2)} \sim AR(1), \phi_1 = 0.6$; $\xi_{t4}^{(2)} \sim AR(1), \phi_1 = 0.5$\\
& $\xi_{t5}^{(1)} \sim AR(1), \phi_1 = 0.5$; $\xi_{t6}^{(1)} \sim AR(1), \phi_1 = 0.4$ & $\xi_{t5}^{(2)} \sim AR(1), \phi_1 = 0.4$; $\xi_{t6}^{(2)} \sim AR(1), \phi_1 = 0.3$\\
\midrule
\multirow{2}{*}{$\zeta_{itl}^{(c)}$} & $\zeta_{it1}^{(1)} \sim AR(1), \phi_1 = 0.5$; $\zeta_{it2}^{(1)} \sim AR(1), \phi_1 = 0.4$ & $\zeta_{it1}^{(2)} \sim AR(1), \phi_1 = 0.4$; $\zeta_{it2}^{(2)} \sim AR(1), \phi_1 = 0.3$\\
& $\zeta_{it3}^{(1)} \sim AR(1), \phi_1 = 0.3$ & $\zeta_{it3}^{(2)} \sim AR(1), \phi_1 = 0.2$\\
\bottomrule \\[-1.8ex]  
\end{tabular} 
}
\end{table}

We also examine the proposed clustering method's ability to handle more complex cases with relatively slow decaying eigenvalues. In this case, we have generated two groups of data assuming six basis functions for the common time trend and $3$ basis functions for the country-specific time trend. 

The data are generated from the following model:
\begin{equation*}
  Y_{it}^{(c)} (x)= \mu^{(c)}(x) + \sum_{k=1}^{6}\xi_{tk}^{(c)} \rho_k^{(c)} (x) + \sum_{l=1}^{3}\zeta_{itl}^{(c)} \psi_l^{(c)} (x) + \upsilon_{it}^{(c)} (x), 
\end{equation*}
where each terms follows exactly the same specification described in Section ~\ref{subsec:5.1}, with the selected candidates for parameters of both clusters summarized in Table~\ref{tab:9}. More specifically, $\xi_{tk}^{(c)}$ and $\zeta_{itl}^{(c)}$ follow auto-regressive time series model of order $1$ (AR(1)) with different parameters for various $c$, $k$ and $l$. The eigenfunctions for both levels are not necessarily orthogonal. 

Competitive clustering methods, together with the proposed clustering method, are applied to the data. Comparisons of cRate and aRand are tabulated in Table~\ref{tab:10}. As we can see, the proposed clustering method performs the best under this complicated scenario, followed by funFEM and $k$CFC. Although the performance of all these three methods stands out, our method has very high cRate and aRand, which are close to $1$ and simultaneously demonstrate the clustering quality. Moreover, our method uses fewer iterations than that of $k$CFC to converge. Therefore our method is faster in terms of the convergence speed. 

\begin{table}[!htbp] \centering 
  \caption{A comparison of cluster qualities of different methods for the complicated case} 
  \label{tab:10} 
  \setlength{\tabcolsep}{6.5pt}
\renewcommand{\arraystretch}{1.5}
\begin{tabular}{@{}llllllllll@{}} 
\toprule
 & & $k$-means & hclust & funFEM & $k$CFC  &  funHDDC & funLBM & $\text{MFPCA}_k$  & MFTSC \\
\midrule
\multirow{3}{*}{} & cRate & 0.669 & 0.573 & 0.811 & 0.745 & 0.569 & 0.571 & 0.589 & \textbf{0.999}  \\
  & aRand & 0.141 & 0.015 & 0.473 & 0.391 & 0.044 & 0.025 & 0.038 & \textbf{0.996} \\
  & Iter. No. & - & - & - & 4.88 & - & - & - & \textbf{2.99} \\
\bottomrule
\end{tabular} 
\end{table}

In summary, our proposed method is more satisfactory in clustering functional time series objects with very high accuracy unless the objects to be clustered are very similar, which is deemed to be robust. 

\section{Application to age-specific mortality rates}\label{sec:6}

We apply the proposed clustering algorithm to the age-specific mortality rates. Once the cluster membership is determined, a functional panel data model with fixed effects is applied to each cluster to produce point and interval forecasts. We then evaluate and compare the point and interval forecasts accuracies with independent functional time series forecasts. The forecasting results based on applying our proposed model to the initial cluster membership are also compared.

We extract the mortality data of $32$ countries from the years $1960$ to $2010$. Due to the sparsity of the exposures and deaths at advanced ages, we aggregate the female and total data aged over $100$ and the male data aged over $98$ before smoothing them \citep[we use \texttt{demography} in \texttt{R} of][to process the mortality data]{demography19}. After taking the logarithms of the smoothed central mortality rates for all those $32$ countries, we performed the clustering procedure on the data. We use the data from $1960$ to $2000$ to perform clustering and modeling and data from $2001$ to $2010$ to evaluate the forecasting accuracy. 

\subsection{Mortality clustering results}

\begin{figure}[!htbp]
\centering
\includegraphics[width=10.2cm]{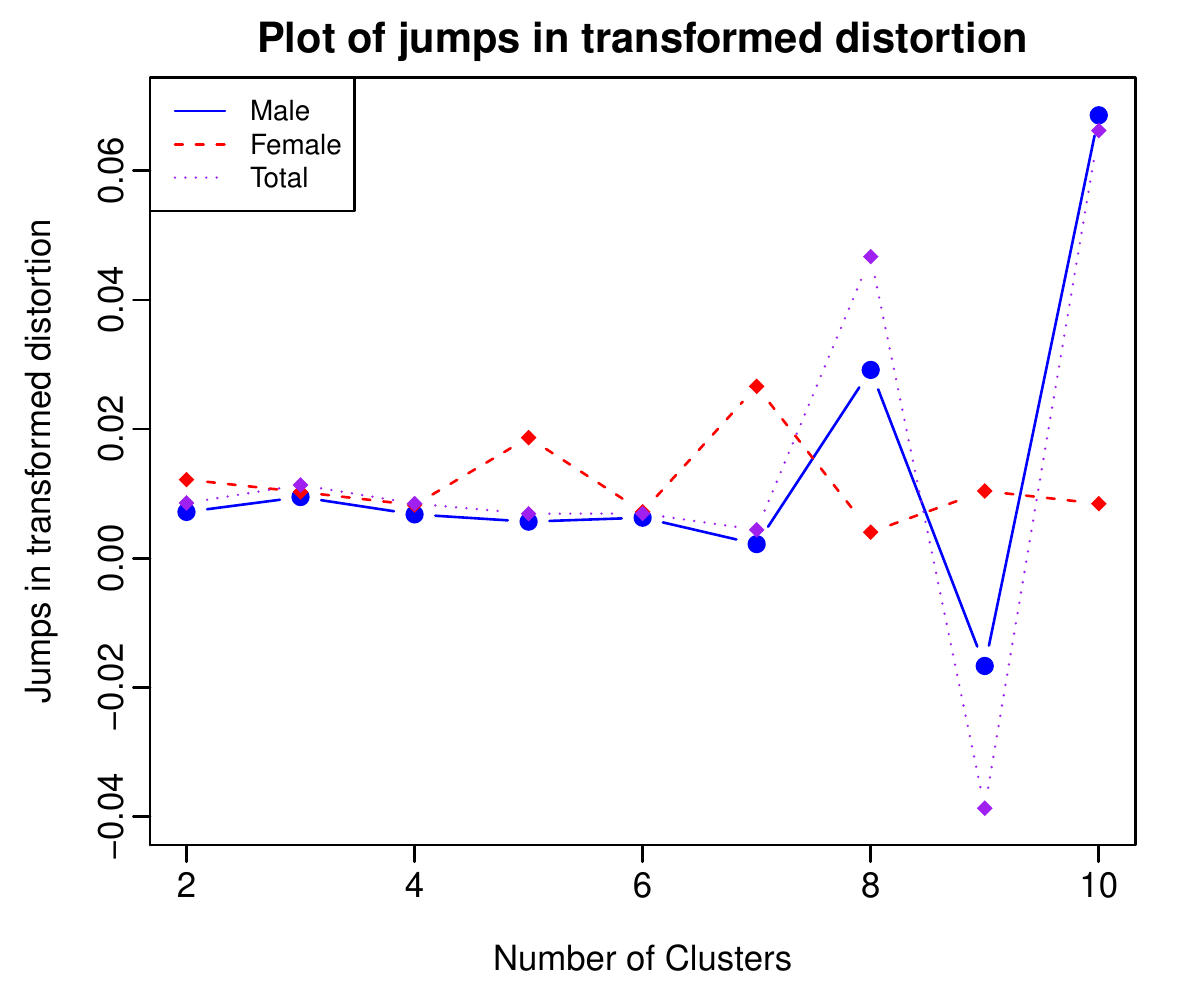}
\caption{A plot of the pairwise difference in negative power transformed distortion to determine the optimal number of clusters. Seven clusters are recognized as optimal for female age-specific mortality, Ten clusters are optimal for male age-specific mortality, and ten clusters are optimal for total age-specific mortality}
\label{fig:6}
\end{figure}

In the initial step, seven clusters are recognized as optimal for females, and ten clusters are optimal for male and total mortality. Figure~\ref{fig:6} shows a plot of the pairwise difference in the negative power transformed distortion to determine the optimal number of clusters. In the iterative step in clustering the age-specific mortality rates, full convergence is achieved after a few iterations. At convergence, we have three clusters for females, six for males, and five for the total.

\begin{figure}[!ht]
    \centering
    \subfloat[Initial clustering result, Female]{{\includegraphics[width=7.15cm]{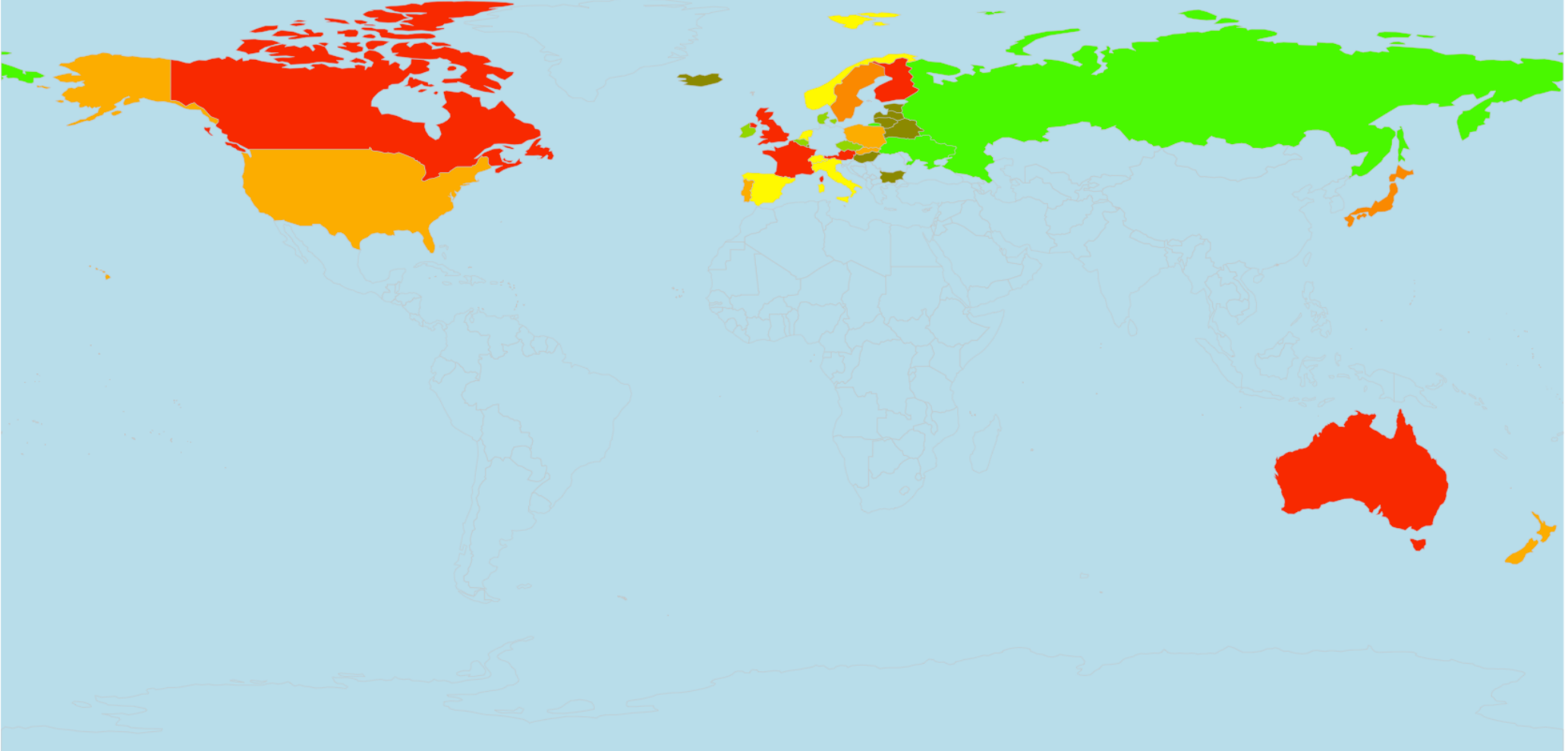} }}
    \qquad
    \subfloat[Final clustering result, Female]{{\includegraphics[width=7.15cm]{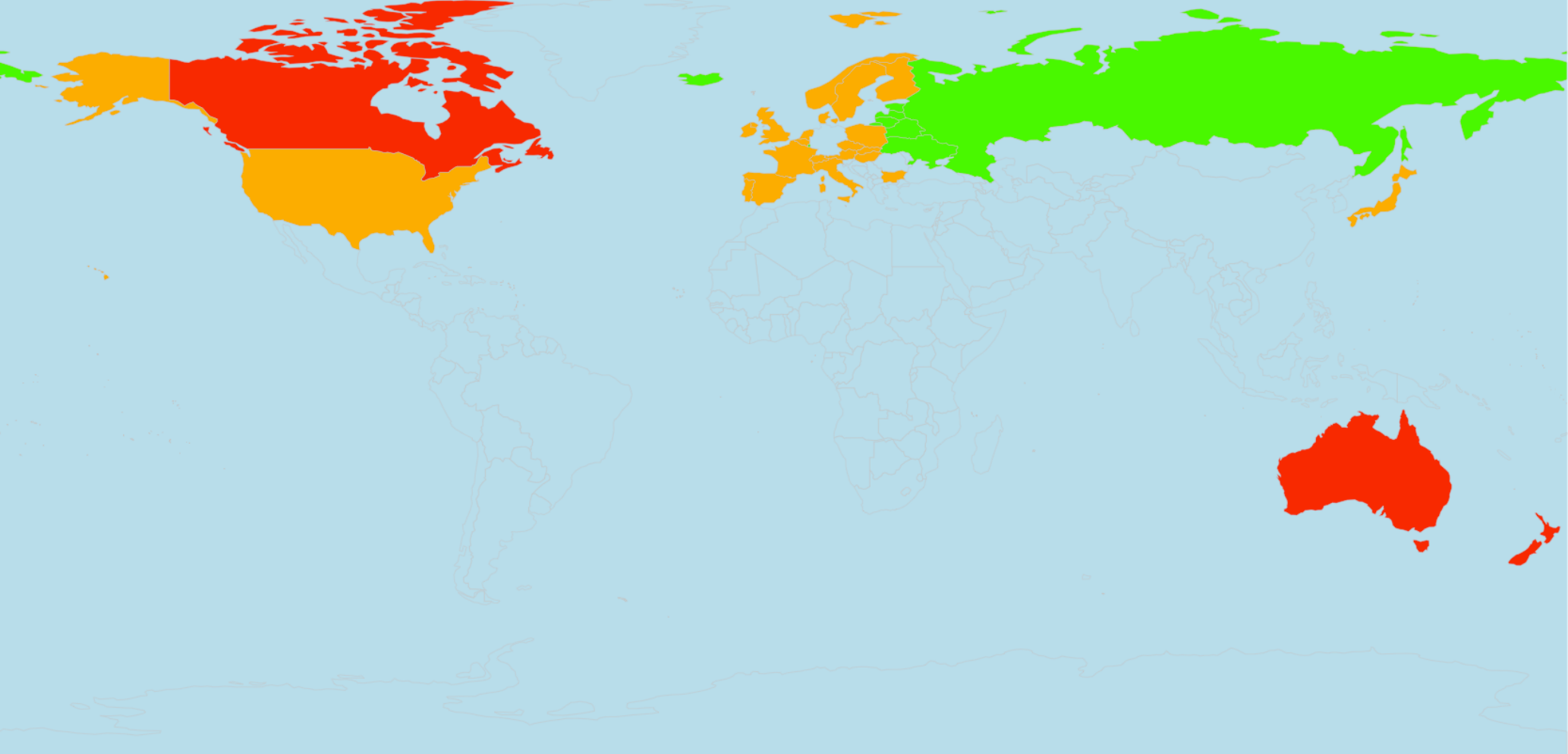} }}
    \qquad
    \subfloat[Initial clustering result, Male]{{\includegraphics[width=7.15cm]{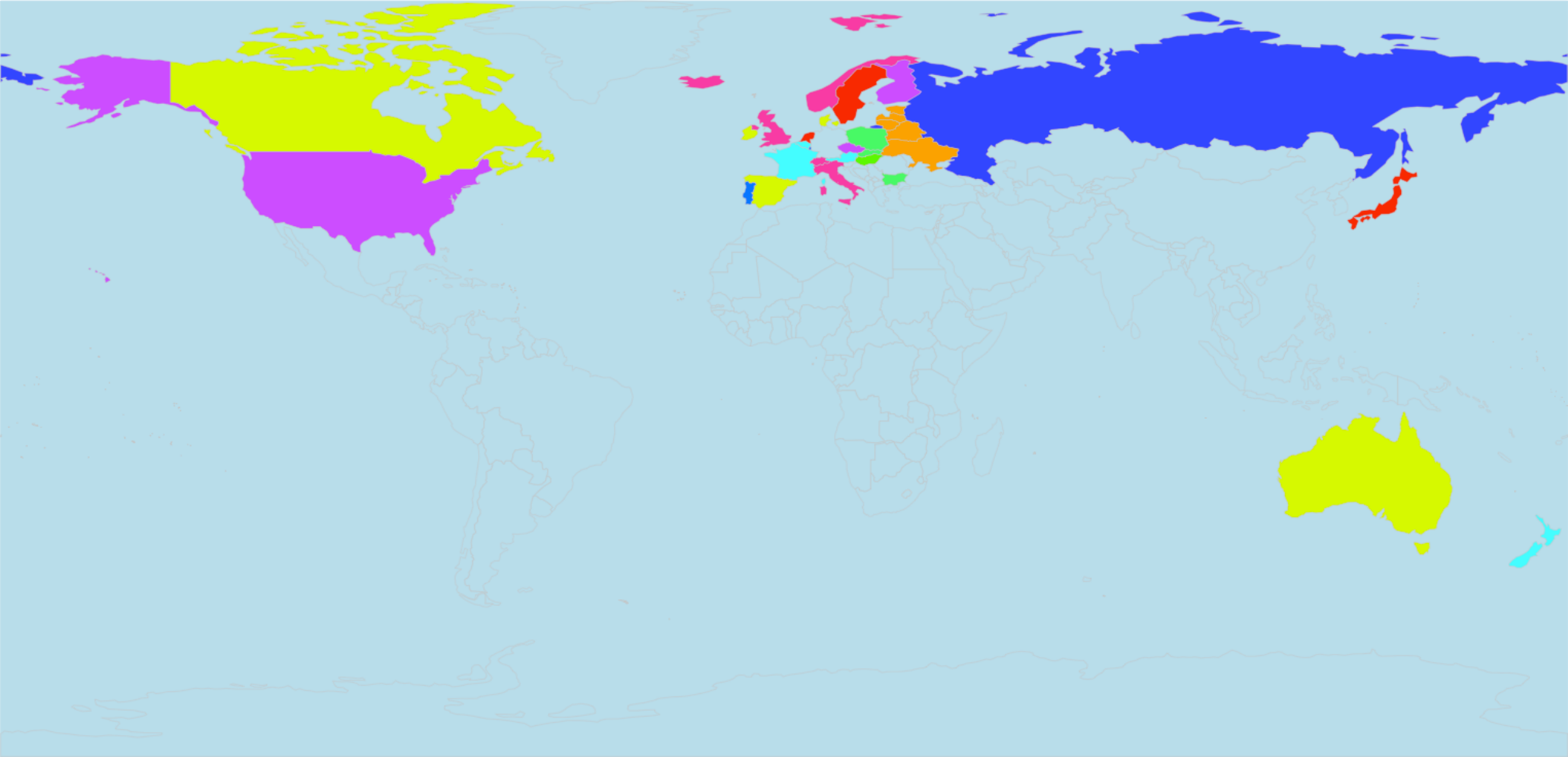} }}
    \qquad
    \subfloat[Final clustering result, Male]{{\includegraphics[width=7.15cm]{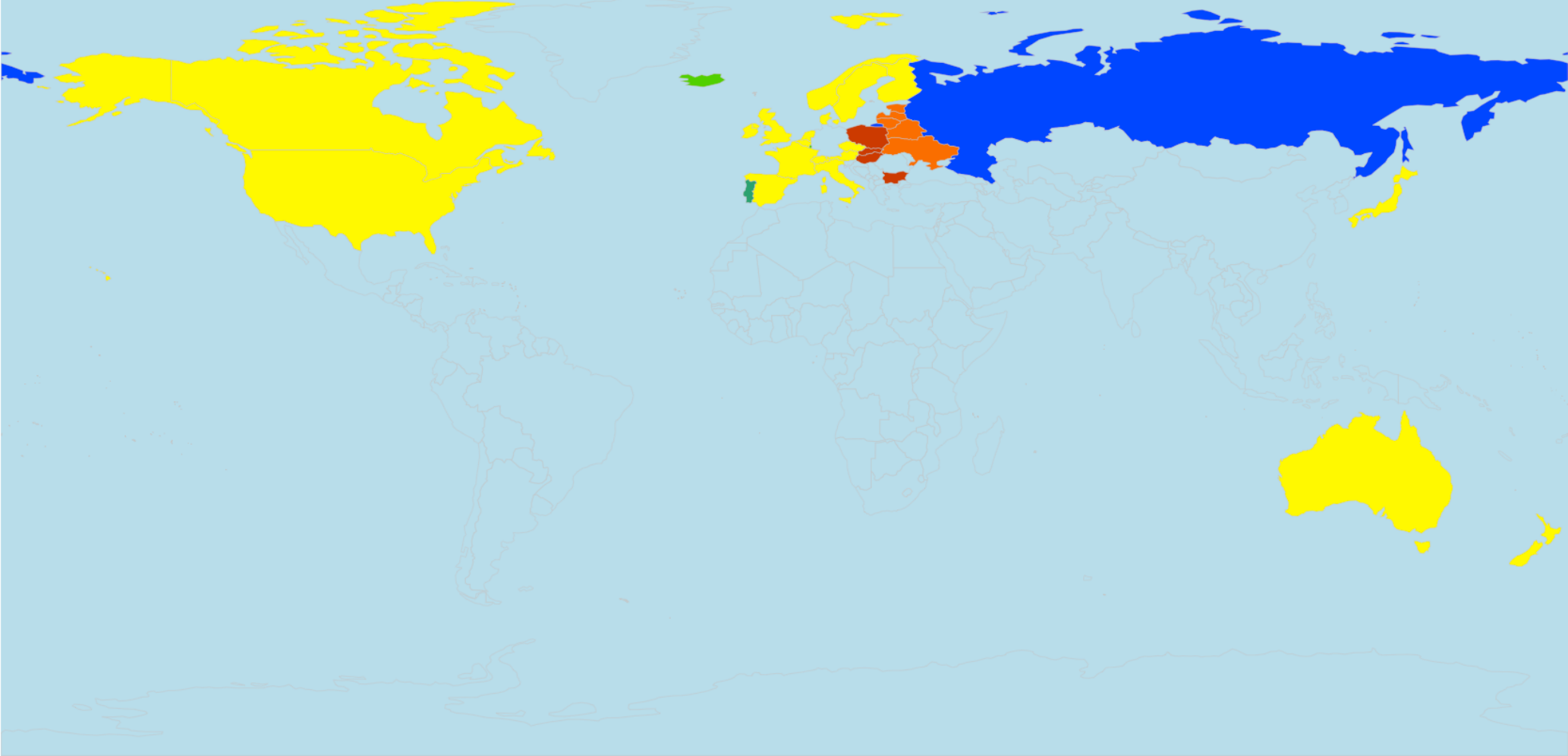} }}
    \qquad
    \subfloat[Initial clustering result, Total]{{\includegraphics[width=7.15cm]{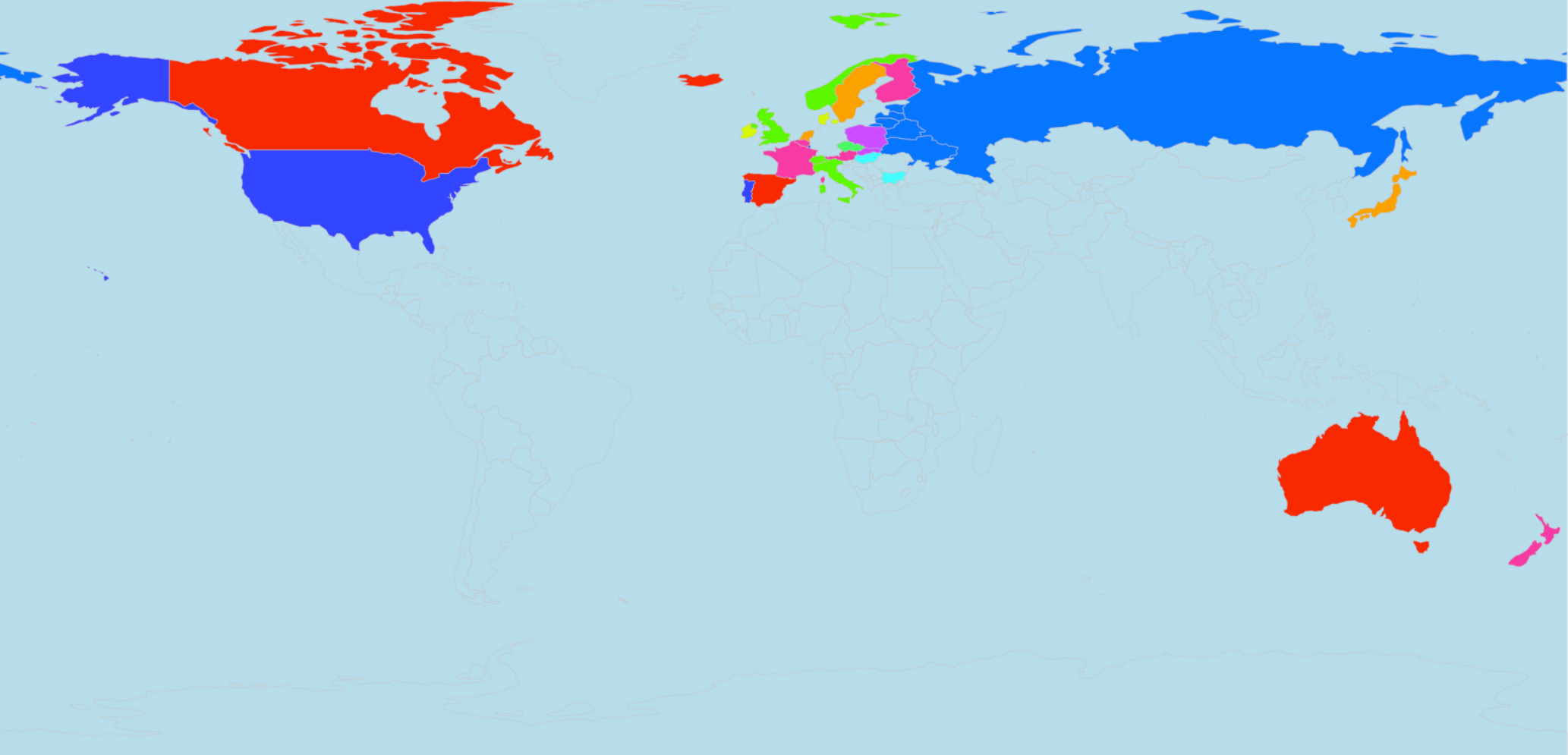} }}
    \qquad
    \subfloat[Final clustering result, Total]{{\includegraphics[width=7.15cm]{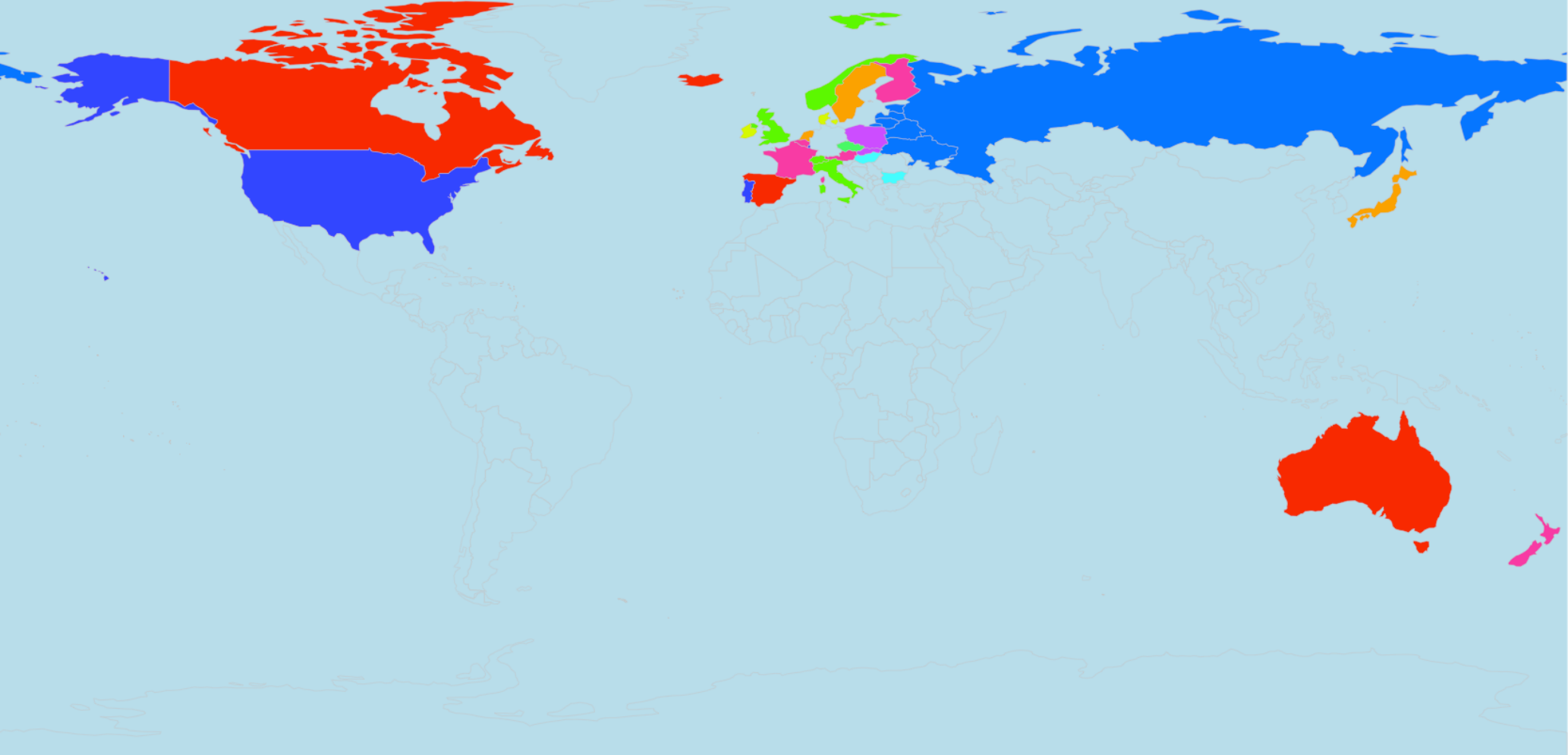} }}
    \caption{A map showing the locations of the clustered countries and areas for female, male, and total age-specific mortality for both of the initial and final clustering results}%
    \label{fig:7}
\end{figure}

Figure~\ref{fig:7} shows a general location map of the clustered countries in different colors for female, male, and total age-specific mortality for the initial and final clustering results. At first glance, the clustering results are highly related to geographical regions.

More information could be obtained if we investigate the country list of various clusters. Table~\ref{tab:5} shows the initial and final clustering results for female age-specific mortality rates for those $32$ countries. We use different colors for different clusters in the initial step to show the change in cluster memberships at convergence. At convergence, we have $3$ clusters. The former Soviet Union member countries tend to group (as in cluster $1$). Most countries in the initial cluster $2$ and $3$, and all countries in the initial cluster $4$, $5$ and $6$ are combined. Note that Australia and New Zealand are bound together, which may be due to geographical reasons. However, Japan tends to be in the same cluster as the Northern European countries, such as Norway, Sweden, and Finland. This may be due to similarities in diet or lifestyle. As a final note, the Eastern European countries tend to bind together, which may be credited to socio-economic status.
\begin{table}[!htbp] \centering 
  \caption{Clustering results for male age-specific mortality. For the initial step, ten clusters are recognized as optimal, while six clusters are obtained at the convergence} 
  \label{tab:4} 
    \setlength{\tabcolsep}{11pt}
\renewcommand{\arraystretch}{0.8}
\begin{tabular}{@{} lcccccccccc@{}} 
\toprule
 Cluster & & \multicolumn{4}{c}{Initial cluster members} &  &\multicolumn{4}{c}{Final cluster members}    \\
\cline{1-1} \cline{3-6} \cline{8-11}\\
\multirow{2}{*}{1} & & \color{ao(english)}BLR & \color{ao(english)}EST & \color{ao(english)}LTU & \color{ao(english)}LVA &  & \color{ao(english)}BLR & \color{ao(english)} EST &\color{ao(english)}LTU & \color{ao(english)}LVA \\
 & & \color{ao(english)}UKR &  &  & & & \color{ao(english)}UKR  & &  & \\  
 \midrule \\[-1.8ex]
 \multirow{5}{*}{2} 
  & & &  &  & & & \color{red}AUS  & \color{blue}AUT & \color{blue}BEL & \color{red}CAN  \\
  & & \color{red}AUS & \color{red}CAN & \color{red}DNK & \color{red}IRL & & \color{darkblue}CZE & \color{red}{DNK}  & \color{darkblue}{FIN} & \color{blue}{FRA}  \\  
 & & \color{red}ESP &  &  &  & &  \color{red}IRL&  \color{purple}{ITA} & \color{orange}{JPN} & \color{orange}{NED}  \\
 & &  & &  &  & & \color{blue}NZL & \color{purple}NOR & \color{red}ESP & \color{orange}{SWE} \\
 & &  &  &  & &  & \color{purple}SUI& \color{purple}GBR &\color{darkblue}USA &  \\
 \midrule \\[-1.8ex]
 \multirow{2}{*}{3} 
 & & \color{purple}{ISL} & \color{purple}{ITA} & \color{purple}{NOR} &  \color{purple}{SUI}& & BGR&\color{magenta}HUN &POL  &SVK \\
 & &  \color{purple}{GBR} & & & & &   &  &  &  \\
  \midrule \\[-1.8ex]

4& & \color{blue}AUT & \color{blue}BEL &\color{blue} FRA & \color{blue}NZL & & \color{purple}ISL & \color{olive}{LUX} &  &  \\
  \midrule \\[-1.8ex]
5 & & BGR & POL & SVK &  & & \color{olive}{PRT} &  &  &  \\
  \midrule \\[-1.8ex]
6 & & \color{orange}{JPN} & \color{orange}{NED} &\color{orange}{SWE} & & &\color{brown}RUS & &  & \\
  \midrule \\[-1.8ex]
7 & & \color{darkblue}{CZE} & \color{darkblue}{FIN} & \color{darkblue}{USA} & & && &  & \\
  \midrule \\[-1.8ex]
8 & & \color{olive}{LUX} &\color{olive}{PRT}  &  & & & & &  & \\
  \midrule \\[-1.8ex]
9 & & \color{magenta}{HUN} & &  & & & & &  & \\
  \midrule \\[-1.8ex]
10 & & \color{brown}{RUS} &  &  & & & & &  & \\
  \bottomrule
\end{tabular} 
\end{table}

Tables~\ref{tab:4} and~\ref{tab:6} present the initial and final clustering results for male and total age-specific mortality rates for these $32$ countries, respectively. At convergence, we have six clusters for males and five clusters for total mortality rates. Similar findings are also observed for the male and total clustering results, i.e., the former Soviet Union member countries tend to cluster, Australia and New Zealand are bound together, the Eastern European countries tend to cluster. 

\begin{table}[!htbp] \centering 
  \caption{Clustering results for female age-specific mortality. For the initial step, seven clusters are recognized as optimal, while three clusters are obtained at the convergence} 
  \label{tab:5} 
  \setlength{\tabcolsep}{11pt}
\renewcommand{\arraystretch}{0.8}
\begin{tabular}{@{} lcccccccccc@{}} 
\toprule
 Cluster & & \multicolumn{4}{c}{Initial cluster members} &  &\multicolumn{4}{c}{Final cluster members}    \\
\cline{1-1} \cline{3-6} \cline{8-11}\\
\multirow{2}{*}{1} & & \color{ao(english)}BLR & \color{ao(english)}BGR & \color{ao(english)}EST & \color{ao(english)}HUN &  & \color{ao(english)}BLR & \color{ao(english)}EST &\color{ao(english)}ISL& \color{ao(english)}LTU \\
 & & \color{ao(english)}ISL & \color{ao(english)}LTU & \color{ao(english)}LVA & \color{ao(english)}LUX & &\color{ao(english)}LVA &\color{ao(english)}LUX & \color{orange}RUS & \color{orange}UKR\\  
 \midrule \\[-1.8ex]
 \multirow{6}{*}{2} 
 & & &  &  & & & \color{blue}AUT  &BEL & \color{ao(english)}BGR & CZE  \\
 & &  &  &  & & &  DNK &\color{blue}FIN  & \color{blue}FRA  &\color{ao(english)}{HUN} \\
  & & \color{blue}AUS  & \color{blue}AUT  &\color{blue}CAN & \color{blue}FIN &  &IRL & \color{red}ITA & \color{darkblue}JPN & \color{red}{NED}\\  
 & & \color{blue}FRA & \color{blue}GBR&  &  & & \color{red}{NOR} & \color{purple}{POL} & \color{purple}PRT & \color{purple}{SVK}\\
 & &  & &  &  & & \color{red}ESP & \color{darkblue}{SWE} & \color{red}{SUI} & \color{blue}{GBR}\\
 & &  &  &  & & & \color{purple}USA &  &  & \\

 \midrule \\[-1.8ex]
 \multirow{2}{*}{3} & & \color{purple}{NZL} & \color{purple}{POL} & \color{purple}{PRT} & \color{purple}{SVK} &  & \color{blue} AUS & \color{blue}{CAN} &\color{purple}{NZL} &  \\
 & & \color{purple}{USA} &  &  &  & & & &  & \\  
 \midrule \\[-1.8ex]
 \multirow{2}{*}{4} & & \color{red}{ITA} & \color{red}{NED} & \color{red}{NOR} &\color{red}{ESP} &  &  &  & &  \\
 & & \color{red}{SUI} &  &  &  & & & &  & \\  
  \midrule \\[-1.8ex]
5 & & BEL & CZE & DNK & IRL & &  & &  &  \\
  \midrule\\[-1.8ex]
6 & & \color{darkblue}{JPN} &\color{darkblue}{SWE}  & & & & & &  & \\
  \midrule \\[-1.8ex]
7 & & \color{orange}{RUS} &\color{orange}{UKR} && & & & &  & \\
  \bottomrule
\end{tabular} 
\end{table} 

\begin{table}[!htbp] \centering 
  \caption{Clustering results for total age-specific mortality. For the initial step, ten clusters are recognized as optimal, while five clusters are obtained at the convergence} 
  \label{tab:6} 
    \setlength{\tabcolsep}{11pt}
\renewcommand{\arraystretch}{0.8}
\begin{tabular}{@{} lcccccccccc@{}} 
\toprule
 Cluster & & \multicolumn{4}{c}{Initial cluster members} &  &\multicolumn{4}{c}{Final cluster members}    \\
\cline{1-1} \cline{3-6} \cline{8-11}\\
1 & & \color{red}AUS & \color{red}CAN & \color{red}ISL &\color{red}ESP  &  & \color{olive} BGR & \color{olive}HUN &\color{magenta}POL &\color{magenta}SVK \\  
 \midrule \\[-1.8ex]
 \multirow{5}{*}{2} 
  & & &  &  & & & \color{red}AUS  & \color{blue}AUT & \color{blue}BEL & \color{red}CAN  \\
  & & \color{ao(english)}BLR & \color{ao(english)}EST & \color{ao(english)}LTU & \color{ao(english)}LVA & & \color{brown}CZE & \color{darkblue}DNK & \color{blue}{FIN} & \color{blue}{FRA} \\  
 & & \color{ao(english)}RUS & \color{ao(english)}UKR &  &  & & \color{darkblue}{IRL}& ITA & \color{purple}{JPN} & \color{purple}NED\\
 & &  & &  &  & & \color{blue}NZL & NOR &  \color{red}ESP & \color{purple}{SWE}\\
 & &  &  &  & & & SUI & GBR & \color{orange}USA & \\
 \midrule \\[-1.8ex]
 \multirow{2}{*}{3} 
 & & \color{blue}AUT & \color{blue}BEL &\color{blue} FIN& \color{blue}FRA & & \color{ao(english)}BLR & \color{ao(english)}EST & \color{ao(english)}LTU& \color{ao(english)}LVA \\
 & &  \color{blue}NZL & & & & & \color{ao(english)}RUS &\color{ao(english)}UKR &  &  \\
  \midrule \\[-1.8ex]
 4 & & ITA & NOR & SUI& GBR & &  \color{red}{ISL} &\color{orange}{LUX} &  &  \\
  \midrule\\[-1.8ex]
5 & & \color{purple}{JPN} & \color{purple}{NED} & \color{purple}{SWE} &  & & \color{orange}PRT& &  & \\
  \midrule \\[-1.8ex]
6 & & \color{orange}{LUX} & \color{orange}{PRT} &\color{orange}{USA} & & & & &  & \\
  \midrule \\[-1.8ex]
7 & & \color{darkblue}{DNK} & \color{darkblue}{IRL} &  & & & & &  & \\
  \midrule \\[-1.8ex]
8 & & \color{olive}{BGR} & \color{olive}{HUN} &  & & & & &  & \\
  \midrule \\[-1.8ex]
9 & & \color{magenta}{POL} &\color{magenta}{SVK} &  & & & & &  & \\
  \midrule \\[-1.8ex]
10 & & \color{brown}{CZE} & &  & & & & &  & \\

  \bottomrule

\end{tabular} 
\end{table}

The cluster results reflect a number of factors, such as geography, lifestyle, ethnic group, socio-economic status, etc. A single factor is not enough to group these $32$ countries, and this shows the importance of cluster analysis before undertaking any joint modeling.

\begin{table}[!htbp] \centering 
  \caption{The number of FPCs selected for each cluster level for female, male, and total mortality series} 
  \label{tab:11} 
  \setlength{\tabcolsep}{7pt}
\renewcommand{\arraystretch}{1.2}
\begin{tabular}{@{} ccccccccccccccc@{}} 
\toprule
& &  \multicolumn{3}{c}{Female clusters} & & \multicolumn{4}{c}{Male clusters} & &  \multicolumn{4}{c}{Total clusters} \\
\cline{3-5} \cline{7-10} \cline{12-15}\\
Parameter & & \textbf{(1)} & \textbf{(2)} & \textbf{(3)} & & \textbf{(1)} & \textbf{(2)} & \textbf{(3)} &\textbf{(4)} & & \textbf{(1)} & \textbf{(2)} & \textbf{(3)}&\textbf{(4)}\\
\midrule
$M$ & & $2$ & $3$  & $2$ & & $3$ & $3$ & $2$ & $2$ &  & $2$ & $4$ & $2$&$2$ \\
\midrule
$N_1$ &  & $2$ & $2$ & $2$ &  & $2$ & $2$ & $2$ & $2$ &  & $2$ & $2$ & $2$&$2$ \\
\midrule
$N_2$ &  & $8$ & $6$ & $7$ &  & $7$ & $9$ & $6$ & $10$ &  & $5$ & $8$ & $6$& $9$\\
\bottomrule
\end{tabular} 
\end{table} 

Table~\ref{tab:11} tabulates the number of FPCs selected for each level for each cluster of the female, male, and total mortality series. We ignore the clusters with a single country as the proposed model can only apply to clusters with multiple countries. Note that countries included in cluster $2$ for all three series are very similar; they have a similar number of selected FPCs. Moreover, we find out that the number of FPCs selected for the first two levels of the model is quite close. The numbers of FPCs selected for the third level are different for various clusters of different series. This confirms our observation in Section~\ref{sec:2} that the third level, $U_{it}$, serves as a supplement. 

\subsection{Forecasting based on the functional panel data model with fixed effects}

Once the cluster memberships are determined, we can use the functional panel data model to re-estimate each component for each cluster and produce forecasts. Since in the forecasting step, we are not using the $\eta_i$ as a pattern recognition component as what we did in the clustering step, we would rather keep this term instead of reducing its dimension. The mortality rates for the $i^{\textsuperscript{th}}$ population at the $t^{\textsuperscript{th}}$ year can be expressed as
\begin{equation*}
  \begin{split}
    y_{it}^{(c)} (x) & = \mu^{(c)} (x) + \eta_i^{(c)}(x) + \sum_{k=1}^{\infty}\xi_{tk}^{(c)} \phi_k^{(c)} (x) + \sum_{l=1}^{\infty}\zeta_{itl}^{(c)} \psi_l^{(c)} (x)\\
    & \approx \mu^{(c)} (x) + \eta_i^{(c)}(x) + \sum_{k=1}^{N_1}\xi_{tk}^{(c)} \phi_k^{(c)} (x) + \sum_{l=1}^{N_2}\zeta_{itl}^{(c)} \psi_l^{(c)} (x),
  \end{split}
\end{equation*}
where $\mu^{(c)}$, $\eta_i^{(c)}$,  $\phi_k^{(c)}$, and $ \psi_l^{(c)}$ are the structural components for any given cluster $c$ and the $i^{\textsuperscript{th}}$ population belongs to the cluster $c$.

In forecasting the functional time series, \cite{hyndman2009forecasting} apply univariate time series forecasting methods (e.g., autoregressive integrated moving average models) to the FPC scores to produce forecasts. \cite{aue2015prediction} suggest using multivariate time series forecasting methods, e.g., a vector autoregressive (VAR) model in forecasting FPC scores. We here adopt the multivariate time series forecasting methods to generate forecasts of the dynamic FPC scores since the dynamic FPC scores we obtain still exhibit correlations. 

The $h$-step-ahead forecast of FPC score vector $\widehat{\bb{\xi}}_{\kappa+h|\kappa} = \left(\widehat\xi_{(\kappa+h|\kappa)1}, \ldots,\widehat\xi_{(\kappa+h|\kappa)N_1}\right)$ and  $\widehat{\bb\zeta}_{i,\kappa+h|\kappa} = \left(\widehat\zeta_{i,(\kappa+h|\kappa)1}, \ldots,\widehat\zeta_{i,(\kappa+h|\kappa)N_2}\right)$ can be obtained by applying a VAR\footnote{ Here we use \texttt{MTS} package in \texttt{R} of \cite{Tsay21} to fit VAR models.} model to the score vectors $\{\bb{\widehat\xi}_t^{(c)}, t= 1, \ldots, \kappa\}$ and $\{\bb{\widehat\zeta_{itl}}^{(c)}, t= 1, \ldots, \kappa\}$, respectively, where $\kappa$ is the number of observations used in forecasting. With a set of holdout sample, the $h$-step-ahead corresponding forecasts can be expressed as
\begin{equation*}
    \widehat y_{i(\kappa+h|\kappa)} (x) = \widehat\mu^{(c)} (x) + \widehat\eta_i^{(c)}(x) + \sum_{k=1}^{N_1}\widehat\xi_{(\kappa+h|\kappa)k}^{(c)} \widehat\phi_k^{(c)} (x) + \sum_{l=1}^{N_2}\widehat\zeta_{i(\kappa+h|\kappa)l}^{(c)}\widehat\psi_l^{(c)} (x),
\end{equation*}
where $\widehat\mu^{(c)}$, $\widehat\eta_i^{(c)}$, $\widehat\phi_k^{(c)}$, and $\widehat\psi_l^{(c)}$ are derived with samples from cluster $c$, and $\kappa$ is the number of observations used in generating the point forecasts.

\subsection{Point forecast evaluation}

We use the root mean square forecast error (RMSFE) of the $h$-step-ahead forecasts to evaluate the point forecast accuracy. The RMSFE measures how close the forecast results are to the actual values of the data under forecast.

The $h$-step-ahead point forecasts are generated using an expanding window analysis, commonly used in time series models to evaluate model stability. By the expanding window analysis, we firstly use the first $n$ observations to generate the $h$-step-ahead point forecasts for $h = 1, 2, \ldots, N-n$. The forecast process is then iterated by increasing the sample size by one year until reaching the data's end period. By doing so, we can produce one $(N-n)$-step-ahead forecast, two $(N-n-1)$-step-ahead forecasts, $\ldots$, and $N-n$ one-step-ahead forecasts. The RMSFE for the $h$-step-ahead forecasts can be written as
\begin{equation*}
  \text{RMSFE}(h) = \sqrt{\frac{1}{J\times(N-n-h+1)}\sum_{\kappa=n+h-1}^{N-h}\sum_{i=1}^{J}\big[y_{\kappa+h}(x_i)-\widehat y_{\kappa+h}(x_i)\big]^2},
\end{equation*}
where $\kappa$ is the number of observations used in generating the point forecasts, $y_{\kappa+h}(x_i)$ is the actual value for the ${(\kappa+h)}^{\textsuperscript{th}}$ observation and $\widehat y_{\kappa+h}(x_i)$ is the $h$-step-ahead point forecast based on the first $\kappa$ observations, and $J$ is the number of grid points of ages. We have aggregated age $100^{+}$ for female and total mortality, while due to the sparsity in male data, we aggregate age $98^{+}$ for male mortality, and hence the number of grid points of ages for female and total mortality is $101$ and $99$ for male mortality.

We applied the functional panel data model with fixed effects to different cluster memberships, i.e., clustering using $k$-means, the initial step of our clustering method, and our proposed clustering method. Forecast results are compared to examine whether our clustering technique can help to improve forecasts. Moreover, results are also compared with the univariate functional time series forecasting model \citep[see, e.g.,][]{hyndman2007robust}. Table~\ref{tab:7} presents the averaged RMSFE values across all countries ($\times 100$) in the holdout sample for the various forecast methods. The bold entries highlight the method that produces the most accurate point forecast.

\begin{table}[!htbp] \centering 
\caption{Average RMSFE values ($\times 100$) in the holdout sample based on various forecasting methods are presented. Forecasts based on univariate functional time series are labeled as ``UTS'', functional panel data model forecasts based on the initial cluster membership are labeled as ``FPCA'', and functional panel data model forecasts based on the final cluster membership (at convergence) are labeled as ``MFTSC''. For comparison, the percentage change in the smallest RMSFE with respect to the second smallest RMSFE are labeled as ``Change (\%)''} 
  \label{tab:7} 
    \setlength{\tabcolsep}{4pt}
{\scriptsize{
\begin{tabular}{@{}lcccccccccccc@{}} 
\toprule
&\multicolumn{4}{c}{Female} & \multicolumn{4}{c}{Male} & \multicolumn{3}{c}{Total}\\
$h$ & UTS & FPCA  & MFTSC & Change(\%)& UTS & FPCA  & MFTSC& Change(\%) & UTS & FPCA & MFTSC& Change(\%)\\ 
\midrule 
$1$ & $1.178$ & $1.146$ & $\textbf{1.130}$ & $-0.086$ & $1.345$ & $1.314$ & $\textbf{1.307}$ &$-0.533$& $1.124$ & $1.102$ & $\textbf{1.095}$ & $-0.635$\\ 
$2$ & $1.207$ & $1.188$ & $\textbf{1.166}$ & $-1.852$ & $1.402$ & $1.356$ & $\textbf{1.344}$ &$-0.885$& $1.170$ & $1.120$ & $\textbf{1.114}$ &$-0.536$\\ 
$3$ & $1.247$ & $1.235$ & $\textbf{1.211}$ & $-1.943$& $1.459$ & $1.395$ & $\textbf{1.394}$&$-0.072$ & $1.222$ & $1.160$ & $\textbf{1.146}$&$-1.207$ \\ 
$4$ & $1.283$ & $1.265$ & $\textbf{1.232}$ &$-2.609$& $1.509$ & $1.442$ & $\textbf{1.438}$ &$-0.277$& $1.254$ & $1.197$ & $\textbf{1.162}$ & $-2.924$\\ 
$5$ & $1.326$ & $1.328$ & $\textbf{1.273}$ &$-3.997$ &$1.573$ & $1.495$ & $\textbf{1.478}$ &$-1.137$& $1.305$ & $1.263$ & $\textbf{1.224}$ &$-3.088$\\ 
$6$ & $1.366$ & $1.378$ & $\textbf{1.300}$ & $-4.832$&$1.630$ & $1.510$ & $\textbf{1.498}$ & $-0.795$&$1.372$ & $1.318$ & $\textbf{1.255}$ & $-4.780$\\ 
$7$ & $1.400$ & $1.447$ & $\textbf{1.327}$ & $-5.214$&$1.687$ & $1.598$ & $\textbf{1.567}$ & $-1.940$&$1.431$ & $1.363$ & $\textbf{1.297}$ &$-4.842$\\ 
$8$ & $1.417$ & $1.485$ & $\textbf{1.315}$ & $-7.198$ &$1.715$ & $1.623$ & $\textbf{1.595}$ &$-1.725$ &$1.466$ & $1.400$ & $\textbf{1.325}$&$-5.357$\\ 
$9$ & $1.427$ & $1.502$ & $\textbf{1.291}$ & $-9.530$& $1.720$ & $1.647$ & $\textbf{1.542}$ & $-6.375$&$1.476$ & $1.394$ & $\textbf{1.332}$ &$-4.448$\\ 
$10$ & $1.433$ & $1.569$ & $\textbf{1.378}$ & $-3.832$&$1.653$ & $1.439$ & $\textbf{1.405}$ &$-5.957$& $1.465$ & $1.454$ & $\textbf{1.426}$ &$-1.926$\\ 

\midrule
Mean & $1.328$ & $1.352$ & $\textbf{1.263}$& $-6.583$ & $1.569$ & $1.482$ & $\textbf{1.457}$ &$-1.687$& $1.328$ & $1.278$ & $\textbf{1.238}$ &$-3.130$\\ 
\bottomrule
\end{tabular} 
}}
\end{table} 

The multilevel functional forecast based on our proposed clustering method outperforms either of the competitive methods uniformly. The average RMSFE values have less variation than those of the other two methods, which means that our method is more robust in forecasting. Additionally, the improvements in point forecast accuracy are generally more significant as the forecasting horizon increases; one possible explanation is that by grouping those homogeneous countries, we can benefit from borrowing information from series with similar patterns in making forecasts. This benefit is not so apparent in the short term. However, as the forecasting horizon increases, the benefit will magnify where there are more considerable uncertainties. 

Moreover, suppose we use different clustering methods to generate clusters and produce forecasts using the functional panel data model. In that case, we find that the forecasts based on cluster membership derived by classical clustering methods do not always outperform the univariate functional time series model. This demonstrates the efficiency of our clustering technique. Efficient clustering will reduce variation and thus improve forecasts, but the clustering must be efficient and reliable to generate improved forecasts. Inefficient or unreliable clustering may deteriorate instead of improving forecasts. In summary, the forecasts based on the functional panel data model and our model-clustering method provide a more robust forecast with less variation. 

To facilitate comparison in the RMSFE values, the percentage change in the smallest RMSFE with respect to the second smallest RMSFE is calculated. As we can see, the percentage change in forecasting improvements ranges from $-0.1\%$ to $-6.5\%$, with an average around $-3\%$. The implication of such improvements on actuarial practice is significant. The financial impact of accurately forecasting on mortality rate is demonstrated in Appendix~\ref{appnc}, where examples of annuity pricing are presented. 

\subsection{Interval forecast evaluation}

To capture the uncertainties in the point forecasts, we also construct the prediction intervals. \cite{aue2015prediction} proposed a parametric approach for constructing uniform prediction intervals, which can be extended to point-wise prediction intervals after considering the nonparametric bootstrap approach of \cite{shang2018bootstrap}. Based on the in-sample-forecast errors, $\widehat e_{\kappa +h|\kappa}^{(l)} = y_{\kappa+h}^{(l)}(x_i)-\widehat y_{\kappa+h}^{(l)}(x_i)$, for the $l^{\text{th}}$ curve, we use sampling with replacement to generate a series of bootstrapped forecast errors to obtain the upper bound and lower bound, $\gamma_{\text{lb}}^{(l)}(x_i)$ and $\gamma_{\text{ub}}^{(l)}(x_i)$, respectively. Then a tuning parameter, $\theta_\alpha$, can be determined, such that 
\begin{equation*}
  \mathbb{P}\big\{\theta_\alpha \times \gamma_{\text{lb}}^{(l)}(x_i)\leq \widehat e_{\kappa +h|\kappa}^{(l)}\leq\theta_\alpha \times \gamma_{\text{ub}}^{(l)}(x_i)\big\} = (1-\alpha) \times 100\%.
\end{equation*}
Then, the $h$-step-ahead pointwise prediction intervals are as follows: 
\begin{equation*}
  \widehat y_{\kappa+h}^{(l)}(x_i) + \theta_\alpha \times \gamma_{\text{lb}}^{(l)}(x_i)\leq y_{\kappa+h}^{(l)}(x_i)\leq \widehat y_{\kappa+h}^{(l)}(x_i) + \theta_\alpha \times \gamma_{\text{ub}}^{(l)}(x_i).
\end{equation*}
We use the interval scoring rule of \cite{gneiting2007strictly} to evaluate the pointwise interval forecast accuracy. The interval score for the pointwise interval forecast at time point $x_{i}$ is 
\begin{equation*}
\begin{split}
  S_\alpha\Big[\widehat y_{\kappa+h}^{\text{lb}}(x_i), \widehat y_{\kappa+h}^{\text{ub}}(x_i); y_{\kappa+h}(x_i)\Big] & = \Big[\widehat y_{\kappa+h}^{\text{ub}}(x_i) - \widehat y_{\kappa+h}^{\text{lb}}(x_i)\Big] \\
  & + \frac{2}{\alpha}\Big[\widehat y_{\kappa+h}^{\text{lb}}(x_i) - y_{\kappa+h}(x_i)\Big]\mathbbm{1}\Big\{ y_{\kappa+h}(x_i)<\widehat y_{\kappa+h}^{\text{lb}}(x_i)\Big\}\\
  & + \frac{2}{\alpha}\Big[y_{\kappa+h}(x_i) -\widehat y_{\kappa+h}^{\text{ub}}(x_i)\Big]\mathbbm{1}\Big\{ y_{\kappa+h}(x_i)>\widehat y_{\kappa+h}^{\text{ub}}(x_i)\Big\},
\end{split}
\end{equation*}
where the level of significance $\alpha$ can be chosen conventionally as $0.2$. It is not difficult to find that the smaller the interval score is, the more accurate the interval forecast. An optimal (which is also minimal) interval score value can be achieved if $y_{\kappa+h}(x_i)$ lies between $\widehat y_{\kappa+h}^{\text{lb}}(x_i)$ and $\widehat y_{\kappa+h}^{\text{ub}}(x_i)$. Then the mean interval score for the $h$-step-ahead forecast can be written as
\begin{equation*}
  \overline S_\alpha(h) = \frac{1}{J\times(N-n-h+1)}\sum_{\kappa=n+h-1}^{N-h}\sum_{i=1}^{J}S_\alpha\Big[\widehat y_{\kappa+h}^{\text{lb}}(x_i), \widehat y_{\kappa+h}^{\text{ub}}(x_i); y_{\kappa+h}(x_i)\Big].
\end{equation*}

\begin{table}[!htbp] \centering 
\caption{Average Interval Score values ($\times 100$) in the holdout sample based on various forecasting methods are presented. Forecasts based on univariate functional time series are labeled as ``UTS'', functional panel data model forecasts based on the initial cluster membership are labeled as ``FPCA'', and functional panel data model forecasts based on the final cluster membership (at convergence) are labeled as ``MFTSC''. For comparison, the percentage change in the smallest interval score with respect to the second smallest interval score is labeled as ``Change (\%)''} 
  \label{tab:8} 
    \setlength{\tabcolsep}{4pt}
{\scriptsize{
\begin{tabular}{@{}lcccccccccccc@{}} 
\toprule
&\multicolumn{4}{c}{Female} & \multicolumn{4}{c}{Male} & \multicolumn{4}{c}{Total}\\
$h$ & UTS & FPCA  & MFTSC & Change(\%)& UTS & FPCA  & MFTSC & Change(\%)& UTS & FPCA  & MFTSC& Change(\%)\\ 
\midrule 

$1$ & $0.900$ & $0.827$ & $\textbf{0.815}$ & $-1.451$& $1.023$ & $0.943$ & $\textbf{0.758}$&$-19.618$ & $0.925$ & $0.939$ & $\textbf{0.854}$ &$-7.676$\\ 
$2$ & $1.230$ & $1.110$ & $\textbf{1.081}$&$-2.613$ & $1.370$ & $1.227$ & $\textbf{1.088}$&$-11.328$ & $1.243$ & $1.256$ & $\textbf{1.158}$ & $-6.838$\\ 
$3$ & $1.469$ & $1.293$ & $\textbf{1.265}$& $-2.166$ & $1.662$ & $1.447$ & $\textbf{1.237}$&$-14.513$ & $1.506$ & $1.460$ & $\textbf{1.348}$& $-7.671$ \\ 
$4$ & $1.773$ & $1.522$ & $\textbf{1.448}$ & $-4.862$& $2.028$ & $1.703$ & $\textbf{1.376}$& $-19.201$ & $1.787$ & $1.688$ & $\textbf{1.609}$&$-4.680$ \\ 
$5$ & $2.108$ & $1.853$ & $\textbf{1.698}$ &$-8.365$& $2.390$ & $1.991$ & $\textbf{1.521}$ & $-23.606$& $2.084$ & $1.992$ & $\textbf{1.935}$ &$-2.861$\\ 
$6$ & $2.485$ & $2.050$ & $\textbf{1.901}$ &$-7.268$& $2.716$ & $2.184$ & $\textbf{1.688}$ &$-22.711$& $2.378$ & $2.203$ & $\textbf{2.170}$& $-1.498$ \\ 
$7$ & $2.823$ & $2.399$ & $\textbf{2.148}$ &$-10.463$& $3.195$ & $2.500$ & $\textbf{1.875}$ &$-25.000$& $2.770$ & $2.458$ & $\textbf{2.409}$ &$-1.993$\\ 
$8$ & $3.398$ & $2.849$ & $\textbf{2.424}$ & $-14.918$&$3.706$ & $2.656$ & $\textbf{2.032}$ &$-23.494$& $3.159$ & $2.812$ & $\textbf{2.706}$ &$-3.770$\\ 
$9$ & $3.670$ & $3.341$ & $\textbf{2.502}$ &$-25.112$& $4.098$ & $2.924$ & $\textbf{2.273}$ & $-22.264$& $3.516$ & $3.021$ & $\textbf{2.935}$ &$-2.847$\\ 
$10$ & $3.561$ & $3.219$ & $\textbf{2.457}$ &$-23.672$& $4.203$ & $2.990$ & $\textbf{2.691}$ & $-10.334$& $3.686$ & $3.044$ & $\textbf{3.003}$& $-1.347$ \\ 

\hline \\[-1.8ex] 
Mean & $2.342$ & $2.046$ & $\textbf{1.774}$&$-13.294$ & $2.639$ & $2.057$ & $\textbf{1.654}$&$-19.592$ & $2.305$ & $2.087$ & $\textbf{2.013}$ & $-3.546$\\  
\bottomrule
\end{tabular} 
}}
\end{table} 

Table~\ref{tab:8} presents the averaged interval score values ($\times 100$) across all countries ($\times 100$) in the holdout sample for the different forecast methods. The bold entries highlight the technique that produces better interval forecasts. The percentage change in the smallest interval score value with respect to the second smallest interval score value is also calculated. We observe that the forecasts based on our proposed model and clustering method have the smallest mean interval score values. Our model and the clustering method perform the best in producing interval forecasts, while forecasts using our model and a classical clustering method rank in second place, and the univariate functional forecasting performs the worst. This indicates that joint modeling of mortality rates from multiple countries could borrow information across countries and reduce variations. Efficient and reliable clustering could further reduce the variations in the interval forecasts. 

After comparing the various methods' point and interval forecast results, our proposed method outperforms other competitive methods. Moreover, the improvement of forecast in the long term of our proposed method is significant. This benefit arises from clustering, where we are extracting the common feature of the same cluster. This characteristic is not significant compared with the individual feature in the short term. Still, as the time horizon extends, the individual characteristic vanishes, and the common feature begins to dominate.

\section{Conclusion and discussion}\label{sec:7}

Extending the panel data model to multiple functional data allows a novel method in clustering and forecasting multiple sets of functional time series. The functional panel data model with fixed effects and model-based clustering techniques is employed to analyze mortality data from $32$ countries to obtain several homogeneous groups with the same common time trend and common functional pattern. Countries in each cluster are modeled via the proposed model, and forecasts are made separately. With the functional panel data model, we can capture the time trend and functional patterns common to countries within one cluster. From another point of view, the proposed functional panel data model extends the two-way functional ANOVA model proposed by \cite{di2009multilevel}, which is established for functional data, to functional time series. 

We use a simulation study to demonstrate our proposed clustering method's clustering performance and compare the clustering results with competitive methods under various scenarios. We have considered similar, moderately distinct, and very distinct data by changing our data generating process parameters. Our proposed clustering method performs the best for all designs where the data are not similar, proving that our method can accurately determine the homogeneous data structure. 

The study on the age-specific mortality rates of $32$ countries illustrates our model's merits as it produces more accurate and robust forecasts. More interestingly, we find that this superior performance is more evident for male data. This concurs with the finding of \cite{shang2016mortality}, which reported that multilevel functional data could achieve higher forecast accuracy for populations with more considerable variability over age and year, as male data generally display greater variability than female and total data.

Following the idea of \cite{Bai2009panel}, a possible future study would be to generalize our model to a functional panel data model with interactive effects. Since interactive effects are more popular than additive effects (i.e., fixed effects) in the panel data literature, interactive effects could consider the multiplicity of time trend effects and country-specific effects instead of only their addition in additive effects.

\section*{Acknowledgments} 

The authors would like to thank the Editor, Professor Jeffrey S. Morris, and the Associate Editor and reviewers for their insightful comments and suggestions, which led to a much-improved manuscript. The authors are grateful for the insightful discussions with the 12th International Conference of the ERCIM WG participants on Computational and Methodological Statistics 2019. The first author would also like to acknowledge the financial support of a Ph.D. scholarship from the Australian National University. 

\newpage
\begin{appendix}
\section*{Appendix A: Smoothing the mortality rates}\label{appna}
Following \cite{hyndman2007robust}, we smooth mortality rates using weighted penalized regression splines with a partial monotonic constraint for ages above $65$.

Therefore the penalized regression spline smoothing estimates of the underlying continuous and smooth function can be written as
\begin{equation*}
  \widehat{Y}_{it}(x)=\underset{a_{it}(x)}{\operatorname{argmin}}\sum_{j=1}^{J}w_{it}(x_j)|f_{it}(x_j)-a_{it}(x_j)| + \tau_0 \sum_{j=1}^{J-1}|a^{\prime}_{it}(x_{j+1})-a^{\prime}_{it}(x_{j})|,
\end{equation*}
where $j=1, 2, \ldots, J$ represent different ages (grid points) with a total of $J$ grid points, $\tau_0$ is a smoothing parameter, $a_{it}(x_j)$ is the value taken at grid point $j$ of the smoothing spline $a_{it}(x)$, $^{\prime}$ is the symbol of the first derivative of a function and the weights $w_{it}(x_j)$ are chosen to be the ``inverse variances", $w_{it}(x_j) = 1/[\delta_{it}^2(x_j)]$ with $\delta_{it}^2(x_j)$ measuring the variability in mortality at each age $j$ in the year $t$ for the population $i$, such that we can model the heterogeneity in mortality rates across different ages. The monotonic increasing constraint helps to reduce the noise from the estimation of older ages \citep{shang2016mortality}.

The smoothed mortality rates, $Y_{it}(x_j)$ at different ages $j$ can be obtained by scrutinizing $Y_{it}(x)$ at discrete data points $j$. We estimate $\delta_{it}^2(x_j)$ in weights $w_{it}(x_j)$ as follows.

Let $m_{it}(x_j) = \text{exp}\big(f_{it}(x_j)\big)$ be the observed central mortality rates for age $x_j$ in year $t$ for the $i^{\textsuperscript{th}}$ population. The observed mortality rate approximately follows a binomial distribution with variance $\frac{m_{it}(x_j)\times [1 - m_{it}(x_j)]}{\text{pop}_{it}(x_j)}$, where $\text{pop}_{it}(x_j)$ is the total $j^{\textsuperscript{th}}$ population of age $x_j$. Based on the Taylor's series expansion, the estimated variance associated with the log mortality rate is approximated by $\frac{1 - m_{it}(x_j)}{m_{it}(x_j)\times \text{pop}_{it}(x_j)}$. As the mortality rates $m_{it}(x_j)$ are close to $0$, the term $\delta_{it}^2(x_j)$ can be approximated by  $\hat{\delta}_{it}^2(x_j)=\frac{1}{m_{it}(x_j)\times \text{pop}_{it}(x_j)}$.

\newpage

\section*{Appendix B: Calculating the principal component scores}\label{appnb} 

Given the structure components $\rho_k^{(c)}$ and $\psi_l^{(c)}$ of cluster $c$, $c = 1, 2, \ldots, K$, we can calculate the corresponding functional principal component scores $\xi_{tk}^{(c)}$ and $\zeta_{itl}^{(c)}$ of curve $y_{it}$ using the projection method \citep[see][for a similar approach]{di2009multilevel}. 

Projecting the demeaned functions, $y_{it} - \mu_i$ onto space spanned by the eigenfunctions, $\rho^{(c)}_{k}$ and $\psi^{(c)}_{l}$, respectively, we obtain
\begin{equation}\label{eq:4}
  \begin{split}
    A_{itk}^{(c)} & = \int_{0}^{1}\{y_{it}(x) - \mu_i(x)\}\rho^{(c)}_{k}(x)dx\\ 
    & = \xi_{tk}^{(c)} + \sum_{l=1}^{N_2}\zeta_{itl}^{(c)}q_{kl}^{(c)} + \upsilon_{itk}^{(c)(1)},
  \end{split}
\end{equation}
and 
\begin{equation}\label{eq:5}
  \begin{split}
    B_{itl}^{(c)} & = \int_{0}^{1}\{y_{it}(x) - \mu_i(x)\}\psi^{(c)}_{l}(x)dx\\ 
    & = \zeta_{itl}^{(c)} + \sum_{k=1}^{N_1}\xi_{tk}^{(c)}q_{kl}^{(c)} + \upsilon_{itl}^{(c)(2)},
  \end{split}
\end{equation}
where $q_{kl}^{(c)} = \int_{0}^{1}\rho^{(c)}_{k}(x)\psi^{(c)}_{l}(x)dx$, the inner product of two eigenfunctions at different levels, $\upsilon^{(c)(1)}_{itk}$ and $\upsilon_{itl}^{(c)(2)}$ are the corresponding residuals due to truncation. $A_{itk}^{(c)}$ and $B_{itl}^{(c)}$ can be estimated by numerical integration.

Let $\mathbf{A}_{it}^{(c)} = (A_{it1}^{(c)}, A_{it2}^{(c)}, \ldots, A_{itN_1}^{(c)})^\top$, $\mathbf{B}_{it}^{(c)} = (B_{it1}^{(c)}, B_{it2}^{(c)}, \ldots, B_{itN_2}^{(c)})^\top$, $\bm{\xi}_{t}^{(c)} = (\xi_{t1}^{(c)}, \xi_{t2}^{(c)},$ $\ldots, \xi_{tN_1}^{(c)})^\top$, $\bm{\zeta}_{it}^{(c)} = (\zeta_{it1}^{(c)}, \zeta_{it2}^{(c)}, \ldots, \zeta_{itN_2}^{(c)})^\top$, $\bm{\upsilon}^{(c)(1)}_{it} = (\upsilon^{(c)(1)}_{it1}, \upsilon^{(c)(1)}_{it2}, \ldots, \upsilon^{(c)(1)}_{itN_1})^\top$ and $\bm{\upsilon}^{(c)(2)}_{it} = (\upsilon^{(c)(2)}_{it1}, \upsilon^{(c)(2)}_{it2}, \ldots, $ $\upsilon^{(c)(2)}_{itN_2})^\top$. Equations~\eqref{eq:4} and~\eqref{eq:5} can be further written into matrix format
\begin{equation}\label{eq:6}
  \bm{A}_{it}^{(c)} = \bm{\xi}_{t}^{(c)} + \bm{Q\zeta}_{it}^{(c)} + \bm{\upsilon}^{(c)(1)}_{it},
\end{equation}
and 
\begin{equation}\label{eq:7}
  \bm{B}_{it}^{(c)} = \bm{\zeta}_{it}^{(c)} + \bm{C^\top\xi}_{t}^{(c)} + \bm{\upsilon}^{(c)(2)}_{it}.
\end{equation}

It is easy to see that~\eqref{eq:6} can be rewritten into multivariate linear regression model format $\bm{Y} = \bm{Z}\bm{\beta}+\bm{\epsilon}$, where $\bm{Y}$ is $\bm{A}_{it}$, $\bm{\beta}$ is $(\bm{\xi}_{t}^{(c)}, \bm{\zeta}_{it}^{(c)})^\top = (\xi_{t1}^{(c)}, \xi_{t2}^{(c)},$ $\ldots, \xi_{tN_1}^{(c)}, \zeta_{it1}^{(c)},\zeta_{it2}^{(c)},$ $\ldots, \zeta_{itN_2}^{(c)})^\top$, a vector of length $N_1 + N_{2}$ and $\bm{Z} = [\bm{I}, \bm{Q}]$, is an $N_1 \times (N_1 + N_2)$ matrix such that
\begin{equation*}
\bm{Z}= 
\left.
\begin{bmatrix}
\overmat{N_1}{1 & 0 & 0 &\dots & 0} & \overmat{N_2}{q_{11} & q_{12} & q_{13} &\dots & q_{1N_2} \\}
0 & 1 & 0 &\dots & 0 & q_{21} & q_{22} & q_{23} &\dots & q_{2N_2}  \\
0 & 0 & 1 & \dots &0 & q_{31} & q_{32} & q_{33} &\dots & q_{3N_2}  \\
\vdots &   &  & \ddots & &\vdots &   &  & \ddots \\
0 & 0 & 0 & \dots & 1 & q_{N_11} & q_{N_12} & q_{N_13} &\dots & q_{N_1N_2}\\
\end{bmatrix}
\right\}{N_1},
\end{equation*}
where $\bm{I}$ is an $N_1 \times N_1$ identity matrix and $\bm{Q}$ is an $N_1 \times N_2$ matrix.

The least-squares estimates of $\bm{\beta} = (\bm{\xi}_{t}^{(c)}, \bm{\zeta}_{it}^{(c)})^\top$ for each object $i$ can be expressed as
\begin{equation*}  \bm{\beta_{it}} = (\bm{Z}^\top\bm{Z})^{\dagger}\bm{Z}^\top\bm{A_{it}^{(c)}},
\end{equation*}
where $(\bm{Z}^\top\bm{Z})^{\dagger}$ is the Moore-Penrose generalized inverse of $\bm{Z}^\top\bm{Z}$. We use a generalized inverse to guarantee invertibility. Then, the estimate of $\bm{\xi_{t}^{(c)}}$ is the first $N_{1}$ terms of $\bm{\beta_{it}}$ and the $\bm{\zeta_{it}^{(c)}}$ is the last $N_{2}$ terms of $\bm{\beta_{it}}$. Similarly, $\bm{\xi_{t}^{(c)}}$ and $\bm{\zeta_{it}^{(c)}}$ can be calculated using~\eqref{eq:7}.

\newpage
\section*{Appendix C: Life annuity pricing}\label{appnc} 

To illustrate the impact of the forecasting improvements in Section~\ref{sec:6}, we use the mortality forecasts to price the life annuities, i.e., the amount of money that an individual pays for life insurer in return for annual payments after retirement until death. Life annuities have been one of the typical longevity insurance products for people to finance their retirements. Rapid improvements in mortality have exposed life insurers with longevity risk \citep{ngai2011longevity}. Accurate mortality forecasts could enable life insurers to manage longevity risk effectively without holding excessive levels of capital. We compare the present values of the life annuities based on mortality forecasts from different methods. The present values of the life annuities represent how much capital that the life insurer should reserve.

In the life annuity comparison, we calculate the present value of the life annuity with \$1 annual payments. More specifically, the price of a life annuity for an individual aged $x$ at year $t$ is the present value of the annual payments of \$1 that the individual receives after retirement until death or a pre-agreed age (which one occurs first). The retirement age is set to be $65$, and the pre-agreed age that the annuity terminates is assumed to be $90$ \citep{he2021data}. Then the annuity price can be calculated as: 
\begin{align*}
  PV_{x, t} = \left\{ \begin{array}{ll}
                \sum_{n=1}^{90-x} \frac{_np_{x,t}}{(1+i)^n}, & x \geq 65 \\ 
                \\
                \sum_{n=1}^{25} \frac{_np_{65,t+(65-x)}}{(1+i)^{n+(65-x)}}, &  x < 65,\\
                \end{array} \right.
\end{align*}
where $PV_{x, t}$ is the present value of the life annuity for an individual aged $x$ at year $t$, $_np_{x,t}$ is the survival probability for an individual aged $x$ at year $t$ to survive after $n$ years, and $i$ is the interest rate used for discounting. For an individual older than $65$-year-old, he/she receives payment for each year of survival, and for an individual younger than $65$-year-old, the annuity is deferred with the first payment paid out at the year that he/she survives his/her $66^{\text{th}}$ birthday. 

To compare the annuity prices of different methods, we use the mortality data of $32$ countries from the years $1960$ to $2000$ as a training dataset used for forecasting and the data from the years $2001$ to $2010$ the holdout dataset. We forecast the mortality rates for the testing data based on the training data using different methods. Then, we calculate the annuity prices, $PV_{x,t}$ using the forecasts of mortality rates from different methods as well as the holdout actual mortality rates. 

Table~\ref{tab:a1} exhibits the average prices of annuities with annual payment $\$1$ and interest rate $2\%$ for some selected ages and years. The bold entries highlight the method that produces annuity prices closest to the estimated annuity price based on the true holdout mortality. It is clear to see that all the forecasting methods tend to underestimate the annuity prices, which is a common phenomenon in actuarial studies, which corresponds to the underestimated longevity risk \citep{ngai2011longevity}. Further investigation of the annuity prices reveals that the pricing errors of the proposed method are much lower than those of the univariate functional time series forecasting approach and those of the functional panel data model with initial clustering membership. More specifically, the pricing errors of the univariate functional time series forecasting approach are around \$0.013 to \$0.136 for male and female mortality rates and \$0.11 to \$0.69 for total mortality rates per \$1 payment. The figure is around \$0.01 to \$0.1 for male and female mortality rates and \$0.005 to \$0.02 for total mortality rates for the functional panel data model with initial clustering membership. The figure is around \$0.005 to \$0.07 for male and female mortality rates and \$0.0005 to \$0.002 for total mortality rates for the functional panel data model with the proposed clustering method. Although the figures appear to be very small, the magnitudes of the underpricing of the univariate functional time series forecasting approach and the functional panel data model with initial clustering membership is around $2$ - $3$ times larger that that of the proposed method for female and male mortality rates and this magnitude is around $10$ -$20$ times larger for total mortality rates. 

{\scriptsize{
\begin{table}[!htbp] \centering 
\caption{Average annuity prices with annual payment $\$1$ and interest rate $2\%$ for some selected ages and years. The estimated annuity prices based on holdout true mortality rates are labeled as "TRUE", annuity prices based on univariate functional time series are labeled as ``UTS'', annuity prices based on the initial cluster membership are labeled as ``FPCA'' and annuity prices based on proposed methods are labeled as ``MFTSC''} 
  \label{tab:a1} 
    \setlength{\tabcolsep}{2pt}
   \renewcommand{\arraystretch}{1.2}
\begin{tabular}{@{}l|rrrr|rrrr|rrrr@{}} 
\toprule
&\multicolumn{4}{c}{Female} & \multicolumn{4}{c}{Male} & \multicolumn{4}{c}{Total}\\
\cmidrule{2-13}
(Year, Age) & TRUE & UTS &FPCA & MFTSC & TRUE & UTS&FPCA  & MFTSC & TRUE & UTS&FPCA  & MFTSC\\ 
\midrule 
1960, 40 & $7.398$ & $7.385$ & $7.389$ & $\textbf{7.394}$ & $5.155$ & $5.137$ & $5.142$ & $\textbf{5.145}$ & $6.347$ & $6.336$ & $6.342$ & $\textbf{6.347}$ \\ 
1970, 50 & $9.301$ & $9.285$ & $9.290$ & $\textbf{9.295}$ & $6.637$ & $6.614$ & $6.621$ & $\textbf{6.624}$ & $8.069$ & $8.054$ & $8.064$ & $\textbf{8.069}$ \\ 
1980, 60 & $12.126$ & $12.105$ & $12.112$ & $\textbf{12.119}$ & $9.292$ & $9.259$ & $9.269$ & $\textbf{9.272}$ & $10.865$ & $10.846$ & $10.858$ & $\textbf{10.865}$ \\ 
1990, 70 & $10.593$ & $10.563$ & $10.573$ & $\textbf{10.584}$ & $8.528$ & $8.474$ & $8.491$ & $\textbf{8.497}$ & $9.751$ & $9.721$ & $9.741$ & $\textbf{9.751}$ \\ 
2000, 80 & $5.612$ & $5.551$ & $5.573$ & $\textbf{5.593}$ & $4.702$ & $4.566$ & $4.616$ & $\textbf{4.631}$ & $5.297$ & $5.228$ & $5.272$ & $\textbf{5.295}$ \\ 
\bottomrule
\end{tabular} 
\end{table} 
}}

To illustrate the financial impact of the mispricing on life insurers, consider the annuity pricing for individuals\footnote{We consider both male and female. Hence the total mortality rates are in use.} aged $50$ at year $1970$. The pricing errors for the univariate functional time series forecasting approach and the functional panel data model with initial clustering membership are \$0.015 and \$0.05 per \$1 payment. The figure for the proposed method is \$0.0003 per \$1 payment. Suppose the annual payment for each individual is \$10,000, and $80,000$ people purchased this product. Then based on the univariate functional time series forecasting approach, the life insurer will face a \$$12$ million shortfalls ($\$0.015\times10000\times80000 =\$12$ million). The shortfall based on the functional panel data model with initial clustering membership is \$4 million, compared with a shortfall of \$$0.24$ million if the proposed forecasting method is used. To reserve for such shortfall, if the insurer could forecast the mortality rates more accurately, the capital they reserve is significantly reduced and cost.

\end{appendix}

\newpage

\bibliographystyle{agsm}
\bibliography{newref}

@Article{abraham2003unsupervised,
  author =    {Abraham, Christophe and Cornillon, Pierre-Andr{\'e} and Matzner-L{\o}ber, ERIC and Molinari, Nicolas},
  title =     {Unsupervised curve clustering using {B}-splines},
  journal =   {Scandinavian Journal of Statistics},
  year =      {2003},
  volume =    {30},
  number =    {3},
  pages =     {581--595},
  publisher = {Wiley Online Library}
}

@Article{andrews1991heteroskedasticity,
  author =  {Andrews, D},
  title =   {Heteroskedasticity and Autocorrelation Consistent Covariant Matrix Estimation},
  journal = {Econometrica},
  year =    {1991},
  volume =  {59},
  number =  {3},
  pages =   {817--858}
}

@Article{aue2015prediction,
  author =    {Aue, Alexander and Norinho, Diogo Dubart and H{\"o}rmann, Siegfried},
  title =     {On the prediction of stationary functional time series},
  journal =   {Journal of the American Statistical Association: Theory and Methods},
  year =      {2015},
  volume =    {110},
  number =    {509},
  pages =     {378--392},
  publisher = {Taylor \& Francis}
}

@Article{bai2009panel,
  author =    {Bai, Jushan},
  title =     {Panel data models with interactive fixed effects},
  journal =   {Econometrica},
  year =      {2009},
  volume =    {77},
  number =    {4},
  pages =     {1229--1279},
  publisher = {Wiley Online Library}
}

@Article{boivin2006more,
  author =    {Boivin, Jean and Ng, Serena},
  title =     {Are more data always better for factor analysis?},
  journal =   {Journal of Econometrics},
  year =      {2006},
  volume =    {132},
  number =    {1},
  pages =     {169--194},
  publisher = {Elsevier}
}

@article{booth2008mortality,
  title={Mortality modelling and forecasting: A review of methods},
  author={Booth, Heather and Tickle, Leonie},
  journal={Annals of Actuarial Science},
  volume={3},
  number={1-2},
  pages={3--43},
  year={2008},
  publisher={Cambridge University Press}
}

@article{bouveyron2007high,
  title={High-dimensional data clustering},
  author={Bouveyron, Charles and Girard, St{\'e}phane and Schmid, Cordelia},
  journal={Computational Statistics \& Data Analysis},
  volume={52},
  number={1},
  pages={502--519},
  year={2007},
  publisher={Elsevier}
}

@article{bouveyron2011model,
  title={Model-based clustering of time series in group-specific functional subspaces},
  author={Bouveyron, Charles and Jacques, Julien},
  journal={Advances in Data Analysis and Classification},
  volume={5},
  number={4},
  pages={281--300},
  year={2011},
  publisher={Springer}
}

@article{bouveyron2015discriminative,
  title={The discriminative functional mixture model for a comparative analysis of bike sharing systems},
  author={Bouveyron, Charles and C{\^o}me, Etienne and Jacques, Julien},
  journal={The Annals of Applied Statistics},
  volume={9},
  number={4},
  pages={1726--1760},
  year={2015},
  publisher={Institute of Mathematical Statistics}
}

@Article{chiou2007functional,
  author =    {Chiou, Jeng-Min and Li, Pai-Ling},
  title =     {Functional clustering and identifying substructures of longitudinal data},
  journal =   {Journal of the Royal Statistical Society: Series B},
  year =      {2007},
  volume =    {69},
  number =    {4},
  pages =     {679--699},
  publisher = {Wiley Online Library}
}

@Article{chiou2009modeling,
  author =    {Chiou, Jeng-Min and M{\"u}ller, Hans-Georg},
  title =     {Modeling hazard rates as functional data for the analysis of cohort lifetables and mortality forecasting},
  journal =   {Journal of the American Statistical Association: Applications \& Case Studies},
  year =      {2009},
  volume =    {104},
  number =    {486},
  pages =     {572--585},
  publisher = {Taylor \& Francis}
}

@Article{chiou2012dynamical,
  author =    {Chiou, Jeng-Min},
  title =     {Dynamical functional prediction and classification, with application to traffic flow prediction},
  journal =   {The Annals of Applied Statistics},
  year =      {2012},
  volume =    {6},
  number =    {4},
  pages =     {1588--1614},
  publisher = {Institute of Mathematical Statistics}
}

@article{crainiceanu2009generalized,
  title={Generalized multilevel functional regression},
  author={Crainiceanu, Ciprian M and Staicu, Ana-Maria and Di, Chong-Zhi},
  journal={Journal of the American Statistical Association: Theory and Methods},
  volume={104},
  number={488},
  pages={1550--1561},
  year={2009},
  publisher={Taylor \& Francis}
}

@article{crainiceanu2010bayesian,
  title={Bayesian functional data analysis using WinBUGS},
  author={Crainiceanu, Ciprian M and Goldsmith, A Jeffrey},
  journal={Journal of Statistical Software},
  volume={32},
  number={11},
  year={2010},
  publisher={NIH Public Access}
}

@article{currie2004smoothing,
  title={Smoothing and forecasting mortality rates},
  author={Currie, Iain D and Durban, Maria and Eilers, Paul HC},
  journal={Statistical Modelling},
  volume={4},
  number={4},
  pages={279--298},
  year={2004},
  publisher={Sage Publications Sage CA: Thousand Oaks, CA}
}

@Manual{demography19,
  title =  {demography: Forecasting Mortality, Fertility, Migration and Population Data},
  author = {Rob J Hyndman},
  year =   {2019},
  note = {Available at \url{https://CRAN.R-project.org/package=demography}, R package version 1.21}
}

@Article{di2009multilevel,
  author =    {Di, Chong-Zhi and Crainiceanu, Ciprian M and Caffo, Brian S and Punjabi, Naresh M},
  title =     {Multilevel functional principal component analysis},
  journal =   {The Annals of Applied Statistics},
  year =      {2009},
  volume =    {3},
  number =    {1},
  pages =     {458-488},
  publisher = {NIH Public Access}
}

@Book{gallant2009nonlinear,
  title =     {Nonlinear Statistical Models},
  publisher = {John Wiley \& Sons},
  year =      {2009},
  author =    {Gallant, A Ronald},
  address =   {Hoboken, New Jersey}
}

@Article{garcia2005proposal,
  author =    {Garcia-Escudero, Luis Angel and Gordaliza, Alfonso},
  title =     {A proposal for robust curve clustering},
  journal =   {Journal of Classification},
  year =      {2005},
  volume =    {22},
  number =    {2},
  pages =     {185--201},
  publisher = {Springer}
}

@book{girosi2008demographic,
  title={Demographic forecasting},
  author={Girosi, Federico and King, Gary},
  year={2008},
  publisher={Princeton University Press},
  address =   {Princeton, New Jersey}
}

@Article{gneiting2007strictly,
  author =    {Gneiting, Tilmann and Raftery, Adrian E},
  title =     {Strictly proper scoring rules, prediction, and estimation},
  journal =   {Journal of the American Statistical Association: Review Article},
  year =      {2007},
  volume =    {102},
  number =    {477},
  pages =     {359--378},
  publisher = {Taylor \& Francis}
}

@incollection{greven2011longitudinal,
  title={Longitudinal functional principal component analysis},
  author={Greven, Sonja and Crainiceanu, Ciprian and Caffo, Brian and Reich, Daniel},
  editor      = {Ferraty, Fr{\'e}d{\'e}ric},
  booktitle={Recent Advances in Functional Data Analysis and Related Topics},
  pages={149--154},
  year={2011},
  publisher={Springer Science \& Business Media},
  address={Berin, Germany}
}

@Article{hall2006assessing,
  author =    {Hall, Peter and Vial, C{\'e}line},
  title =     {Assessing the finite dimensionality of functional data},
  journal =   {Journal of the Royal Statistical Society: Series B},
  year =      {2006},
  volume =    {68},
  number =    {4},
  pages =     {689--705},
  publisher = {Wiley Online Library}
}

@Article{hansen1982large,
  author =    {Hansen, Lars Peter},
  title =     {Large sample properties of generalized method of moments estimators},
  journal =   {Econometrica: Journal of the Econometric Society},
  year =      {1982},
  volume =    {50},
  number =    {4},
  pages =     {1029--1054},
  publisher = {JSTOR}
}

@Manual{hmd,
  title  = {University of {California, Berkeley (USA), and Max Planck Institute for Demographic Research (Germany).}},
  author = {{Human Mortality Database}},
  note   = {Available at \url{http://www.mortality.org} (data downloaded on 2018-12-24)},
  year   = {2021},
}

@InCollection{hormann2012functional,
  author =    {H{\"o}rmann, Siegfried and Kokoszka, Piotr},
  title =     {Functional time series},
  booktitle = {Handbook of Statistics},
  publisher = {Elsevier},
  year =      {2012},
  editor =    {Tata Subba Rao and Suhasini Subba Rao and C.R. Rao},
  volume =    {30},
  pages =     {157--186},
  address =   {North Holland, Amsterdam}
}

@Article{hormann2015dynamic,
  author =    {H{\"o}rmann, Siegfried and Kidzi{\'n}ski, {\L}ukasz and Hallin, Marc},
  title =     {Dynamic functional principal components},
  journal =   {Journal of the Royal Statistical Society: Series B},
  year =      {2015},
  volume =    {77},
  number =    {2},
  pages =     {319--348},
  publisher = {Wiley Online Library}
}

@Article{hormann2015note,
  author =    {H{\"o}rmann, Siegfried and Kidzi{\'n}ski, {\L}ukasz},
  title =     {A note on estimation in {H}ilbertian linear models},
  journal =   {Scandinavian Journal of Statistics},
  year =      {2015},
  volume =    {42},
  number =    {1},
  pages =     {43--62},
  publisher = {Wiley Online Library}
}

@Book{horvath2012inference,
  title =     {Inference For Functional Data with Applications},
  publisher = {Springer Science \& Business Media},
  year =      {2012},
  author =    {Horv{\'a}th, Lajos and Kokoszka, Piotr},
  address =   {New York}
}

@article{horvath2013estimation,
  title={Estimation of the mean of functional time series and a two-sample problem},
  author={Horv{\'a}th, Lajos and Kokoszka, Piotr and Reeder, Ron},
  journal={Journal of the Royal Statistical Society: Series B},
  volume={75},
  number={1},
  pages={103--122},
  year={2013},
  publisher={Wiley Online Library}
}

@Book{Hsiao2014,
  title =     {Analysis of Panel Data},
  publisher = {Cambridge University Press},
  year =      {2014},
  author =    {Hsiao, Cheng},
  address =   {Cambridge}
}

@Article{hubert1985comparing,
  author =    {Hubert, Lawrence and Arabie, Phipps},
  title =     {Comparing partitions},
  journal =   {Journal of Classification},
  year =      {1985},
  volume =    {2},
  number =    {1},
  pages =     {193--218},
  publisher = {Springer}
}

@article{hyndman2007robust,
  title={Robust forecasting of mortality and fertility rates: {A} functional data approach},
  author={Hyndman, Rob J and Ullah, Md Shahid},
  journal={Computational Statistics \& Data Analysis},
  volume={51},
  number={10},
  pages={4942--4956},
  year={2007},
  publisher={Elsevier}
}

@Article{hyndman2009forecasting,
  author =    {Hyndman, Rob J and Shang, Han Lin},
  title =     {Forecasting functional time series},
  journal =   {Journal of the Korean Statistical Society},
  year =      {2009},
  volume =    {38},
  number =    {3},
  pages =     {199--221},
  publisher = {Elsevier},
  note = {(With discussion)}
}

@Article{hyndman2010rainbow,
  author =    {Hyndman, Rob J and Shang, Han Lin},
  title =     {Rainbow plots, bagplots, and boxplots for functional data},
  journal =   {Journal of Computational and Graphical Statistics},
  year =      {2010},
  volume =    {19},
  number =    {1},
  pages =     {29--45},
  publisher = {Taylor \& Francis}
}

@inproceedings{jacques2012model,
  title={Model-based clustering of functional data},
  author={Jacques, Julien and Preda, Cristian},
  booktitle={20th European Symposium on Artificial Neural Networks, Computational Intelligence and Machine Learning. Bruges},
  pages={459--464},
  year={2012}
}

@Article{jacques2014functional,
  author =    {Jacques, Julien and Preda, Cristian},
  title =     {Functional data clustering: {A} survey},
  journal =   {Advances in Data Analysis and Classification},
  year =      {2014},
  volume =    {8},
  number =    {3},
  pages =     {231--255},
  publisher = {Springer}
}

@article{jacques2014model,
  title={Model-based clustering for multivariate functional data},
  author={Jacques, Julien and Preda, Cristian},
  journal={Computational Statistics \& Data Analysis},
  volume={71},
  pages={92--106},
  year={2014},
  publisher={Elsevier}
}

@Article{karhunen1946spektraltheorie,
  author =  {Karhunen, Kari},
  title =   {Zur spektraltheorie stochastischer prozesse},
  journal = {Annales Academiae Scientiarum Fennicae. Series A I},
  year =    {1946},
  volume =  {34}
}

@Article{kodinariya2013review,
  author =  {Kodinariya, Trupti M and Makwana, Prashant R},
  title =   {Review on determining number of cluster in {k}-Means clustering},
  journal = {International Journal of Advance Research in Computer Science and Management Studies},
  year =    {2013},
  volume =  {1},
  number =  {6},
  pages =   {90--95}
}

@article{lee1992modeling,
  title={Modeling and forecasting US mortality},
  author={Lee, Ronald D and Carter, Lawrence R},
  journal={Journal of the American Statistical Association: Application \& Case Studies},
  volume={87},
  number={419},
  pages={659--671},
  year={1992},
  publisher={Taylor \& Francis}
}

@Article{li2005coherent,
  author =    {Li, Nan and Lee, Ronald},
  title =     {Coherent mortality forecasts for a group of populations: {A}n extension of the Lee-Carter method},
  journal =   {Demography},
  year =      {2005},
  volume =    {42},
  number =    {3},
  pages =     {575--594},
  publisher = {Springer}
}

@article{li2013extending,
  title={Extending the Lee-Carter method to model the rotation of age patterns of mortality decline for long-term projections},
  author={Li, Nan and Lee, Ronald and Gerland, Patrick},
  journal={Demography},
  volume={50},
  number={6},
  pages={2037--2051},
  year={2013},
  publisher={Springer}
}

@Article{li2013poisson,
  author =    {Li, Jackie},
  title =     {A {P}oisson common factor model for projecting mortality and life expectancy jointly for females and males},
  journal =   {Population Studies},
  year =      {2013},
  volume =    {67},
  number =    {1},
  pages =     {111--126},
  publisher = {Taylor \& Francis}
}

@Article{li2013selecting,
  author =    {Li, Yehua and Wang, Naisyin and Carroll, Raymond J},
  title =     {Selecting the number of principal components in functional data},
  journal =   {Journal of the American Statistical Association: Theory and Methods},
  year =      {2013},
  volume =    {108},
  number =    {504},
  pages =     {1284--1294},
  publisher = {Taylor \& Francis}
}

@Book{loeve1955probability,
  title =     {Probability Theory: Foundations, Random Sequences},
  publisher = {van Nostrand Princeton, New Jersey},
  year =      {1955},
  author =    {Lo{\`e}ve, Michel}
}

@Article{lopez2009concept,
  author =    {L{\'o}pez-Pintado, Sara and Romo, Juan},
  title =     {On the concept of depth for functional data},
  journal =   {Journal of the American Statistical Association: Theory and Methods},
  year =      {2009},
  volume =    {104},
  number =    {486},
  pages =     {718--734},
  publisher = {Taylor \& Francis}
}

@InCollection{macqueen1967,
  author =       {MacQueen, James and others},
  title =        {Some methods for classification and analysis of multivariate observations},
  booktitle =    {Proceedings of the Fifth Berkeley Symposium on Mathematical Statistics and Probability},
  publisher =    {University of California Press},
  year =         {1967},
  volume =       {1},
  pages =        {281--297},
  address =      {Berkeley, California},
  organization = {Oakland, CA, USA.}
}

@Article{muller2005functional,
  author =    {M{\"u}ller, HANS-GEORG},
  title =     {Functional modelling and classification of longitudinal data},
  journal =   {Scandinavian Journal of Statistics},
  year =      {2005},
  volume =    {32},
  number =    {2},
  pages =     {223--240},
  publisher = {Wiley Online Library}
}

@Article{newey1986simple,
  author =    {Newey, Whitney K and West, Kenneth D},
  title =     {A simple, positive semi-definite, heteroskedasticity and autocorrelation consistent covariance matrix},
  journal =   {Econometrica},
  year =      {1987},
  volume =    {55},
  number =    {3},
  pages =     {703--708},
  publisher = {National Bureau of Economic Research Cambridge, Mass., USA}
}

@Article{pampel2005forecasting,
  author =    {Pampel, Fred},
  title =     {Forecasting sex differences in mortality in high income nations: {T}he contribution of smoking},
  journal =   {Demographic Research},
  year =      {2005},
  volume =    {13},
  number =    {18},
  pages =     {455-484},
  publisher = {NIH Public Access}
}

@article{panaretos2013fourier,
  title={Fourier analysis of stationary time series in function space},
  author={Panaretos, Victor M and Tavakoli, Shahin},
  journal={The Annals of Statistics},
  volume={41},
  number={2},
  pages={568--603},
  year={2013},
  publisher={Institute of Mathematical Statistics}
}

@Article{politis1996flat,
  author =    {Politis, Dimitris N and Romano, Joseph P},
  title =     {On flat-top kernel spectral density estimators for homogeneous random fields},
  journal =   {Journal of Statistical Planning and Inference},
  year =      {1996},
  volume =    {51},
  number =    {1},
  pages =     {41--53},
  publisher = {Elsevier}
}

@Article{politis1999multivariate,
  author =    {Politis, Dimitris N and Romano, Joseph P},
  title =     {Multivariate density estimation with general flat-top kernels of infinite order},
  journal =   {Journal of Multivariate Analysis},
  year =      {1999},
  volume =    {68},
  number =    {1},
  pages =     {1--25},
  publisher = {Elsevier}
}

@article{renshaw2003lee,
  title={Lee--Carter mortality forecasting with age-specific enhancement},
  author={Renshaw, Arthur E and Haberman, Steven},
  journal={Insurance: Mathematics and Economics},
  volume={33},
  number={2},
  pages={255--272},
  year={2003},
  publisher={Elsevier}
}

@Article{rice1991estimating,
  author =    {Rice, John A and Silverman, Bernard W},
  title =     {Estimating the mean and covariance structure nonparametrically when the data are curves},
  journal =   {Journal of the Royal Statistical Society: Series B},
  year =      {1991},
  volume =    {53},
  number =    {1},
  pages =     {233--243},
  publisher = {JSTOR}
}

@Article{rice2017plug,
  author =    {Rice, Gregory and Shang, Han Lin},
  title =     {A plug-in bandwidth selection procedure for long-run covariance estimation with stationary functional time series},
  journal =   {Journal of Time Series Analysis},
  year =      {2017},
  volume =    {38},
  number =    {4},
  pages =     {591--609},
  publisher = {Wiley Online Library}
}

@Article{serban2005cats,
  author =    {Serban, Nicoleta and Wasserman, Larry},
  title =     {{CATS}: {C}lustering after transformation and smoothing},
  journal =   {Journal of the American Statistical Association: Theory and Methods},
  year =      {2005},
  volume =    {100},
  number =    {471},
  pages =     {990--999},
  publisher = {Taylor \& Francis}
}

@Article{serban2012multilevel,
  author =    {Serban, Nicoleta and Jiang, Huijing},
  title =     {Multilevel functional clustering analysis},
  journal =   {Biometrics},
  year =      {2012},
  volume =    {68},
  number =    {3},
  pages =     {805--814},
  publisher = {Wiley Online Library}
}

@Article{shang2016mortality,
  author =    {Shang, Han Lin},
  title =     {Mortality and life expectancy forecasting for a group of populations in developed countries: {A} multilevel functional data method},
  journal =   {The Annals of Applied Statistics},
  year =      {2016},
  volume =    {10},
  number =    {3},
  pages =     {1639--1672},
  publisher = {Institute of Mathematical Statistics}
}

@Article{shang2018bootstrap,
  author =    {Shang, Han Lin},
  title =     {Bootstrap methods for stationary functional time series},
  journal =   {Statistics and Computing},
  year =      {2018},
  volume =    {28},
  number =    {1},
  pages =     {1--10},
  publisher = {Springer}
}

@article{slimen2018model,
  title={Model-based co-clustering for functional data},
  author={Slimen, Yosra Ben and Allio, Sylvain and Jacques, Julien},
  journal={Neurocomputing},
  volume={291},
  pages={97--108},
  year={2018},
  publisher={Elsevier}
}

@Article{sugar2003finding,
  author =    {Sugar, Catherine A and James, Gareth M},
  title =     {Finding the number of clusters in a dataset: {A}n information-theoretic approach},
  journal =   {Journal of the American Statistical Association: Theory and Methods},
  year =      {2003},
  volume =    {98},
  number =    {463},
  pages =     {750--763},
  publisher = {Taylor \& Francis}
}

@Article{tarpey2003clustering,
  author =    {Tarpey, Thaddeus and Kinateder, Kimberly KJ},
  title =     {Clustering functional data},
  journal =   {Journal of Classification},
  year =      {2003},
  volume =    {20},
  number =    {1},
  pages =     {93--114},
  publisher = {Springer}
}

@Article{ward1963hierarchical,
  author =    {Ward Jr, Joe H},
  title =     {Hierarchical grouping to optimize an objective function},
  journal =   {Journal of the American Statistical Association},
  year =      {1963},
  volume =    {58},
  number =    {301},
  pages =     {236--244},
  publisher = {Taylor \& Francis Group}
}

@Book{white2014asymptotic,
  title =     {Asymptotic Theory for Econometricians},
  publisher = {Academic press},
  year =      {1984},
  author =    {White, Halbert},
  address =   {Cambridge, Massachusetts}
}

@article{wisniowski2015bayesian,
  title={Bayesian population forecasting: Extending the Lee-Carter method},
  author={Wi{\'s}niowski, Arkadiusz and Smith, Peter WF and Bijak, Jakub and Raymer, James and Forster, Jonathan J},
  journal={Demography},
  volume={52},
  number={3},
  pages={1035--1059},
  year={2015},
  publisher={Springer}
}

@Book{Wooldridge2010,
  title =     {Econometric Analysis of Cross Section and Panel Data},
  publisher = {The MIT Press},
  year =      {2010},
  author =    {Wooldridge, Jeffrey M.},
  address =   {Cambridge, Massachusetts}
}

@Article{yao2005functional,
  author =    {Yao, Fang and M{\"u}ller, Hans-Georg and Wang, Jane-Ling},
  title =     {Functional data analysis for sparse longitudinal data},
  journal =   {Journal of the American Statistical Association: Theory and Methods},
  year =      {2005},
  volume =    {100},
  number =    {470},
  pages =     {577--590},
  publisher = {Taylor \& Francis}
}

@article{ngai2011longevity,
  title={Longevity risk management for life and variable annuities: The effectiveness of static hedging using longevity bonds and derivatives},
  author={Ngai, Andrew and Sherris, Michael},
  journal={Insurance: Mathematics and Economics},
  volume={49},
  number={1},
  pages={100--114},
  year={2011},
  publisher={Elsevier}
}

@article{he2021data,
  title={Data-adaptive dimension reduction for US mortality forecasting},
  author={He, Lingyu and Huang, Fei and Yang, Yanrong},
  journal={arXiv preprint arXiv:2102.04123},
  year={2021},
  url =     {\url{https://arxiv.org/abs/2102.04123}},
}

@Manual{HS21,
  title  = {{ftsa: Functional Time Series Analysis}},
  author = {R. J. Hyndman and H. L. Shang},
  note = {Available at \url{https://CRAN.R-project.org/package=ftsa}, R package version 6.1},
  year   = {2021},
}

@Manual{Tsay21,
  title =  {MTS: All-Purpose Toolkit for Analyzing Multivariate Time Series and Estimating Multivariate Volatility Models},
  author = {Ruey S. Tsay and David Wood},
  year =   {2021},
  note = {Available at \url{ https://CRAN.R-project.org/package=MTS}, R package version 1.03},
}

@article{reiss2007functional,
  title={Functional principal component regression and functional partial least squares},
  author={Reiss, Philip T and Ogden, R Todd},
  journal={Journal of the American Statistical Association: Theory and Methods},
  volume={102},
  number={479},
  pages={984--996},
  year={2007},
  publisher={Taylor \& Francis}
}

\end{document}